\PassOptionsToPackage{unicode}{hyperref}
\PassOptionsToPackage{hyphens}{url}
\PassOptionsToPackage{dvipsnames,svgnames,x11names}{xcolor}
\documentclass[
  12pt]{article}

\usepackage{amsmath,amssymb}
\usepackage{iftex}
\ifPDFTeX
  \usepackage[T1]{fontenc}
  \usepackage[utf8]{inputenc}
  \usepackage{textcomp} 
\else 
  \usepackage{unicode-math}
  \defaultfontfeatures{Scale=MatchLowercase}
  \defaultfontfeatures[\rmfamily]{Ligatures=TeX,Scale=1}
\fi
\usepackage{lmodern}
\ifPDFTeX\else  
\fi
\IfFileExists{upquote.sty}{\usepackage{upquote}}{}
\IfFileExists{microtype.sty}{
  \usepackage[]{microtype}
  \UseMicrotypeSet[protrusion]{basicmath} 
}{}
\makeatletter
\@ifundefined{KOMAClassName}{
  \IfFileExists{parskip.sty}{%
    \usepackage{parskip}
  }{
    \setlength{\parindent}{0pt}
    \setlength{\parskip}{6pt plus 2pt minus 1pt}}
}{
  \KOMAoptions{parskip=half}}
\makeatother
\usepackage{xcolor}
\makeatletter
\ifx\paragraph\undefined\else
  \let\oldparagraph\paragraph
  \renewcommand{\paragraph}{
    \@ifstar
      \xxxParagraphStar
      \xxxParagraphNoStar
  }
  \newcommand{\xxxParagraphStar}[1]{\oldparagraph*{#1}\mbox{}}
  \newcommand{\xxxParagraphNoStar}[1]{\oldparagraph{#1}\mbox{}}
\fi
\ifx\subparagraph\undefined\else
  \let\oldsubparagraph\subparagraph
  \renewcommand{\subparagraph}{
    \@ifstar
      \xxxSubParagraphStar
      \xxxSubParagraphNoStar
  }
  \newcommand{\xxxSubParagraphStar}[1]{\oldsubparagraph*{#1}\mbox{}}
  \newcommand{\xxxSubParagraphNoStar}[1]{\oldsubparagraph{#1}\mbox{}}
\fi
\makeatother

\usepackage{longtable,booktabs,array}
\usepackage{calc} 
\usepackage{etoolbox}
\makeatletter
\patchcmd\longtable{\par}{\if@noskipsec\mbox{}\fi\par}{}{}
\makeatother
\IfFileExists{footnotehyper.sty}{\usepackage{footnotehyper}}{\usepackage{footnote}}
\makesavenoteenv{longtable}
\usepackage{graphicx}
\makeatletter
\def\maxwidth{\ifdim\Gin@nat@width>\linewidth\linewidth\else\Gin@nat@width\fi}
\def\maxheight{\ifdim\Gin@nat@height>\textheight\textheight\else\Gin@nat@height\fi}
\makeatother
\setkeys{Gin}{width=\maxwidth,height=\maxheight,keepaspectratio}
\makeatletter
\def\fps@figure{htbp}
\makeatother

\makeatletter
\@ifpackageloaded{caption}{}{\usepackage{caption}}
\AtBeginDocument{%
\ifdefined\contentsname
  \renewcommand*\contentsname{Table of contents}
\else
  \newcommand\contentsname{Table of contents}
\fi
\ifdefined\listfigurename
  \renewcommand*\listfigurename{List of Figures}
\else
  \newcommand\listfigurename{List of Figures}
\fi
\ifdefined\listtablename
  \renewcommand*\listtablename{List of Tables}
\else
  \newcommand\listtablename{List of Tables}
\fi
\ifdefined\figurename
  \renewcommand*\figurename{Figure}
\else
  \newcommand\figurename{Figure}
\fi
\ifdefined\tablename
  \renewcommand*\tablename{Table}
\else
  \newcommand\tablename{Table}
\fi
}
\@ifpackageloaded{float}{}{\usepackage{float}}
\floatstyle{ruled}
\@ifundefined{c@chapter}{\newfloat{codelisting}{h}{lop}}{\newfloat{codelisting}{h}{lop}[chapter]}
\floatname{codelisting}{Listing}

\makeatother
\makeatletter
\@ifpackageloaded{caption}{}{\usepackage{caption}}
\@ifpackageloaded{subcaption}{}{\usepackage{subcaption}}
\makeatother

\ifLuaTeX
  \usepackage{selnolig}  
\fi
\usepackage[numbers]{natbib}
\usepackage{bookmark}

\AtBeginEnvironment{table}{\spacingset{1}}

\usepackage{amsmath,amsthm,amssymb}
\usepackage{placeins}
\usepackage{float}
\usepackage{enumerate}
\usepackage{graphicx}
\usepackage{xcolor}
\usepackage{array, booktabs, longtable, multirow}

\newtheorem{assumption}{Assumption}
\newtheorem{result}{Result}

\usepackage[font=small,labelfont=bf,labelsep=period]{caption}
\usepackage{authblk}
\usepackage{tikz}
\usetikzlibrary{arrows.meta, positioning}

\usepackage{etoolbox}

\newcommand{\indep}{\perp \!\!\! \perp}

\IfFileExists{xurl.sty}{\usepackage{xurl}}{} 
\hypersetup{
  pdftitle={Title},
  pdfauthor={Author 1; Author 2},
  pdfkeywords={3 to 6 keywords, that do not appear in the title},
  colorlinks=true,
  linkcolor={blue},
  filecolor={Maroon},
  citecolor={Blue},
  urlcolor={Blue},
  pdfcreator={LaTeX via pandoc}}

\makeatletter
\newcommand{\printappendixtoc}{%
  \section*{Appendix Contents}
  \begingroup
    \setcounter{tocdepth}{2}
    \@starttoc{apc}
  \endgroup
}
\makeatother

\newcommand{\appsection}[1]{%
  \section{#1}%
  \addcontentsline{apc}{section}{\protect\numberline{\thesection}#1}%
}
\newcommand{\appsubsection}[1]{%
  \subsection{#1}%
  \addcontentsline{apc}{subsection}{\protect\numberline{\thesubsection}#1}%
}

\newcommand{\anon}{1}

\begin{document}

\def\spacingset#1{\renewcommand{\baselinestretch}%
{#1}\small\normalsize} \spacingset{1}


\if1\anon
{
  \title{\bf Regression-Based Proximal Reconciliation of Conflicting Trials with Unmeasured Effect Modifiers}
  \author[1]{Daniel Xu}
\author[1]{Eric Tchetgen Tchetgen}
\author[1]{Enrique F Schisterman}
\author[2]{Sean C Blackwell}
\author[1]{Ellen C Caniglia \thanks{
    This work was supported by a PCORI grant.}} 

\affil[1]{University of Pennsylvania, Perelman School of Medicine, Department of Biostatistics, Epidemiology, and Informatics}
\affil[2]{UTHealth Houston McGovern Medical School}
  \maketitle
} \fi

\if0\anon
{
  \bigskip
  \bigskip
  \bigskip
  \begin{center}
    {\LARGE\bf Regression-Based Proximal Reconciliation of Conflicting Trials with Unmeasured Effect Modifiers}
\end{center}
  \medskip
} \fi

\bigskip
\begin{abstract}
    Randomized controlled trials with similar protocols may yield conflicting findings when the distribution of relevant effect modifiers differs across study populations. Yet no formal statistical framework exists for defining and assessing whether conflicting trials are reconcilable, despite the importance of this question for evidence synthesis and regulatory decision making. To address this gap, we develop a causal inference framework for evaluating conditional and marginal reconcilability on additive and multiplicative scales in the presence of unmeasured effect modifiers. Within this framework, we use proxy variables for hypothesized unmeasured effect modifiers to develop regression-based tests of conditional reconcilability under parametric structural models. To assess marginal reconcilability, we extend existing transportability methods and develop an equivalence testing framework. We also introduce a reconciliation proportion to quantify the degree of marginal reconciliation. We illustrate these methods using the conflicting Meis and PROLONG trials of 17-alpha-hydroxyprogesterone caproate for preventing recurrent preterm birth. The analyses provided limited evidence that unmeasured effect modifiers such as cervical length, as captured by the selected proxies, were sufficient to marginally reconcile the trials. These findings demonstrate how proximal reconciliation methods may help regulators, researchers, and clinicians evaluate whether differences in study populations explain conflicting trial findings.
\end{abstract}

\noindent%
\vfill

\newpage
\spacingset{1.8} 

\section{Introduction}
When randomized controlled trials (RCTs) with similar protocols yield conflicting results, it becomes essential for regulators and clinicians to understand the reasons for the discrepancy. An illustrative example is the evaluation of 17-alpha-hydroxyprogesterone caproate (17OHP-C) for preventing recurrent preterm birth (PTB), defined as delivery before 37 weeks’ gestation. An initial RCT by Meis et al. reported a clinically meaningful reduction in recurrent PTB with 17OHP-C compared to placebo (risk difference (RD) $-18.6\%$; 95\% confidence interval (CI): $-28.2\%$, $-9.2\%$) \citep{meis2003prevention}. However, a subsequent confirmatory trial -- Progestin’s Role in Optimizing Neonatal Gestation (PROLONG) -- found no evidence of benefit (RD $1.2\%$; 95\% CI: $-3.0\%$, $5.3\%$). The results from PROLONG, together with findings from other prospective cohort studies, led to the market withdrawal of 17OHP-C, the only approved pharmacologic intervention for preventing recurrent PTB \citep{blackwell202017}. Given the substantial burden of PTB, understanding these seemingly conflicting results is of clear public health importance. More broadly, this case underscores the need for principled statistical methods to reconcile conflicting evidence from RCTs.

Although there are several potential reasons for conflicting trial results, including type I or II error, differential loss to follow-up, or lack of adherence, we focus on differences in the underlying study populations. For the Meis and PROLONG trials, a leading hypothesis for the conflicting results is that participants enrolled in Meis had a higher baseline risk of recurrent PTB than those in PROLONG \citep{virkud2025clarifying}. Compared with PROLONG, the Meis cohort had a higher prevalence of established risk factors for recurrent PTB, such as a greater number of prior PTBs and a lower gestational age at prior PTB. If 17OHP-C were more effective among individuals at elevated baseline risk, then the protective effect observed in Meis and the null effect observed in PROLONG could arise solely from differences in the underlying study populations. In other words, variation in the distribution of effect modifiers of 17OHP-C on recurrent PTB could account for conflicting marginal treatment effects, even if the causal effect of 17OHP-C conditional on those modifiers were identical in both trials. Thus, even if the marginal treatment effects in the two trials truly differ, this alone does not necessarily imply that the trials are irreconcilable.

A key conceptual challenge is how to formalize this notion of reconcilability from a statistical perspective, which has not been explicitly considered in the literature. In this work, we propose two related notions of reconcilability, both adapted from the transportability literature. The first is the notion of \textit{conditional reconcilability}, which states that treatment effects (on some scale) conditional on relevant effect modifiers are the same across trials. In other words, participants with the same relevant characteristics would experience the same treatment effect, regardless of which trial they participated in. This condition is identical to the transportability in effect measure assumption in the transportability literature and forms the basis of methods developed to transport treatment effect estimates from an RCT to an external target population, with limited individual-level data in the target population \citep{dahabreh2020extending, dahabreh2020toward, dahabreh2024learning}. Whereas this assumption is taken to hold a priori in the transportability literature, in contrast, our goal is to directly evaluate its credibility given availability of individual-level data in both study samples.

The second notion, \textit{marginal reconcilability}, examines agreement at the level of marginal effects. However, rather than simply comparing the marginal treatment effects across the two trials directly, marginal reconcilability asks whether the marginal treatment effect observed in a given trial can be recovered by transporting the conditional causal effect from the other trial and marginalizing it over the target population. Clearly, marginal reconcilability is weaker than conditional reconcilability. Conditional reconcilability implies marginal reconcilability, but not conversely, because differences in conditional effects may average out when marginalized over a common target population distribution. Whether marginal or conditional reconcilability is the more scientifically relevant target depends on the substantive question of interest. Marginal reconcilability is most relevant when the goal is to determine whether differences in observed (marginal) trial results can be explained by differences in study populations, whereas conditional reconcilability is more appropriate when the goal is to determine whether individuals with the same characteristics would experience the same treatment effect regardless of which trial they participated in.

Previous work by \citet{virkud2025clarifying} has attempted to reconcile these conflicting RCT results by leveraging existing transportability methods to test the null hypothesis of marginal reconcilability. Because marginal reconcilability is implied by conditional reconcilability, the proposed test is also a valid test for the null hypothesis of conditional reconcilability. However, the aforementioned approach has several important limitations. First, it only accounts for differences in measured covariates between trials. If unobserved effect modifiers differ in distribution across studies, then conditional or marginal reconcilability based only on measured effect modifiers is unlikely to hold. In the context of the Meis and PROLONG trials, hypothesized unobserved effect modifiers that may have differed between study populations include cervical length, access to and utilization of quality care, experiences of racism and discrimination, and changes to care practices and 17OHP-C availability.

Second, transportability methods may be significantly underpowered if the primary scientific question concerns conditional reconcilability. Previous work applying these methods to the Meis and PROLONG trials showed that while the transported marginal effect estimates (e.g., from Meis to PROLONG) differed substantially from untransported marginal effect estimates (e.g., in PROLONG), confidence intervals for the difference between transported and untransported estimates were wide and included the null \citep{virkud2025clarifying}. More generally, transportability-based approaches to testing conditional reconcilability would be expected to have limited power, as such procedures reduce the problem of testing equality between two conditional causal effect functions over a rich set of effect modifiers, to a comparison of one-dimensional marginal functionals. Many data-generating processes can have meaningfully different conditional effects but similar or identical transported and untransported marginal effects. Developing more powerful procedures for testing the null hypothesis of conditional reconcilability, rather than marginal reconcilability, is therefore of significant interest.

Third, prior work has focused exclusively on testing the null hypothesis that trials are reconcilable. While rejecting this null hypothesis provides evidence against reconcilability, failing to reject the null hypothesis does not provide evidence in favor of (marginal or conditional) reconcilability and may instead reflect an underpowered test \citep{altman1995statistics, imai2008misunderstandings, wasserstein2016asa}. More broadly, a binary determination of whether two trials are or are not reconcilable may oversimplify the underlying scientific question. In many settings, the extent or degree to which conflicting trial results can be explained by differing effect modifier distributions may itself be of substantive scientific interest.

In this paper, we propose a framework for reconciling RCTs on the basis of both measured and unmeasured effect modifiers. We address each of the limitations described above. First, we leverage proximal causal methods to capture the role of unmeasured effect modifiers in explaining conflicting trial results. Proximal causal methods were originally developed for estimating causal effects in the presence of unmeasured confounding by leveraging proxies of the proposed unmeasured confounder \citep{miao2018identifying, cui2024semiparametric, tchetgen2024introduction}. Recent work has extended these ideas to transportability in nonparametric settings where the distribution of unmeasured effect modifiers may differ across populations \citep{su2025proximal, nilsson2023proxy}. In this setting, the transportability assumption underlying conditional reconcilability requires equality of treatment effects conditional on both measured and unmeasured modifiers.

Although these proximal transportability methods could easily be adapted to test the marginal reconcilability null, as discussed above, they would be substantially underpowered if interest lies primarily in testing the null hypothesis of conditional reconcilability. Therefore, we additionally propose a more powerful approach for testing whether causal effects conditional on both measured covariates and selected hypothesized unmeasured effect modifiers are equal across trials. Following \citet{liu2025regression}, we adopt generalized linear models for the proxies and outcome. Under these structural models, assessing conditional reconcilability reduces to testing whether specific model coefficients are equal. We develop tests for conditional reconcilability on both the additive and multiplicative scales, corresponding to the conditional average treatment effect (CATE) and conditional relative risk (CRR), respectively. The latter is particularly relevant for trials with binary outcomes, and to the best of our knowledge, has not yet been formally studied in the context of reconciliation of RCTs or in the proximal transportability literature.

Lastly, to construct a statistical test for which rejection of the null hypothesis provides evidence in favor of reconcilability, we reframe the problem as an equivalence test. Equivalence tests have traditionally been used in clinical trials to establish that one treatment is sufficiently similar to another, e.g., that a generic formulation has comparable efficacy to an existing therapeutic \citep{schuirmann1987comparison, westlake1972use}. Equivalence tests require prespecifying an equivalence margin, $\delta$, which represents the smallest meaningful effect size of interest. In contrast to the null hypothesis in a standard null-hypothesis significance test (NHST), the null in an equivalence test is that the contrast of interest, more specifically the difference in causal effects we wish to establish as equivalent, is at least $\delta$ in magnitude. Rejection of this null therefore provides evidence that the two causal effects are equivalent up to a margin $\delta$, and consequently that the trials themselves are marginally reconcilable up to that margin.

Because equivalence tests have traditionally been studied at the marginal level, we similarly propose an equivalence test for which rejection provides evidence in favor of marginal reconcilability. Doing so circumvents several challenges associated with formulating the equivalence test directly in terms of the underlying conditional treatment effect functions, including difficulties in interpreting meaningful equivalence margins and the need for stronger assumptions than those proposed in this paper to account for unmeasured covariates \citep{liu2009assessing, gsteiger2011simultaneous, dette2018equivalence, mollenhoff2020equivalence, mollenhoff2024testing, hagemann2025overcoming}. To further enhance interpretation, we introduce the \textit{reconciliation proportion} to quantify the degree to which two trials are marginally reconcilable, analogous to the mediation proportion \citep{mackinnon1994analysis, ditlevsen2005mediation, nevo2017estimation} or the proportion of treatment effect explained by surrogates \citep{freedman1992statistical, wang2002measure, parast2016robust}.

The remainder of this paper is organized as follows. Section \ref{s:notation} formalizes the setup and notation. Section \ref{s:hypothesis} develops hypothesis tests for the null hypothesis that two trials are conditionally or marginally reconcilable on the additive or multiplicative scale in the presence of unmeasured effect modifiers. In Section \ref{s:equiv}, we extend the proposed proximal reconciliation framework by introducing equivalence tests and the reconciliation proportion. Section \ref{s:application} applies the methods to the conflicting Meis and PROLONG trials, and Section \ref{s:discussion} concludes. Proofs for the main results are presented in Appendix \ref{app:proofs}, and additional supporting materials (including simulations) are contained in Appendix \ref{app:additional}.

\section{Notation and setup}
\label{s:notation}
Let $Y$ denote the primary outcome of interest, $A \in \{0,1\}$ an indicator of treatment assignment, $X$ a set of measured baseline covariates, $U$ a set of unmeasured baseline covariates known to be predictive of the outcome and to potentially modify the treatment effect on a given scale (additive vs. multiplicative), and $S \in \{0,1\}$ a binary indicator denoting which RCT a patient participated in. Additionally, assume that we observe a pair of proxies $(Z,W)$, where $Z$ denotes the reweighting proxy and $W$ the adjustment proxy, following terminology introduced in \citep{su2025proximal}. In total, the full data consist of $n$ independent and identically distributed samples $(Y,A,X,Z,W,U,S)$. For clarity of exposition, we assume that all variables are scalars, though the proposed methods could immediately be extended to multivariate $X$ and $Z$.

Let $Y(a)$ denote the potential outcome that would be observed if, possibly contrary to fact, an individual were assigned treatment $A = a$. We make the following standard causal assumptions.
\begin{assumption} 
\label{assump:causal} (Causal consistency, exchangeability, and positivity). 
    \begin{enumerate}[(i)]
        \item (Consistency) $Y(a) = Y$ whenever $A = a \in \{0,1\}$
        \item (Conditional Exchangeability) $\{Y(1),Y(0),U\} \indep A \mid (X, S = s)$ for $s \in \{0,1\}$
        \item (Positivity of Treatment Assignment) $P(A = a \mid X = x, S = s) >0$ for $a \in \{0,1\}$, $s \in \{0,1\}$ and each $x$ with positive density $f_{X,S}(x,s) > 0$
        \item (Positivity of Trial Participation) $P(S = s \mid X = x) >0$ for $s \in \{0,1\}$ and each $x$ with positive density $f_{X}(x) > 0$
    \end{enumerate}
\end{assumption}

Assumption \ref{assump:causal} is standard in the causal inference literature. Assumption \ref{assump:causal}(i) states that the observed outcome corresponds to the potential outcome under the assigned treatment. Assumption \ref{assump:causal}(ii) states that the effect of treatment is unconfounded given $(X,S=s)$, which holds by design in randomized or conditionally randomized trials; the inclusion of $U$ in the independence statement reflects that these unmeasured baseline covariates are also balanced by randomization. Assumption \ref{assump:causal}(iii) is the standard treatment positivity assumption in each population and would be expected to hold by design in a randomized trial. Assumption \ref{assump:causal}(iv) is an extension of the standard positivity of trial participation in the transportability literature to both populations \citep{dahabreh2020extending}. Under the parametric models introduced in Section \ref{s:hypothesis}, these positivity conditions are not strictly necessary, but violations require identification to rely partly on model-based extrapolation.

We first review the assumptions governing the relationships between the proxies and other variables. The following assumptions are used by \citet{su2025proximal} to nonparametrically identify transported causal effects in the presence of unmeasured effect modifiers and would similarly be required to develop tests of conditional and marginal reconcilability under nonparametric models.
\begin{assumption}
\label{assump:exchangeability} (Proxy structure). We assume that 
    \begin{enumerate}[(i)]
        \item $(Z,W,U) \indep A \mid (X,S=s)$
        \item $Z \indep W \mid (X,U,S=s)$
        \item $Z \indep Y \mid (A,X,U,S=s)$
        \item $W \indep S \mid (X,U)$
    \end{enumerate}
\end{assumption}
Assumption \ref{assump:exchangeability} formalizes the role of the proxy variables. Assumption \ref{assump:exchangeability}(i) states that treatment assignment is independent of unmeasured baseline covariates and measured proxies, conditional on $(X,S=s)$. This assumption will hold by design in (conditionally) randomized trials, as long as the proxies are measured at baseline. If proxies are not measured at baseline, then this assumption would generally require that treatment does not causally impact the proxies. Assumption \ref{assump:exchangeability}(ii) requires that the reweighting proxy $Z$ and adjustment proxy $W$ share no common causes other than $(X,U)$ and that neither proxy has a causal effect on the other. Assumption \ref{assump:exchangeability}(iii) requires $Z$ to have no causal effect on $Y$ and to be associated with $Y$ only through $(X,U)$. Assumption \ref{assump:exchangeability}(iv) requires that differences in the distribution of $W$ across trials are fully explained by $(X,U)$. We discuss whether these assumptions are likely to hold in the context of the Meis and PROLONG trials in Section \ref{s:application}.

\begin{figure}[ht]
\caption{The following DAG (a) and interaction DAG (b) would be compatible with Assumption \ref{assump:exchangeability} \citep{nilsson2021directed, nilsson2023proxy} under the null hypothesis of conditional reconcilability. $\Delta Y(a)$ denotes the causal effect of $A$ on $Y$ on a given scale.}
\centering
\label{fig:DAGs}
\includegraphics[width=.8\textwidth]{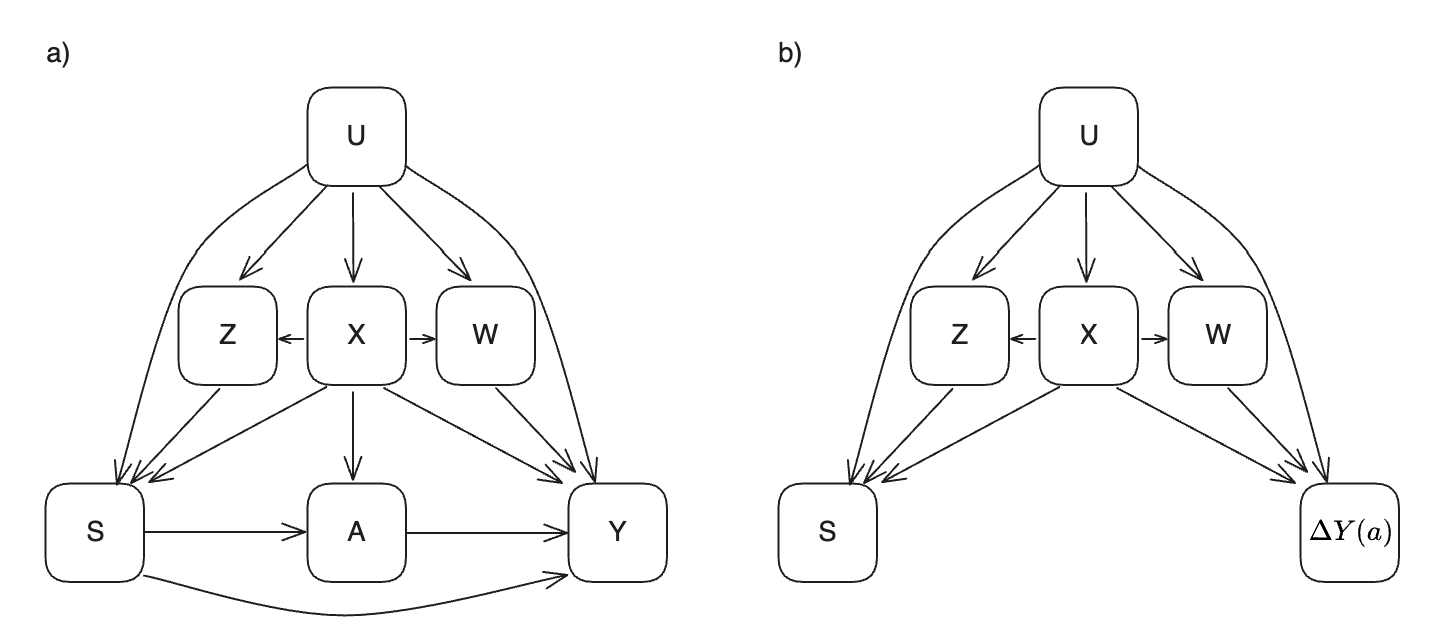}
\end{figure}

Figure \ref{fig:DAGs}(a) depicts one directed acyclic graph (DAG) encoding Assumption \ref{assump:exchangeability}; Appendix \ref{app:altDAGs} provides alternatives. As shown in Figure \ref{fig:DAGs}(b), Assumption \ref{assump:exchangeability}(iii) may be relaxed if, conditional on $(X,U)$, $Z$ does not modify the treatment effect on the scale of interest \citep{su2025proximal}.

Lastly, define
\begin{align*}
    \mu_a(s,x,u)= \mathbb{E}[Y(a) \mid S = s, X =x,U=u], \qquad \mu_a(s,x)= \mathbb{E}[Y(a) \mid S = s, X =x]
\end{align*}
We refer to $\mu_1(s,x,u)-\mu_0(s,x,u)$ as the \textit{latent CATE} and $\mu_1(s,x,u)/\mu_0(s,x,u)$ as the \textit{latent CRR} in trial $s$. Similarly, we refer to $\mu_1(s,x)-\mu_0(s,x)$ as the \textit{observed data CATE} and $\mu_1(s,x)/\mu_0(s,x)$ as the \textit{observed data CRR} in trial $s$.

\section{Additive/multiplicative-reconcilability null hypothesis significance testing (NHST)}
\label{s:hypothesis}

In this section, we develop methods for testing the null hypothesis that the two RCTs are conditionally reconcilable on a given scale. As introduced previously, conditional reconcilability holds when, after conditioning on both measured and unmeasured baseline covariates, both trials agree on the causal effect of treatment on a given scale. On the additive scale, this null hypothesis can be expressed as
\begin{align}
\label{eq:nullCATE}
    H_0: \mu_1(1,x,u)-\mu_0(1,x,u) = \mu_1(0,x,u) - \mu_0(0,x,u) \text{ for a.e. } (x,u),
\end{align}
which states that the latent CATE is equal across trials. Similarly, on the multiplicative scale, the null hypothesis of interest is
\begin{align}
\label{eq:nullCRR}
    H_0: \frac{\mu_1(1,x,u)}{\mu_0(1,x,u)} = \frac{\mu_1(0,x,u)}{\mu_0(0,x,u)} \text{ for a.e. } (x,u),
\end{align}
which states that the latent CRR is equal across trials. We also propose transportability-based methods for testing the null hypothesis of marginal reconcilability. On the additive scale, the null hypothesis of marginal reconcilability can be expressed for $s \in \{0,1\}$ as
\begin{align}
\label{eq:nullATE}
    H_0: \mathbb{E}[\mu_1(1-s,X,U)-\mu_0(1-s,X,U) \mid S=s] = \mathbb{E}[\mu_1(s,X,U)-\mu_0(s,X,U) \mid S=s],
\end{align}
which states that the ATE (average treatment effect) in trial $s$ can be recovered by transporting the CATE from the opposite trial population and marginalizing over the distribution of $(X,U)$ in trial $s$. Likewise, on the multiplicative scale, the null hypothesis of marginal reconcilability can be expressed for $s \in \{0,1\}$ as
\begin{align}
\label{eq:nullRR}
    H_0: \frac{\mathbb{E}[\mu_0(s,X,U) \frac{\mu_1(1-s,X,U)}{\mu_0(1-s,X,U)} \mid S=s]}{\mathbb{E}[\mu_0(s,X,U) \mid S = s]} = \frac{\mathbb{E}[\mu_1(s,X,U) \mid S=s]}{\mathbb{E}[\mu_0(s,X,U) \mid S = s]},
\end{align}
which states that the RR (relative risk) in trial $s$ can similarly be recovered by transporting the CRR from the opposite trial population and appropriately marginalizing over the distribution of $(X,U)$ in trial $s$.

Although the nonparametric proximal transportability framework of \citet{su2025proximal} provides a natural test for the null hypothesis of marginal reconcilability, such an approach would be substantially underpowered when the primary interest lies in testing the null hypothesis of conditional reconcilability, for reasons previously discussed. While one could attempt to develop tests of conditional reconcilability directly within a nonparametric proximal framework by formulating testing procedures based on nonparametric estimates of so-called bridge functions required by the approach, this approach would likely have poor finite-sample performance due to the slower convergence rates associated with nonparametric estimation of bridge functions, which are typically defined as solutions to challenging ill-posed integral equations. Motivated by these considerations, we instead adopt a parametric regression-based framework that assesses conditional reconcilability by testing the equality of specific regression coefficients. Such a procedure is straightforward to implement, easily interpretable, and may have improved finite-sample performance relative to purely nonparametric approaches. Although model misspecification is a potential concern with any parametric framework, we discuss the robustness of the proposed procedures to potential model misspecification in Appendix \ref{app:robustness}.

\subsection{NHST on the additive scale}
\label{ss:hypothesisadditive}

When testing the null hypothesis that the latent CATE is equal across trials (Equation \ref{eq:nullCATE}), we assume that the following linear structural models hold.
\begin{assumption} \label{assump:CATEstructural}
    (Additive structural models). For $s \in \{0,1\}$,
\begin{align*}
    &\mathbb{E}[Y \mid A,U,X,Z,S=s]= \beta_{0,s} + \beta_{a,s} A + \beta_{au,s} AU + \beta_{u,s} U +\beta_{ax,s}AX + \beta_{x,s} X\\
    &\mathbb{E}[W \mid A, U, X,Z, S = s] = \alpha_0 + \alpha_u U + \alpha_x X,
\end{align*}
where $|\mathbb{E}[U \mid A, X, Z,S=s]| < \infty$ and $\alpha_u \neq 0$.
\end{assumption}
Assumption \ref{assump:CATEstructural} specifies linear structural models compatible with Assumption \ref{assump:exchangeability}. Notably, the conditional mean of $Y$ does not depend on $Z$ once conditioned on $(A,U,X,S=s)$, as stated in Assumption \ref{assump:exchangeability}(iii). Likewise, the conditional mean of $W$ does not depend on $A$, $Z$, or $S$ once conditioned on $(X,U)$, as implied by Assumptions \ref{assump:exchangeability}(i), (ii), and (iv). 

Alternative structural models (e.g., with higher-order interaction terms) could be considered for greater modeling flexibility. One could also consider a structural model in which the conditional mean of $Y$ depends on $Z$, thereby relaxing Assumption \ref{assump:exchangeability}(iii), or a structural model in which a log link is used to model the conditional mean of $W$; we consider such models in Appendices \ref{app:robustness} and \ref{app:logproxy}, respectively. After introducing the structural models in Assumption \ref{assump:CATEstructural}, the proposed approach and subsequent results rely on these parametric models rather than the nonparametric proxy independence restrictions in Assumption \ref{assump:exchangeability}, similar to \citet{liu2025regression}.

Under Assumptions \ref{assump:causal} and \ref{assump:CATEstructural}, the latent CATE in each trial can be identified as a function of the full data as follows:
\begin{align*}
    &\mathbb{E}[Y(1) - Y(0) \mid U = u,X=x,S=s]\\
    &= \mathbb{E}[ \mathbb{E}[Y \mid A=1,U=u,X=x,S=s,Z] \mid A=1,U=u,X=x,S=s]\\
    &\qquad -\mathbb{E}[ \mathbb{E}[Y \mid A=0,U=u,X=x,S=s,Z] \mid A=0,U=u,X=x,S=s]\\
    &= \beta_{a,s} + \beta_{au,s} u +\beta_{ax,s}x
\end{align*}
Thus, under Assumptions \ref{assump:causal} and \ref{assump:CATEstructural}, testing the equality of the latent CATE across trials is equivalent to jointly testing
\begin{align}
\label{eq:equalcoefs}
    \beta_{a,0} = \beta_{a,1}, \qquad \beta_{au,0} = \beta_{au,1}, \qquad \beta_{ax,0} = \beta_{ax,1}
\end{align}
If $U$ were observed, this could be accomplished by fitting a regression corresponding to the model for $\mathbb{E}[Y | A,U,X,Z,S]$ in Assumption \ref{assump:CATEstructural} pooling data from both samples, while including interaction terms between $S$ and all other regressors. Testing the joint null hypothesis above in Equation \ref{eq:equalcoefs} would then be equivalent to testing that the coefficients for the interaction terms involving $S$ with $A$, $AU$, and $AX$ (corresponding to $\beta_{a,1} - \beta_{a,0}$, $\beta_{au,1} - \beta_{au,0}$, and $\beta_{ax,1} - \beta_{ax,0}$ respectively) are all equal to 0. However, because $U$ is not observed, direct estimation of these coefficients is not straightforward. To this end, we derive the following expression for the conditional mean of the outcome given the observed data.
\begin{result} \label{result:CATE}
    Under Assumption \ref{assump:CATEstructural},
\begin{align*}
    \mathbb{E}[Y \mid A, X, Z, S = s] = \beta_{0,s}^* + \beta_{a,s}^*A + \beta_{ax,s}^* AX + \beta_{u,s}^* h_s(A,X,Z) + \beta_{x,s}^* X + \beta_{au,s}^* A h_s(A,X,Z)
\end{align*}
where $h_s(A,X,Z) = \mathbb{E}[W \mid S = s, A,X, Z]$ and
\begin{align*}
\beta_{0,s}^*  &= \beta_{0,s} - \frac{\beta_{u,s}\alpha_0}{\alpha_u}
&
\beta_{a,s}^*  &= \beta_{a,s} - \frac{\beta_{au,s}\alpha_0}{\alpha_u}
&
\beta_{ax,s}^* &= \beta_{ax,s} - \frac{\beta_{au,s}\alpha_x}{\alpha_u}
\\
\beta_{u,s}^*  &= \frac{\beta_{u,s}}{\alpha_u}
&
\beta_{au,s}^* &= \frac{\beta_{au,s}}{\alpha_u}
&
\beta_{x,s}^*  &= \beta_{x,s} - \frac{\beta_{u,s}\alpha_x}{\alpha_u}.
\end{align*}
\end{result}
The above result leverages proxies to eliminate dependence on the unobserved covariate $U$ and is similar in spirit to \citet{liu2025regression}. Importantly, Result \ref{result:CATE} together with Equation \ref{eq:equalcoefs} imply that under Assumptions \ref{assump:causal} and \ref{assump:CATEstructural}, equality of the latent CATE across trials holds if and only if
\begin{align}
\label{eq:equalstarcoefs}
    \beta_{a,0}^* = \beta_{a,1}^*, \qquad \beta_{au,0}^* = \beta_{au,1}^*, \qquad \beta_{ax,0}^* = \beta_{ax,1}^*.
\end{align}
Hence, under these assumptions, a test of Equation \ref{eq:equalstarcoefs} will be a valid test of Equation \ref{eq:equalcoefs}. This motivates the following ``coefficient testing'' procedure.
\begin{enumerate}
    \item Fit a linear regression for $h_S(A,X,Z) = \mathbb{E}[W \mid A,X,Z, S]$, say $W \sim (A+X+Z +AX + AZ) \times S$.
    \item Obtain predictions for $\widehat{h}_S(A,X,Z)$ using the estimated coefficients from the first-stage regression.
    \item Fit a pooled linear regression for $\mathbb{E}[Y \mid A, X, Z, S]$ based on Result \ref{result:CATE} using $\widehat{h}_S(A,X,Z)$ and including interaction terms between $S$ and all other regressors.
    \begin{align*}
    &\mathbb{E}[Y \mid A, X, Z, S]\\
    &= \beta_{0,0}^* + (\beta_{0,1}^* - \beta_{0,0}^*)S + \beta_{a,0}^*A + (\beta_{a,1}^* - \beta_{a,0}^*)AS + \beta_{ax,0}^* AX\\
    &\quad + (\beta_{ax,1}^* - \beta_{ax,0}^*) AXS + \beta_{u,0}^* h_S(A,X,Z) + (\beta_{u,1}^* - \beta_{u,0}^*) h_S(A,X,Z) S\\
    &\quad + \beta_{x,0}^* X  + (\beta_{x,1}^* - \beta_{x,0}^*)XS + \beta_{au,0}^* A h_S(A,X,Z) + (\beta_{au,1}^* - \beta_{au,0}^*)Ah_S(A,X,Z) S
    \end{align*}
    \item Test Equation \ref{eq:equalstarcoefs} using a Wald test by jointly testing that the coefficients for the interaction terms involving $S$ with $A$, $A \cdot h_S(A,X,Z)$, and $AX$ (corresponding to $\beta_{a,1}^* - \beta_{a,0}^*$, $\beta_{au,1}^* - \beta_{au,0}^*$, and $\beta_{ax,1}^* - \beta_{ax,0}^*$) are 0. Let $\widehat{\beta}^*$ denote the $p$-dimensional vector of second-stage coefficient estimates. This null hypothesis can be expressed as $H_0:R \beta^* = 0$, where $R$ is a $q \times p$ constraint matrix selecting the coefficients of interest. The Wald test statistic is
    \begin{align}
    T_1=(R \widehat{\beta}^*)^\top \left[ R \widehat{\text{Var}}(\widehat{\beta}^*) R^\top\right]^{-1} (R \widehat{\beta}^*). \label{eq:coefficienttest}
    \end{align}
    The covariance matrix $\widehat{\text{Var}}(\widehat{\beta}^*)$ can be estimated using either the nonparametric bootstrap or from its asymptotic distribution (Appendix \ref{app:asymp}). We reject the null hypothesis in Equation \ref{eq:nullCATE} at level $\alpha$ when $T_1 > \chi_{q,1-\alpha}^2$.
\end{enumerate}
While the above approach may raise concerns regarding specification of $\mathbb{E}[W \mid A,X,Z,S]$, we show in Appendix \ref{app:robustness} that, when ordinary least squares is used, the proposed procedure remains valid under Assumption \ref{assump:CATEstructural} even if the first-stage regression $W \sim (A + X + Z + AX + AZ) \times S$ is misspecified. Furthermore, under the sufficient assumption that $A$ is marginally randomized in both trials and that $(U,Z,W,X)$ are baseline covariates, the first-stage regression may be simplified to $W \sim (X + Z)\times S$, and the resulting procedure remains valid even if this model does not correctly specify $\mathbb{E}[W \mid X,Z,S]$. Robustness to misspecification of the first-stage regression in our setting follows analogously to robustness in two-stage least squares for instrumental variables \citep{wooldridge2010econometric}.

In fact, we further show in Appendix \ref{app:robustness} that, under the assumption that $A$ is marginally randomized in both trials and that $(U,Z,W,X)$ are baseline covariates, the proposed procedure remains valid under substantially more general versions of Assumption \ref{assump:CATEstructural}. In particular, the baseline outcome model can be left completely unspecified, provided that the treatment effect remains linear in $(U,X)$.

Building on the approach of \citet{virkud2025clarifying}, we also extend transportability-based methods for testing marginal reconcilability (Equation \ref{eq:nullATE}) to our setting with both measured and unmeasured effect modifiers.

\begin{result} \label{result:transportATE} (Marginal reconcilability on the additive scale). Suppose Assumptions \ref{assump:causal} and \ref{assump:CATEstructural} hold. Then for $s \in \{0,1\}$,
\begin{align*}
&\mathbb{E}[\mu_1(s,X,U)-\mu_0(s,X,U) \mid S=s]\\
&= \beta_{a,s}^* +\beta_{ax,s}^* \mathbb{E}[X \mid S = s] +\beta_{au,s}^* \mathbb{E}[W \mid S = s]\\
&\mathbb{E}[\mu_1(1-s,X,U)-\mu_0(1-s,X,U) \mid S=s]\\
&= \beta_{a,1-s}^* +\beta_{ax,1-s}^* \mathbb{E}[X \mid S = s] +\beta_{au,1-s}^* \mathbb{E}[W \mid S = s]
\end{align*}
\end{result}
Result \ref{result:transportATE} identifies both the transported ATE (the left-hand side of Equation \ref{eq:nullATE}) and the corresponding observed target trial ATE (the right-hand side of Equation \ref{eq:nullATE}) as functionals of the observed data under the causal assumptions in Assumption \ref{assump:causal} and the structural models proposed in Assumption \ref{assump:CATEstructural}. This naturally motivates a test of marginal reconcilability based on comparing plug-in estimates of these two quantities. Under the null hypothesis of marginal reconcilability, the transported and observed ATEs should coincide, whereas a substantial discrepancy between the two estimates provides evidence against the null. We explicitly outline this procedure when the target trial is $S = 1$, noting that an analogous approach can be taken when the target trial is $S = 0$. Specifically, letting $\widehat{\theta}_{ATE,1}$ denote the observed data estimator and $\widehat{\phi}_{ATE,1}$ denote the transported estimator,
\begin{align*}
    \widehat{\theta}_{ATE,1} &= \widehat{\beta}_{a,1}^* + \widehat{\beta}_{ax,1}^* \left(\sum_iS_i \right)^{-1} \sum_{i} S_iX_i + \widehat{\beta}_{au,1}^* \left(\sum_iS_i \right)^{-1} \sum_{i} S_iW_i\\
    \widehat{\phi}_{ATE,1} &= \widehat{\beta}_{a,0}^* + \widehat{\beta}_{ax,0}^* \left(\sum_iS_i \right)^{-1} \sum_{i} S_iX_i + \widehat{\beta}_{au,0}^* \left(\sum_iS_i \right)^{-1} \sum_{i} S_iW_i,
\end{align*}
we construct the test statistic
\begin{align}
    T_2 = \frac{\widehat{\theta}_{ATE,1} - \widehat{\phi}_{ATE,1}}{\sqrt{\widehat{\text{Var}}(\widehat{\theta}_{ATE,1} - \widehat{\phi}_{ATE,1})}} \label{eq:transporttest}
\end{align}
One may readily obtain a nonparametric estimate $\widehat{\text{Var}}(\widehat{\theta}_{ATE,1} - \widehat{\phi}_{ATE,1})$ via the nonparametric bootstrap, and reject the null hypothesis in Equation \ref{eq:nullATE} at level $\alpha$ when $|T_2| > z_{1 - \alpha/2}$. In practice, if $A$ is marginally randomized as is common in RCTs, then an unadjusted ITT difference-in-means estimator can also be used for $\widehat{\theta}_{ATE,1}$.

Though the transportability approach also provides a valid test of the null hypothesis of conditional reconcilability, the proposed coefficient testing approach is expected to have greater power for testing conditional reconcilability. This is because the coefficient testing approach jointly tests equality of the full set of relevant parameters, rather than a linear combination of the parameters, as in the transportability approach. In other words, there exist alternatives under which Equations \ref{eq:equalcoefs} and \ref{eq:equalstarcoefs} do not hold, yet the transported effect is exactly equal to the untransported effect in a given trial. Against such alternatives, the transportability approach has no power, whereas the proposed coefficient testing approach does.

Next, we interpret the proposed hypothesis tests. If either the transportability-based procedure or coefficient testing procedure rejects conditional reconcilability on the additive scale for measured covariates $X$ and hypothesized unmeasured covariates $U$ represented by $(Z,W)$, then:
\begin{enumerate}
    \item There exist unmeasured or otherwise omitted effect modifiers $V$ on the additive scale beyond those included in $X$ and the hypothesized $U$ which are not proxied by $(Z,W)$, and
    \item $V$ must be conditionally imbalanced between trials in the sense that $V \not\perp \! \! \! \perp S \mid X,U$.
\end{enumerate}
Importantly, $V$ may represent baseline population differences or differences induced by study enrollment itself, for example through differences in study protocols, standards of care, or other trial engagement effects \citep{dahabreh2019generalizing, ung2025generalizing}; Appendix \ref{app:DAGs_alternative} provides an illustrative DAG. Moreover, significance of the transportability procedure additionally rejects the null hypothesis of marginal reconcilability. Because conditional reconcilability implies marginal reconcilability, rejection of marginal reconcilability is inherently a stronger claim. While the above conclusions still stand, this additionally indicates that differences in conditional treatment effects across trials persist even after marginalizing over a common distribution of effect modifiers.

\subsection{NHST on the multiplicative scale}
\label{ss:hypothesismultiplicative}

We next test the null hypothesis that the latent CRR is equal across trials (Equation \ref{eq:nullCRR}). We outline the key assumptions here and defer the full identification results and implementation details to Appendix \ref{app:multiplicative_nhst}.
\begin{assumption}
\label{assump:CRRstructural} (Multiplicative structural models)
    For $s \in \{0,1\}$,
\begin{align*}
    &\log(\mathbb{E}(Y \mid A,U,X,Z,S=s))= \beta_{0,s} + \beta_{a,s} A + \beta_{au,s} AU + \beta_{u,s} U + \beta_{ax,s} AX + \beta_{x,s} X\\
    &\mathbb{E}(W \mid A, U, X, Z, S = s)= \alpha_0 + \alpha_u U + \alpha_x X
\end{align*}
where $U \mid A, X, Z, S = s \sim \mathbb{E}[U \mid A, X, Z, S = s] + \epsilon$ for $\mathbb{E}[\epsilon] = 0$ and $\epsilon \indep A,X,Z,S$, and $\alpha_u \neq 0$.
\end{assumption}
Assumption \ref{assump:CRRstructural} replaces the identity-link outcome model in Assumption \ref{assump:CATEstructural} with a log-link model. Under Assumptions \ref{assump:causal} and \ref{assump:CRRstructural}, the latent CRR is
\begin{align*}
    \frac{\mu_1(s,x,u)}{\mu_0(s,x,u)}
    =\exp(\beta_{a,s}+\beta_{au,s}u+\beta_{ax,s}x).
\end{align*}
Thus, conditional reconcilability is equivalent to equality across trials of $\beta_{a,s}$, $\beta_{au,s}$, and $\beta_{ax,s}$, as in Equation \ref{eq:equalcoefs}.

To test these restrictions without observing $U$, Result \ref{result:CRR} in Appendix \ref{app:multiplicative_nhst} derives an expression for the observed data conditional outcome mean $\mathbb{E}[Y \mid A, X, Z, S = s]$ involving $h_s(A,X,Z)=\mathbb{E}[W\mid A,X,Z,S=s]$ and transformed coefficients $\beta_s^*$. From this, we propose a similar two-stage coefficient-testing procedure to the one developed on the additive scale. We detail some key differences in terms of power between the coefficient-testing procedures on both scales in Appendix \ref{app:multiplicative_nhst}.

Alternatively, marginal reconcilability can be tested by comparing transported and untransported marginal counterfactual means under treatment. Identification of the transported mean (Result \ref{result:transportRR} in Appendix \ref{app:multiplicative_nhst}) additionally requires $\beta_{u,0}=\beta_{u,1}$, which asserts that the effect of $U$ on the baseline risk is invariant across trials.

Lastly, under the assumptions of the relevant test, rejection of conditional or marginal reconcilability implies the presence of omitted effect modifiers $V$ on the multiplicative scale beyond $(X,U)$ that are not captured by $(Z,W)$. However, in contrast to the additive scale, $V$ need not be conditionally imbalanced across trials. Differences in conditional baseline risks can themselves prevent reconciliation, which we formalize in Appendix \ref{app:multiplicative_nhst} \citep{huitfeldt2019collapsibility}.

\section{Equivalence testing and reconciliation proportion}
\label{s:equiv}

Because failure to reject reconcilability may reflect weak proxies or limited power rather than true reconcilability, in this section we develop an equivalence test whose rejection supports marginal reconcilability within a prespecified margin. We also introduce a measure quantifying the degree to which two trials can be marginally reconciled, which we term the reconciliation proportion.

For ease of notation, we define the following quantities for $s \in \{0,1\}$:
\begin{align*}
    &\theta_{ATE,s}=\mathbb{E}[\mu_1(s,X,U)-\mu_0(s,X,U) \mid S=s]\\
    &\theta_{RR,s}=\frac{\mathbb{E}[\mu_1(s,X,U) \mid S=s]}{\mathbb{E}[\mu_0(s,X,U) \mid S = s]}\\
    &\phi_{ATE,s}=\mathbb{E}[\mu_1(1-s,X,U)-\mu_0(1-s,X,U) \mid S=s]\\
    &\phi_{RR,s}=\frac{\mathbb{E}[\mu_0(s,X,U) \frac{\mu_1(1-s,X,U)}{\mu_0(1-s,X,U)} \mid S=s]}{\mathbb{E}[\mu_0(s,X,U) \mid S = s]}
\end{align*}
which denote untransported and transported ATEs and RRs in target trial $s$. We then consider the following equivalence null hypotheses on the additive and multiplicative scales, analogous to Equations \ref{eq:nullATE} and \ref{eq:nullRR}. For $s \in \{0,1\}$ and $\delta_s >0$,
\begin{align}
    &H_{0,equiv,ATE,s}: \left| \theta_{ATE,s} - \phi_{ATE,s} \right| \geq \delta_s,\label{eq:equivnullATE}\\
    &H_{0,equiv,RR,s}: \left| \log \theta_{RR,s} - \log \phi_{RR,s} \right| \geq \delta_s \label{eq:equivnullRR}
\end{align}
Rather than comparing two treatments \citep{lumley2002network, snapinn2011indirect}, these hypotheses compare transported and trial-specific causal effects within target trial $s$. We allow $\delta_s$, which represents the largest difference between treatment effects that is still clinically insignificant, to vary across trials, as the determination of the margin may depend on the underlying study population. For example, if Meis enrolled a higher-risk population than PROLONG, a smaller margin for Meis may be chosen. In general, specification of $\delta_s$ should be guided by substantive and clinical considerations.

Rejecting the above equivalence null hypothesis would allow one to claim marginal reconcilability (on either the additive or multiplicative scale), up to the margin $\delta_s$, for the direction corresponding to transport into trial $s$. One may also consider the union equivalence null hypotheses
\begin{align}
    &H_{0,equiv,ATE,union}: \left| \theta_{ATE,0} - \phi_{ATE,0} \right| \geq \delta_0 \cup \left| \theta_{ATE,1} - \phi_{ATE,1} \right| \geq \delta_1,\label{eq:equivnullATEunion}\\
    &H_{0,equiv,RR,union}: \left| \log \theta_{RR,0} - \log \phi_{RR,0} \right| \geq \delta_0 \cup \left| \log \theta_{RR,1} - \log \phi_{RR,1} \right| \geq \delta_1 \label{eq:equivnullRRunion}
\end{align}
for which rejection would imply marginal reconcilability (up to the equivalence margins) in both transport directions. We additionally consider the following equivalence null hypotheses
\begin{align}
    &H_{0,equiv,ATE,mean}: \frac{\left| \theta_{ATE,0} - \phi_{ATE,0} \right| + \left| \theta_{ATE,1} - \phi_{ATE,1} \right|}{2} \geq \frac{\delta_0 + \delta_1}{2} ,\label{eq:equivnullATEmean}\\
    &H_{0,equiv,RR,mean}: \frac{\left| \log \theta_{RR,0} - \log \phi_{RR,0} \right| + \left| \log \theta_{RR,1} - \log \phi_{RR,1} \right|}{2} \geq \frac{\delta_0 + \delta_1}{2} ,\label{eq:equivnullRRmean}
\end{align}
which assess the average degree of marginal reconcilability across the two transport directions. Rejection provides evidence that the mean discrepancy between transported and untransported estimands is smaller than the average margin. The mean notion of marginal reconcilability is weaker than the corresponding union notion, which requires marginal reconcilability separately in both directions, because mean reconcilability can be satisfied as long as a large discrepancy in one direction is offset by a small discrepancy in the other. Which notion of reconcilability is more appropriate depends on the substantive question of interest. Conceptually, this distinction is analogous to the use of co-primary versus composite endpoints in clinical trials, where requiring efficacy for each endpoint represents a stronger criterion than assessing efficacy with respect to a single composite endpoint \citep{freemantle2003composite}. More generally, weighted averages could also be considered in Equations \ref{eq:equivnullATEmean} and \ref{eq:equivnullRRmean}, allowing greater emphasis to be placed on one transport direction than the other.

To test the above hypotheses, one may construct plug-in estimators of $\theta$ and $\phi$ using the identification results in Sections \ref{ss:hypothesisadditive} and \ref{ss:hypothesismultiplicative} and evaluate the equivalence test using the two one-sided tests (TOST) framework \citep{schuirmann1987comparison, westlake1972use}. For example, to test the equivalence null hypothesis in Equation \ref{eq:equivnullATE} for $S=1$, one may estimate $\widehat{\theta}_{ATE,1}-\widehat{\phi}_{ATE,1}$, as defined in Section \ref{ss:hypothesisadditive}, and construct a corresponding $(1-2\alpha)$ confidence interval using the nonparametric bootstrap. Rejection of the equivalence null occurs whenever the resulting confidence interval lies entirely within $(-\delta_1,\delta_1)$. For the union equivalence null hypotheses (Equations \ref{eq:equivnullATEunion} and \ref{eq:equivnullRRunion}), we propose a simple intersection-union test (IUT). Specifically, the union null is rejected only if both trial-specific equivalence tests reject for $s\in\{0,1\}$.

Rather than evaluating equivalence at a single prespecified margin $\delta_s$, one may also report the least equivalent allowable difference (LEAD) \citep{meyners2007least}. The LEAD is the smallest equivalence margin for which the corresponding equivalence test would reject the null hypothesis. Specifically, if $(L,U)$ denotes the $(1-2\alpha)$ confidence interval for the discrepancy of interest, then the LEAD would be equal to $\max \{|L|,|U|\}$. Reporting the LEAD provides a useful complement to equivalence testing, as it quantifies the degree of similarity supported by the data without requiring specification of a single equivalence margin in advance.

To complement the equivalence testing framework, we also consider a descriptive measure of the degree of marginal reconciliation, which we call the reconciliation proportion. The reconciliation proportion rescales the remaining discrepancy between the transported and trial-specific effects by the original discrepancy in marginal effects between the two trials.
\begin{align*}
    &\text{RP}_{s,ATE}= 1 - \frac{|\theta_{ATE,s} - \phi_{ATE,s}|}{|\theta_{ATE,1} - \theta_{ATE,0}|}, \qquad \text{RP}_{s,RR}= 1 - \frac{|\log \theta_{RR,s} - \log \phi_{RR,s}|}{|\log \theta_{RR,1} - \log \theta_{RR,0}|}.
\end{align*}
Similarly, analogous to the mean equivalence null, we define
\begin{align*}
    &\text{RP}_{mean,ATE} = 1 - \frac{\{|\theta_{ATE,1} - \phi_{ATE,1}|+|\phi_{ATE,0}-\theta_{ATE,0}|\}}{2|\theta_{ATE,1}-\theta_{ATE,0}|},\\
    &\text{RP}_{mean,RR} = 1 - \frac{\{|\log \theta_{RR,1} - \log \phi_{RR,1}|+|\log \phi_{RR,0}-\log \theta_{RR,0}|\}}{2|\log \theta_{RR,1}-\log \theta_{RR,0}|}.
\end{align*}
When marginal reconcilability holds in direction $s$, $\theta_s=\phi_s$ and therefore $\text{RP}_{s}=1$. Values closer to one indicate greater reconciliation, whereas smaller values indicate larger remaining discrepancies. As with the mediation proportion or proportion of treatment effect explained by a surrogate, the reconciliation proportion can be negative. This occurs when the transported estimand is farther from the untransported estimand than the original discrepancy in marginal effects between the two trials, indicating complete failure of marginal reconciliation.

Reconciliation proportions may be estimated by plug-in with nonparametric bootstrap confidence intervals. Because they are ratio estimands, inference may be unstable when the discrepancy between the marginal treatment effects in the two trials is small, resulting in a denominator close to zero. Fieller-type intervals or bootstrap procedures applied to a log-transformed ratio are possible alternative approaches. More broadly, the reconciliation proportion is most informative when marginal effects differ meaningfully across trials.

Lastly, we note that non-proximal analogues of the equivalence tests and reconciliation proportions can be constructed using standard transportability estimators based only on measured covariates $X$, but they assess a different notion of reconcilability that omits latent $U$. Accordingly, comparisons of LEADs or reconciliation proportions between proximal and non-proximal procedures are not formal tests and may reflect the different notions of reconcilability considered.

\section{Application to 17P Trials}
\label{s:application}
We illustrate the proposed methods by examining whether the Meis and PROLONG trials are reconcilable on the additive scale, using de-identified clinical data obtained from AMAG Pharmaceuticals. Unlike the original analyses of the trials, which assessed the binary outcome of delivery before 37 weeks of gestation in Meis and before 35 weeks of gestation in PROLONG, we focus in this section on the outcome of gestational age at delivery (in weeks). Corresponding analyses of the original binary outcome on the multiplicative scale are deferred to Appendix \ref{app:mult}.

Relevant baseline covariates included maternal age, maternal race, gestational age at qualifying prior spontaneous PTB, number of prior spontaneous PTBs, marital status, pre-pregnancy BMI, years of education, smoking during pregnancy, alcohol use during pregnancy, and substance use during pregnancy. After excluding observations with missing outcomes (4 observations in Meis, 24 observations in PROLONG) or covariates (16 observations in Meis, 8 observations in PROLONG), the study population consisted of $443$ patients in Meis and $1676$ patients in PROLONG. Compared to participants in Meis, those in PROLONG had fewer prior PTBs, were more likely to be non-Hispanic White than non-Hispanic Black, had lower pre-pregnancy BMI, and were less likely to report smoking, alcohol use, or substance use during pregnancy. A complete summary of measured baseline characteristics is provided in Table \ref{table1} in Appendix \ref{app:databaseline}.

We consider baseline cervical length as a hypothesized unmeasured effect modifier whose distribution may have differed between the two trials. Baseline cervical length was assessed by transvaginal ultrasound at enrollment in 73\% of participants in PROLONG but was not measured in Meis. Cervical shortening is an established risk factor for recurrent PTB \citep{o2013cervical}. Although risk factors need not necessarily be effect modifiers, subgroup analyses from PROLONG suggest that the effect of 17OHP-C may vary by cervical length, even though the underlying mechanism remains unclear \citep{blackwell202017}. Additionally, although cervical length was not measured in Meis, the very low rate of shortened cervix ($<2\%$) measured in PROLONG may suggest potential differences in its distribution across the two trials \citep{blackwell202017}.

We consider two primary analyses with different pairs of candidate proxies. In the first analysis, we take the adjustment proxy $W$ to be the number of prior spontaneous PTBs and the reweighting proxy $Z$ to be gestational age at the qualifying prior PTB. In the second analysis, we take $W$ to be gestational age at the qualifying prior PTB and $Z$ to be pre-pregnancy BMI. In both analyses, $Z$ and $W$ are taken to be continuous variables, and $X$ denotes the measured baseline covariates excluding the variables selected as $Z$ and $W$.

We briefly discuss the plausibility of Assumption \ref{assump:exchangeability}, which motivates the proposed structural equation models. Because all proposed proxies were measured prior to treatment assignment and treatment was randomized, Assumption \ref{assump:exchangeability}(i) is likely satisfied. Conditions (ii)--(iv) are plausible if $W$ and $Z$ are viewed as proxies for an underlying biological process $U$, such as cervical length, that is more directly relevant to trial participation and effect modification. Although these conditions are untestable, we believe they are scientifically reasonable for the proxy choices considered here. Appendix \ref{app:data} considers alternative proxies among the measured baseline covariates.

For each analysis, we test the null hypothesis that the two trials are conditionally reconcilable on the additive scale, using the coefficient testing and transportability approaches described in Section \ref{ss:hypothesisadditive}, with the transportability approach also providing a valid test for the null hypothesis of marginal reconcilability. For the transportability method, transported estimates were compared against an unadjusted difference-in-means estimator, as in Section \ref{ss:hypothesisadditive}. In all cases, the first-stage linear regression was specified as $W \sim (X + Z + XZ) \times S$ and estimated using linear regression. Because of randomization, it was not necessary to include $A$ in the first-stage regression. Results are presented in Table \ref{tab:table2}. Although not presented here, transportability-based tests for the non-proximal analogs of the null hypotheses of conditional and marginal reconcilability are available in \citet{virkud2025clarifying}.

\begin{table}[!h]
\centering
\caption{\label{tab:table2}Results for testing the null hypothesis of conditional and marginal reconcilability on the additive scale for the Meis and PROLONG trials. $\widehat{\phi}_{ATE,s}$ and $\widehat{\theta}_{ATE,s}$ denote the transported and untransported estimates for the ATE in trial $s$, respectively, as defined in Section \ref{ss:hypothesisadditive}. $\widehat{\theta}_{ATE,s}$ was taken to be the standard difference-in-means estimator. 95\% confidence intervals and p-values were computed using the nonparametric bootstrap with 5,000 resamples. All estimates are in weeks (gestational age at delivery).}
\centering
\resizebox{\ifdim\width>\linewidth\linewidth\else\width\fi}{!}{
\begin{tabular}[t]{>{\raggedright\arraybackslash}p{13em}>{\raggedright\arraybackslash}p{4em}>{\raggedright\arraybackslash}p{6em}>{\raggedright\arraybackslash}p{8em}>{\raggedright\arraybackslash}p{8em}>{\raggedright\arraybackslash}p{8em}>{\raggedright\arraybackslash}p{4em}}
\toprule
 & Method & Target & $\widehat{\phi}_{ATE,s}$ & $\widehat{\theta}_{ATE,s}$ & $\widehat{\phi}_{ATE,s} - \widehat{\theta}_{ATE,s}$ & p-value\\
\midrule
 &  & Meis & 0.91 (-1.18, 2.99) & 1.09 (0.15, 2.03) & -0.18 (-2.45, 2.08) & 0.873\\

 & \multirow[t]{-2}{4em}{\raggedright\arraybackslash Transport} & PROLONG & 0.02 (-2.45, 2.49) & 0.10 (-0.24, 0.44) & -0.08 (-2.58, 2.41) & 0.948\\

\multirow[t]{-3}{13em}{\raggedright\arraybackslash Analysis 1: W is number of prior PTBs, Z is gestational age at prior PTB} & Coefficient & -- & -- & -- & -- & 0.207\\

\cmidrule{1-7}
 &  & Meis & -0.40 (-1.71, 0.90) & 1.09 (0.15, 2.03) & -1.49 (-3.11, 0.12) & 0.069\\

 & \multirow[t]{-2}{4em}{\raggedright\arraybackslash Transport} & PROLONG & 0.43 (-1.19, 2.05) & 0.10 (-0.24, 0.44) & 0.33 (-1.33, 1.99) & 0.695\\

\multirow[t]{-3}{13em}{\raggedright\arraybackslash Analysis 2: W is gestational age at prior PTB, Z is pre-pregnancy BMI} & Coefficient & -- & -- & -- & -- & 0.022\\
\bottomrule
\end{tabular}}
\end{table}


In the first analysis, none of the methods yields a statistically significant result at the $\alpha = 0.05$ level, suggesting that we fail to reject either the conditional or marginal reconcilability null hypothesis. Notably, transportability estimates and corresponding unadjusted ITT estimates were similar in both Meis (0.91 vs. 1.09) and PROLONG (0.02 vs. 0.10). While these results may suggest reconcilability on the additive scale, such conclusions should be interpreted cautiously and complemented with the proposed equivalence test analyses. In fact, measures of proxy relevance, namely coefficient estimates and standard errors for terms involving $\widehat{h}_S(X,Z)$ in the second-stage (Appendix \ref{app:data}) suggest that the proposed proxies may be weak.

In the second analysis, the coefficient testing approach yields a statistically significant result ($p = 0.022$), whereas the transportability approach fails to reject either the null hypothesis of marginal or conditional reconcilability on the additive scale. Measures of proxy relevance (Appendix \ref{app:data}) again suggest that the proposed proxies may be relatively weak. Nevertheless, this result highlights the practical advantages of the coefficient testing approach when testing conditional reconcilability is the primary goal. Under the proposed structural models and identifying assumptions, the results of the second analysis imply the presence of additional, conditionally imbalanced, unmeasured effect modifiers not adequately captured by the proxies of gestational age at the qualifying prior preterm birth and pre-pregnancy BMI. As discussed in the introduction, such imbalanced unmeasured factors may include differences in access to and utilization of high-quality care, experiences of racism and discrimination, and changes in care practices or availability of 17OHP-C over time.

\begin{table}[!h]
\centering
\caption{\label{tab:table3}Results for equivalence testing for marginal reconcilability on the additive scale for the Meis and PROLONG trials. Estimate denotes the plug-in estimate of the quantity defining the corresponding equivalence null hypothesis on the additive scale (Section \ref{s:equiv}) and is presented with a 90\% confidence interval computed using the nonparametric bootstrap. 95\% confidence intervals for the reconciliation proportion (RP) were also computed using the nonparametric bootstrap. Reported p-values correspond to equivalence tests with equivalence margins of 1 week (gestational age at delivery) in both populations.}
\centering
\resizebox{\ifdim\width>\linewidth\linewidth\else\width\fi}{!}{
\begin{tabular}[t]{>{\raggedright\arraybackslash}p{10em}>{\raggedright\arraybackslash}p{7em}>{\raggedright\arraybackslash}p{5em}>{\raggedright\arraybackslash}p{9em}>{\raggedright\arraybackslash}p{5em}>{\raggedright\arraybackslash}p{5em}>{\raggedright\arraybackslash}p{9em}}
\toprule
Analysis & Method & Hypothesis & Estimate (CI) & p-value & LEAD & RP (CI)\\
\midrule
 &  & Meis & 0.18 (-1.72, 2.08) & 0.240 & 2.08 & 0.81 (-1.06, 0.98)\\

 &  & PROLONG & 0.08 (-2.01, 2.18) & 0.236 & 2.18 & 0.92 (-0.14, 0.99)\\

 &  & Union & -- & 0.240 & 2.18 & --\\

 & \multirow{-4}{7em}{\raggedright\arraybackslash Proximal} & Mean & 0.13 (-0.78, 1.05) & 0.060 & 1.05 & 0.86 (0.16, 0.98)\\
\cmidrule{2-7}
 &  & Meis & 1.34 (0.08, 2.59) & 0.671 & 2.59 & -0.35 (-5.31, 0.71)\\

 &  & PROLONG & -1.10 (-2.48, 0.28) & 0.547 & 2.48 & -0.11 (-6.66, 0.84)\\

 &  & Union & -- & 0.671 & 2.59 & --\\

\multirow{-8}{10em}[0.5\dimexpr\aboverulesep+\belowrulesep+\cmidrulewidth]{\raggedright\arraybackslash Analysis 1: W is number of prior PTBs, Z is gestational age at prior PTB} & \multirow{-4}{7em}{\raggedright\arraybackslash Non-proximal} & Mean & 1.22 (0.24, 2.19) & 0.643 & 2.19 & -0.23 (-3.07, 0.63)\\
\cmidrule{1-7}
 &  & Meis & 1.49 (0.14, 2.85) & 0.726 & 2.85 & -0.51 (-5.53, 0.65)\\

 &  & PROLONG & -0.33 (-1.72, 1.06) & 0.214 & 1.72 & 0.67 (-3.38, 0.97)\\

 &  & Union & -- & 0.726 & 2.85 & --\\

 & \multirow{-4}{7em}{\raggedright\arraybackslash Proximal} & Mean & 0.91 (0.09, 1.74) & 0.431 & 1.74 & 0.08 (-2.50, 0.76)\\
\cmidrule{2-7}
 &  & Meis & 1.40 (0.16, 2.64) & 0.701 & 2.64 & -0.41 (-4.78, 0.66)\\

 &  & PROLONG & -0.25 (-1.52, 1.02) & 0.166 & 1.52 & 0.75 (-2.29, 0.98)\\

 &  & Union & -- & 0.701 & 2.64 & --\\

\multirow{-8}{10em}[0.5\dimexpr\aboverulesep+\belowrulesep+\cmidrulewidth]{\raggedright\arraybackslash Analysis 2: W is gestational age at prior PTB, Z is pre-pregnancy BMI} & \multirow{-4}{7em}{\raggedright\arraybackslash Non-proximal} & Mean & 0.83 (0.07, 1.58) & 0.352 & 1.58 & 0.17 (-2.06, 0.77)\\
\bottomrule
\end{tabular}}
\end{table}


We next apply the equivalence testing procedures developed in Section \ref{s:equiv} to assess whether either analysis provides evidence of marginal reconcilability on the additive scale. Results are summarized in Table \ref{tab:table3}, with p-values corresponding to equivalence tests with equivalence margins of 1 week in both Meis and PROLONG ($\delta_0=\delta_1=1$). For each analysis, we compare the proposed proximal approach with a non-proximal approach that accounts only for measured effect modifiers and evaluates the corresponding non-proximal analog of marginal reconcilability. The non-proximal approach is based on the parametric standardization transportability estimator of \citet{dahabreh2020extending}, where the outcome regression is specified as a linear regression $Y \sim X$ among individuals in trial $S = s$ assigned to treatment $A=a$.

For the first analysis, none of the equivalence tests (proximal or non-proximal) are statistically significant at an equivalence margin of 1 week, suggesting insufficient evidence to claim marginal reconcilability in either transport direction or on average across both transport directions. The LEADs from the proximal procedures indicate that equivalence margins of 2.08 weeks, 2.18 weeks, and 1.05 weeks would be required to claim marginal reconcilability in the Meis direction, PROLONG direction, and on average across both directions, respectively. The smaller LEAD for mean reconcilability is unsurprising and highlights the tradeoff between potentially greater power to establish marginal reconcilability and the interpretability of the resulting claim (since mean reconcilability does not imply reconcilability in each transport direction individually). The estimated proximal reconciliation proportions are relatively high (92\% in the PROLONG direction and 81\% in the Meis direction), although the associated confidence intervals are wide, ranging from no reconciliation to complete reconciliation, potentially reflecting additional uncertainty in the proximal approach. A similar pattern is observed for the non-proximal procedures, although the corresponding LEADs are larger and the estimated reconciliation proportions are smaller.

For the second analysis, none of the equivalence tests (proximal or non-proximal) are statistically significant at an equivalence margin of 1 week, again suggesting insufficient evidence to claim marginal reconcilability in either transport direction or on average across both directions. The LEADs from the proximal procedures indicate that equivalence margins of 2.85 weeks, 1.72 weeks, and 1.74 weeks would be required to claim marginal reconcilability in the Meis direction, PROLONG direction, and on average across both directions, respectively. While the estimated reconciliation proportion is moderately high in the PROLONG direction (67\%), it is substantially lower in the Meis direction ($-51\%$) and on average across both directions (8\%), albeit with wide confidence intervals. A similar pattern is observed for the non-proximal procedures, although the corresponding LEADs are smaller and the estimated reconciliation proportions are larger.

Taken together, the NHST and equivalence testing results suggest different conclusions for the two analyses. In Analysis 1, the transportability point estimates appear more consistent with reconcilability than in previous analyses and suggest that accounting for the proposed latent effect modifier may partially explain the discrepancy between trials. However, the large LEADs and wide confidence intervals make it difficult to draw definitive conclusions regarding either conditional or marginal reconcilability. One possible explanation is that the proposed proxies are insufficiently informative about cervical length (or a related latent effect modifier).

In contrast, Analysis 2 provides direct evidence against conditional reconcilability on the additive scale through the coefficient-testing approach. Under the assumptions of the proposed model, these findings suggest the existence of additional conditionally imbalanced effect modifiers that are not captured by the proposed proxies. Consistent with this interpretation, the equivalence testing analyses provide little evidence in favor of marginal reconcilability at clinically insignificant equivalence margins.

\section{Discussion}
\label{s:discussion}

In this work, we develop a proximal framework for assessing reconcilability of conflicting RCTs in the presence of unmeasured effect modifiers. We formalize reconcilability at both the conditional and marginal levels and on both additive and multiplicative scales. By leveraging proxy variables for hypothesized unmeasured effect modifiers under parametric structural models, we develop hypothesis tests that 1) extend transportability methods to test for marginal reconcilability and 2) more powerfully test for conditional reconcilability by testing equality of certain structural model coefficients. We further highlight several important distinctions between reconcilability on the additive and multiplicative scales, including the additional assumptions required to identify transported marginal estimands under the proposed parametric models and the fact that differences in conditional baseline risks between trials alone may prevent reconciliation on the multiplicative scale. Lastly, we also introduce an equivalence testing framework for which rejection of the null provides evidence in favor of marginal reconcilability, together with a complementary reconciliation proportion metric that quantifies the degree of marginal reconciliation. When applied to Meis and PROLONG, the methods provided mixed evidence and little affirmative support for marginal reconcilability, potentially because the selected proxies were weak.

Several limitations of the methods and directions for future research are worth highlighting. First, the approach relies on parametric structural models for the outcome and proxies. While these models yield interpretable tests, misspecification may induce bias in both estimation and inference. Though we discuss robustness of the proposed approach to potential model misspecification, developing semiparametric or nonparametric extensions of the proposed framework for both NHST and equivalence testing remains an important direction for future work.

Second, the proposed methods depend crucially on the availability and quality of proxies for the hypothesized unmeasured effect modifiers. Extending the proposed framework to accommodate weak proxies would be of interest. Alternatively, sensitivity analyses, which avoid proxy assumptions, could quantify the effect-modification strength and distributional imbalance needed for an unmeasured modifier to explain the between-trial discrepancy \citep{vanderweele2017sensitivity}.

Third, while the proposed equivalence testing framework is able to provide evidence in favor of marginal reconcilability, we are unable to develop a procedure to provide evidence in favor of conditional reconcilability under the proposed assumptions. Nevertheless, establishing conditional reconcilability, particularly after accounting for latent effect modifiers, may be of substantial scientific interest. Investigating stronger but interpretable assumptions under which equivalence tests can be formulated directly for latent conditional causal effects is therefore an important avenue for future research.

Finally, our analysis focuses only on how differences in study populations could account for conflicting results across trials and does not address other potential explanations, such as protocol deviations due to loss to follow-up, differential adherence, or potential unblinding. Extending the reconciliation framework to incorporate such ideas, as well as to settings with more than two trials, would further broaden its relevance for evidence synthesis and regulatory decision making.

\section{Disclosure statement}\label{disclosure-statement}

No conflicts of interest exist.

\section{Data Availability and Software Statement}\label{data-availability-statement}

R code implementing the proposed methods on a simulated dataset is available on GitHub at \texttt{https://github.com/danielxu19/Regression-Proximal-Reconciliation}.

\bibliography{bibliography.bib}

@article{nilsson2021directed,
  title={A directed acyclic graph for interactions},
  author={Nilsson, Anton and Bonander, Carl and Str{\"o}mberg, Ulf and Bj{\"o}rk, Jonas},
  journal={International journal of epidemiology},
  volume={50},
  number={2},
  pages={613--619},
  year={2021},
  publisher={Oxford University Press}
}

@article{nilsson2023proxy,
  title={Proxy variables and the generalizability of study results},
  author={Nilsson, Anton and Bj{\"o}rk, Jonas and Bonander, Carl},
  journal={American journal of epidemiology},
  volume={192},
  number={3},
  pages={448--454},
  year={2023},
  publisher={Oxford University Press}
}

@article{dahabreh2024learning,
  title={Learning about treatment effects in a new target population under transportability assumptions for relative effect measures},
  author={Dahabreh, Issa J and Robertson, Sarah E and Steingrimsson, Jon A},
  journal={European journal of epidemiology},
  volume={39},
  number={9},
  pages={957},
  year={2024}
}

@article{dahabreh2020extending,
  title={Extending inferences from a randomized trial to a new target population},
  author={Dahabreh, Issa J and Robertson, Sarah E and Steingrimsson, Jon A and Stuart, Elizabeth A and Hernan, Miguel A},
  journal={Statistics in medicine},
  volume={39},
  number={14},
  pages={1999--2014},
  year={2020},
  publisher={Wiley Online Library}
}

@article{su2025proximal,
  title={Proximal indirect comparison},
  author={Su, Zehao and Rytgaard, Helene C and Ravn, Henrik and Eriksson, Frank},
  journal={Biometrika},
  pages={asaf044},
  year={2025},
  publisher={Oxford University Press}
}

@article{liu2025regression,
  title={Regression-based proximal causal inference},
  author={Liu, Jiewen and Park, Chan and Li, Kendrick and Tchetgen Tchetgen, Eric J},
  journal={American journal of epidemiology},
  volume={194},
  number={7},
  pages={2030--2036},
  year={2025},
  publisher={Oxford University Press}
}

@article{meis2003prevention,
  title={Prevention of recurrent preterm delivery by 17 alpha-hydroxyprogesterone caproate},
  author={Meis, Paul J and Klebanoff, Mark and Thom, Elizabeth and Dombrowski, Mitchell P and Sibai, Baha and Moawad, Atef H and Spong, Catherine Y and Hauth, John C and Miodovnik, Menachem and Varner, Michael W and others},
  journal={New England Journal of Medicine},
  volume={348},
  number={24},
  pages={2379--2385},
  year={2003},
  publisher={Mass Medical Soc}
}

@article{blackwell202017,
  title={17-OHPC to prevent recurrent preterm birth in singleton gestations (PROLONG study): a multicenter, international, randomized double-blind trial},
  author={Blackwell, Sean C and Gyamfi-Bannerman, Cynthia and Biggio Jr, Joseph R and Chauhan, Suneet P and Hughes, Brenna L and Louis, Judette M and Manuck, Tracy A and Miller, Hugh S and Das, Anita F and Saade, George R and others},
  journal={American journal of perinatology},
  volume={37},
  number={02},
  pages={127--136},
  year={2020},
  publisher={Thieme Medical Publishers}
}

@article{virkud2025clarifying,
  title={Clarifying Contradictions: Transportability in 17OHP-C Trials and Preterm Birth Outcomes Using Doubly Debiased Machine Learning},
  author={Virkud, Arti V and Tchetgen, Eric Tchetgen and Schisterman, Enrique F and Pineles, Beth and Levine, Lisa D and Cole, Stephen R and Hinkle, Stefanie N and Mumford, Sunni and Gerson, Kristin D and Taylor, Brandie D and others},
  journal={American journal of epidemiology},
  pages={kwaf202},
  year={2025},
  publisher={Oxford University Press}
}

@article{dahabreh2020toward,
  title={Toward causally interpretable meta-analysis: transporting inferences from multiple randomized trials to a new target population},
  author={Dahabreh, Issa J and Petito, Lucia C and Robertson, Sarah E and Hern{\'a}n, Miguel A and Steingrimsson, Jon A},
  journal={Epidemiology},
  volume={31},
  number={3},
  pages={334--344},
  year={2020},
  publisher={LWW}
}

@article{miao2018identifying,
  title={Identifying causal effects with proxy variables of an unmeasured confounder},
  author={Miao, Wang and Geng, Zhi and Tchetgen Tchetgen, Eric J},
  journal={Biometrika},
  volume={105},
  number={4},
  pages={987--993},
  year={2018},
  publisher={Oxford University Press}
}

@article{tchetgen2024introduction,
  title={An introduction to proximal causal inference},
  author={Tchetgen Tchetgen, Eric J and Ying, Andrew and Cui, Yifan and Shi, Xu and Miao, Wang},
  journal={Statistical Science},
  volume={39},
  number={3},
  pages={375--390},
  year={2024},
  publisher={Institute of Mathematical Statistics}
}

@article{cui2024semiparametric,
  title={Semiparametric proximal causal inference},
  author={Cui, Yifan and Pu, Hongming and Shi, Xu and Miao, Wang and Tchetgen Tchetgen, Eric},
  journal={Journal of the American Statistical Association},
  volume={119},
  number={546},
  pages={1348--1359},
  year={2024},
  publisher={Taylor \& Francis}
}

@article{altman1995statistics,
  title={Statistics notes: Absence of evidence is not evidence of absence},
  author={Altman, Douglas G and Bland, J Martin},
  journal={Bmj},
  volume={311},
  number={7003},
  pages={485},
  year={1995},
  publisher={British Medical Journal Publishing Group}
}

@article{imai2008misunderstandings,
  title={Misunderstandings between experimentalists and observationalists about causal inference},
  author={Imai, Kosuke and King, Gary and Stuart, Elizabeth A},
  journal={Journal of the Royal Statistical Society Series A: Statistics in Society},
  volume={171},
  number={2},
  pages={481--502},
  year={2008},
  publisher={Oxford University Press}
}

@misc{wasserstein2016asa,
  title={The ASA statement on p-values: context, process, and purpose},
  author={Wasserstein, Ronald L and Lazar, Nicole A},
  journal={The American Statistician},
  volume={70},
  number={2},
  pages={129--133},
  year={2016},
  publisher={Taylor \& Francis}
}

@article{schuirmann1987comparison,
  title={A comparison of the two one-sided tests procedure and the power approach for assessing the equivalence of average bioavailability},
  author={Schuirmann, Donald J},
  journal={Journal of pharmacokinetics and biopharmaceutics},
  volume={15},
  number={6},
  pages={657--680},
  year={1987},
  publisher={Springer}
}

@article{westlake1972use,
  title={Use of confidence intervals in analysis of comparative bioavailability trials},
  author={Westlake, Wilfred J},
  journal={Journal of Pharmaceutical Sciences},
  volume={61},
  number={8},
  pages={1340--1341},
  year={1972},
  publisher={Wiley Online Library}
}

@article{liu2009assessing,
  title={Assessing nonsuperiority, noninferiority, or equivalence when comparing two regression models over a restricted covariate region},
  author={Liu, W and Bretz, F and Hayter, AJ and Wynn, HP},
  journal={Biometrics},
  volume={65},
  number={4},
  pages={1279--1287},
  year={2009},
  publisher={Oxford University Press}
}

@article{gsteiger2011simultaneous,
  title={Simultaneous confidence bands for nonlinear regression models with application to population pharmacokinetic analyses},
  author={Gsteiger, S and Bretz, F and Liu, W},
  journal={Journal of Biopharmaceutical Statistics},
  volume={21},
  number={4},
  pages={708--725},
  year={2011},
  publisher={Taylor \& Francis}
}

@article{dette2018equivalence,
  title={Equivalence of regression curves},
  author={Dette, Holger and M{\"o}llenhoff, Kathrin and Volgushev, Stanislav and Bretz, Frank},
  journal={Journal of the American Statistical Association},
  volume={113},
  number={522},
  pages={711--729},
  year={2018},
  publisher={Taylor \& Francis}
}

@article{mollenhoff2020equivalence,
  title={Equivalence of regression curves sharing common parameters},
  author={M{\"o}llenhoff, Kathrin and Bretz, Frank and Dette, Holger},
  journal={Biometrics},
  volume={76},
  number={2},
  pages={518--529},
  year={2020},
  publisher={Oxford University Press}
}

@article{mollenhoff2024testing,
  title={Testing similarity of parametric competing risks models for identifying potentially similar pathways in healthcare},
  author={M{\"o}llenhoff, Kathrin and Binder, Nadine and Dette, Holger},
  journal={Statistics in Medicine},
  volume={43},
  number={28},
  pages={5316--5330},
  year={2024},
  publisher={Wiley Online Library}
}

@article{hagemann2025overcoming,
  title={Overcoming Model Uncertainty-How Equivalence Tests Can Benefit From Model Averaging},
  author={Hagemann, Niklas and M{\"o}llenhoff, Kathrin},
  journal={Statistics in Medicine},
  volume={44},
  number={6},
  pages={e10309},
  year={2025},
  publisher={Wiley Online Library}
}

@article{meyners2007least,
  title={Least equivalent allowable differences in equivalence testing},
  author={Meyners, Michael},
  journal={Food quality and preference},
  volume={18},
  number={3},
  pages={541--547},
  year={2007},
  publisher={Elsevier}
}

@book{van2000asymptotic,
  title={Asymptotic statistics},
  author={Van der Vaart, Aad W},
  volume={3},
  year={2000},
  publisher={Cambridge university press}
}

@article{o2013cervical,
  title={Cervical length for predicting preterm birth and a comparison of ultrasonic measurement techniques},
  author={O'Hara, Sandra and Zelesco, Marilyn and Sun, Zhonghua},
  journal={Australasian journal of ultrasound in medicine},
  volume={16},
  number={3},
  pages={124--134},
  year={2013},
  publisher={Wiley Online Library}
}

@article{vanderweele2017sensitivity,
  title={Sensitivity analysis in observational research: introducing the E-value},
  author={VanderWeele, Tyler J and Ding, Peng},
  journal={Annals of internal medicine},
  volume={167},
  number={4},
  pages={268--274},
  year={2017},
  publisher={American College of Physicians}
}

@article{nevo2017estimation,
  title={Estimation and inference for the mediation proportion},
  author={Nevo, Daniel and Liao, Xiaomei and Spiegelman, Donna},
  journal={The international journal of biostatistics},
  volume={13},
  number={2},
  pages={20170006},
  year={2017},
  publisher={De Gruyter}
}

@article{wang2002measure,
  title={A measure of the proportion of treatment effect explained by a surrogate marker},
  author={Wang, Yue and Taylor, Jeremy MG},
  journal={Biometrics},
  volume={58},
  number={4},
  pages={803--812},
  year={2002},
  publisher={Wiley Online Library}
}

@article{parast2016robust,
  title={Robust estimation of the proportion of treatment effect explained by surrogate marker information},
  author={Parast, Layla and McDermott, Mary M and Tian, Lu},
  journal={Statistics in medicine},
  volume={35},
  number={10},
  pages={1637--1653},
  year={2016},
  publisher={Wiley Online Library}
}

@article{freedman1992statistical,
  title={Statistical validation of intermediate endpoints for chronic diseases},
  author={Freedman, Laurence S and Graubard, Barry I and Schatzkin, Arthur},
  journal={Statistics in medicine},
  volume={11},
  number={2},
  pages={167--178},
  year={1992},
  publisher={Wiley Online Library}
}

@book{wooldridge2010econometric,
  title={Econometric analysis of cross section and panel data},
  author={Wooldridge, Jeffrey M},
  year={2010},
  publisher={MIT press}
}

@article{huitfeldt2019collapsibility,
  title={On the collapsibility of measures of effect in the counterfactual causal framework},
  author={Huitfeldt, Anders and Stensrud, Mats J and Suzuki, Etsuji},
  journal={Emerging themes in epidemiology},
  volume={16},
  number={1},
  pages={1},
  year={2019},
  publisher={Springer}
}

@article{lumley2002network,
  title={Network meta-analysis for indirect treatment comparisons},
  author={Lumley, Thomas},
  journal={Statistics in medicine},
  volume={21},
  number={16},
  pages={2313--2324},
  year={2002},
  publisher={Wiley Online Library}
}

@article{snapinn2011indirect,
  title={Indirect comparisons in the comparative efficacy and non-inferiority settings},
  author={Snapinn, Steven and Jiang, Qi},
  journal={Pharmaceutical statistics},
  volume={10},
  number={5},
  pages={420--426},
  year={2011},
  publisher={Wiley Online Library}
}

@article{ditlevsen2005mediation,
  title={The mediation proportion: a structural equation approach for estimating the proportion of exposure effect on outcome explained by an intermediate variable},
  author={Ditlevsen, Susanne and Christensen, Ulla and Lynch, John and Damsgaard, Mogens Trab and Keiding, Niels},
  journal={Epidemiology},
  volume={16},
  number={1},
  pages={114--120},
  year={2005},
  publisher={LWW}
}

@article{mackinnon1994analysis,
  title={Analysis of mediating variables},
  author={MacKinnon, David P},
  journal={Scientific methods for prevention intervention research},
  volume={139},
  number={127.1},
  year={1994}
}

@article{freemantle2003composite,
  title={Composite outcomes in randomized trials: greater precision but with greater uncertainty?},
  author={Freemantle, Nick and Calvert, Melanie and Wood, John and Eastaugh, Joanne and Griffin, Carl},
  journal={Jama},
  volume={289},
  number={19},
  pages={2554--2559},
  year={2003},
  publisher={American Medical Association}
}

@article{ung2025generalizing,
  title={Generalizing and transporting causal inferences from randomized trials in the presence of trial engagement effects},
  author={Ung, Lawson and VanderWeele, Tyler J and Dahabreh, Issa J},
  journal={Epidemiology},
  volume={36},
  number={4},
  pages={500--510},
  year={2025},
  publisher={LWW}
}

@article{dahabreh2019generalizing,
  title={Generalizing causal inferences from randomized trials: counterfactual and graphical identification},
  author={Dahabreh, Issa J and Robins, James M and Haneuse, Sebastien JP and Hern{\'a}n, Miguel A},
  journal={arXiv preprint arXiv:1906.10792},
  year={2019}
}

\clearpage
\appendix

\printappendixtoc

\appsection{Appendix A: NHST on the multiplicative scale}
\label{app:multiplicative_nhst}

This section provides the identification results and implementation details underlying the NHSTs on the multiplicative scale summarized in Section \ref{ss:hypothesismultiplicative}. While it is important that the outcome structural model is specified on the log scale as in Assumption \ref{assump:CRRstructural}, it is possible to consider other structural models for the conditional mean of $W$, which we discuss in Appendix \ref{app:logproxy}.
\begin{result}
\label{result:CRR}
    Suppose that $M_{\epsilon}(e)$, the MGF of $\epsilon$, exists for every real $e$. Under Assumption \ref{assump:CRRstructural},
    \begin{align*}
        &\log(\mathbb{E}[Y \mid A, X, Z, S = s])\\
        &= \beta_{0,s}^* + \beta_{a,s}^* A + \beta_{ax,s}^* AX+\beta_{x,s}^*X+\beta_{u,s}^*h_s(A,X,Z)+ \beta_{au,s}^* A h_s(A,X,Z)
    \end{align*}
    where
    \begin{align*}
\beta_{0,s}^* &= \beta_{0,s} + \log M_{\epsilon}(\beta_{u,s}) - \frac{\beta_{u,s}\alpha_0}{\alpha_u}, \qquad \\
\beta_{a,s}^* &= \beta_{a,s} + \log \left( \frac{M_{\epsilon}(\beta_{au,s} + \beta_{u,s})}{M_{\epsilon}(\beta_{u,s})} \right) - \frac{\beta_{au,s} \alpha_0}{\alpha_u}\\
\beta_{ax,s}^* &= \beta_{ax,s} - \frac{\beta_{au,s} \alpha_x}{\alpha_u}\\
\beta_{x,s}^* &= \beta_{x,s}  - \frac{\beta_{u,s} \alpha_x}{\alpha_u}\\
\beta_{u,s}^* &= \frac{\beta_{u,s}}{\alpha_u}\\
\beta_{au,s}^* &= \frac{\beta_{au,s}}{\alpha_u}\\
\end{align*}
\end{result}
Under Assumptions \ref{assump:causal} and \ref{assump:CRRstructural}, Result \ref{result:CRR} provides that equality of the latent CRR across trials (Equation \ref{eq:nullCRR}) implies
\begin{align}
\label{eq:equalstarcoefsCRR}
\beta_{ax,0}^* = \beta_{ax,1}^*, \qquad \beta_{au,0}^* = \beta_{au,1}^*
\end{align}
This is in contrast to the prior section, in which equality of the latent CATE across trials additionally implied that $\beta_{a,0}^* = \beta_{a,1}^*$. In the CRR case, this restriction will generally not hold due to the dependence of $\beta_{a,s}^*$ on $\beta_{u,s}$. Hence, while a valid test of Equation \ref{eq:equalstarcoefsCRR} remains a valid test of the equality of the latent CRR across trials, it would have no power against alternatives for which $\beta_{a,0} \neq \beta_{a,1}$ but $\beta_{ax,0} = \beta_{ax,1}$ and $\beta_{au,0} = \beta_{au,1}$. Nonetheless, we propose a similar ``coefficient testing'' procedure to that developed previously.
\begin{enumerate}
    \item Fit a linear regression or generalized linear model for $h_S(A,X,Z) = \mathbb{E}[W \mid A,X,Z, S]$, say $W \sim A+X+Z +S +XS+ZS+XZS$.
    \item Obtain predictions for $\widehat{h}_S(A,X,Z)$ using the estimated coefficients from the first-stage regression.
    \item Fit a pooled generalized linear model with a log link for $\mathbb{E}[Y \mid A, X, Z, S]$ based on Result \ref{result:CRR} using $\widehat{h}_S(A,X,Z)$ and including interaction terms between $S$ and all other regressors.
    \item Test Equation \ref{eq:equalstarcoefsCRR} using a Wald test by jointly testing that the coefficients for the interaction terms involving $S$ with $A \cdot \widehat{h}_S(A,X,Z)$ and $AX$ (corresponding to $\beta_{au,1}^* - \beta_{au,0}^*$ and $\beta_{ax,1}^* - \beta_{ax,0}^*$) are both equal to 0, using the test statistic in Equation \ref{eq:coefficienttest}.
\end{enumerate}
Alternatively, one can test the null hypothesis of marginal reconcilability (Equation \ref{eq:nullRR}) on the multiplicative scale using a transportability approach.
\begin{result} \label{result:transportRR}
(Marginal reconcilability on the multiplicative scale) Suppose Assumptions \ref{assump:causal} and \ref{assump:CRRstructural} hold. Then for $s \in \{0,1\}$,
\begin{align*}
    &\mathbb{E}[\mu_1(s,X,U) \mid S = s]\\
    &= \mathbb{E}[\exp\{\beta_{0,s}^* + \beta_{a,s}^* + \beta_{ax,s}^* X + \beta_{x,s}^* X +\beta_{u,s}^* h_s(A,X,Z) + \beta_{au,s}^* h_s(A,X,Z) \}\mid S = s]
\end{align*}
If additionally $\beta_{u,0} = \beta_{u,1}$, then
\begin{align}
    &\mathbb{E}[\mu_0(s,X,U) \frac{\mu_1(1-s,X,U)}{\mu_0(1-s,X,U)} \mid S=s] \nonumber\\
    &= \mathbb{E}[\exp\{\beta_{0,s}^* + \beta_{a,1-s}^* + \beta_{ax,1-s}^* X + \beta_{x,s}^* X +\beta_{u,s}^* h_s(A,X,Z) + \beta_{au,1-s}^* h_s(A,X,Z) \}\mid S = s] \label{eq:transportRR}
\end{align}
\end{result}
As the denominators in the null hypothesis of marginal reconcilability on the multiplicative scale (Equation \ref{eq:nullRR}) are identical, it suffices to compare the corresponding numerators, which represent marginal counterfactual means under treatment. Result \ref{result:transportRR} identifies both the transported marginal counterfactual mean under treatment and the corresponding target-trial marginal counterfactual mean under treatment under Assumptions \ref{assump:causal} and \ref{assump:CRRstructural}. Consequently, a substantial discrepancy between the transported and untransported estimates based on Equation \ref{eq:transportRR} would provide evidence against the null hypothesis of marginal reconcilability on the multiplicative scale, provided the remaining assumptions hold. Under marginal randomization, the untransported estimator could be replaced by the sample mean among individuals assigned $A = 1$ in trial $s$. A test statistic may then be constructed analogously to Equation \ref{eq:transporttest}.

Under the additional assumption of transportability of the latent CRR across trials, Result \ref{result:transportRR} provides an identification result for the marginal RR in trial $s$ via transportability. This extends previous work on transportability on the multiplicative scale to allow for unmeasured effect modifiers, albeit under the structural models in Assumption \ref{assump:CRRstructural} \citep{dahabreh2024learning}. An additional assumption required for the multiplicative scale is that $\beta_{u,0} = \beta_{u,1}$, which implies invariance of the effect of the unmeasured covariate $U$ on $Y$ across trials in the individuals assigned $A = 0$. This assumption may be difficult to justify in practice, providing further statistical motivation for the coefficient testing approach when the primary goal is to test conditional reconcilability on the multiplicative scale.

Lastly, we provide guidance on interpreting the proposed hypothesis tests. Suppose that the transportability-based procedure or coefficient testing procedure yields a significant test, thereby rejecting the null hypothesis of conditional reconcilability on the multiplicative scale. Such a conclusion generally implies both of the following (when the assumptions of the test hold):
\begin{enumerate}
    \item There exist unmeasured or otherwise omitted effect modifiers $V$ on the multiplicative scale beyond those included in $X$ and the hypothesized $U$ which are not proxied by $(Z,W)$.
    \item $V$ is conditionally imbalanced between trials such that $V \not\perp \! \! \! \perp S \mid X,U$, or there are distinct baseline risks across trials such that $\mu_0(1,x,u,v) \neq \mu_0(0,x,u,v)$, or both.
\end{enumerate}
While the first implication is identical to that on the additive scale, the second implication is different. In other words, while rejection of conditional reconcilability still implies the existence of unmeasured or otherwise unaccounted-for effect modifiers, these effect modifiers need not be conditionally imbalanced between trials. Instead, differences in baseline risk alone may prevent reconciliation on the multiplicative scale. Such a scenario may arise if the baseline and treated risks as functions of $V$ both differ between trials, but in a manner that preserves equality of the CRR conditional on $(X,U,V)$ across trials. This phenomenon arises because, when marginalizing a conditional risk ratio to obtain a marginal risk ratio, the weights used for marginalization depend not only on the covariate distribution but also on the baseline risk \citep{huitfeldt2019collapsibility}.

We illustrate this point in more detail. Let $V$ denote all effect modifiers of $A$ on $Y$ on the multiplicative scale not already included in $X$ and $U$, allowing $V$ to be empty if $X$ and $U$ together already comprise the full set of effect modifiers. For clarity of exposition, take $V$ to be a discrete scalar. Since $(X,U,V)$ denotes the complete set of effect modifiers on the multiplicative scale, the CRR must satisfy
\begin{align*}
    \frac{\mu_1(1,x,u,v)}{\mu_0(1,x,u,v)} = \frac{\mu_1(0,x,u,v)}{\mu_0(0,x,u,v)} \text{ for a.e. }(x,u,v).
\end{align*}
Taking iterated expectations with respect to $V$, it follows that a violation of reconcilability at the level of $(X,U)$, i.e., $\mu_1(1,x,u)/\mu_0(1,x,u) \neq \mu_1(0,x,u)/\mu_0(0,x,u)$ implies
\begin{align*}
    &\frac{\mu_1(1,x,u)}{\mu_0(1,x,u)}\\
    &=\frac{\sum_{v} \frac{\mu_1(1,x,u,v)}{\mu_0(1,x,u,v)}\mu_0(1,x,u,v)P(V=v \mid S = 1,X=x,U=u)}{\sum_{v}\mu_0(1,x,u,v)P(V=v \mid S = 1,X=x,U=u)} \\
    &\neq \frac{\sum_{v} \frac{\mu_1(0,x,u,v)}{\mu_0(0,x,u,v)}\mu_0(0,x,u,v)P(V=v \mid S = 0,X=x,U=u)}{\sum_{v}\mu_0(0,x,u,v)P(V=v \mid S = 0,X=x,U=u)}\\
    &= \frac{\mu_1(0,x,u)}{\mu_0(0,x,u)}
\end{align*}
Hence, the relative risk in each trial conditional on $(X,U)$ can be expressed as a weighted average of the relative risk conditional on $(X,U,V)$, with weights
\begin{align*}
    w(v \mid s,x,u) \propto \mu_0(s,x,u,v) P(V=v \mid S = s,X=x,U=u)
\end{align*}
The weights depend on both the conditional distribution of $V$ and the conditional baseline risk. Consequently, even if $V$ has the same conditional distribution in both trials, trial-specific baseline risks can yield different CRRs after marginalizing over $V$.

When the transportability-based procedure is significant, then this additionally rejects the null hypothesis of marginal reconcilability on the multiplicative scale, under the assumptions of the test. Interpretation is analogous to rejection of marginal reconcilability on the additive scale. In particular, the preceding conclusions would continue to hold, but rejection further indicates that the differences in the latent CRR between trials are such that they persist even after being marginalized over a common distribution.

\appsection{Appendix B: Proof of main results}
\label{app:proofs}
\label{app:appendixA}
\appsubsection{Proof of Result \ref{result:CATE}}
We closely follow the steps of \citet{liu2025regression}. Taking the expectation with respect to $U$ conditional on $(A,X,Z,S=s)$ for both structural equation models, we have
\begin{align*}
    \mathbb{E}[Y \mid A, X, Z, S = s] &= \beta_{0,s} + \beta_{a,s} A + \beta_{au,s} A \mathbb{E}[U \mid A, X, Z, S = s] + \beta_{u,s} \mathbb{E} [U \mid A, X, Z, S = s]\\
    &\quad + \beta_{x,s} X + \beta_{ax,s} AX\\
    \mathbb{E}[W \mid A, X, Z, S = s] &= \alpha_0 + \alpha_u \mathbb{E}[U \mid A, X, Z, S = s] + \alpha_x X.
\end{align*}
Solving for $\mathbb{E}[U \mid A, X,Z, S = s]$ in the second equation and plugging into the first equation yields:
\begin{align*}
&\mathbb{E}[Y \mid A, X, Z, S = s]\\
&= \beta_{0,s} + \beta_{a,s} A + \beta_{au,s} A \left(\frac{\mathbb{E}[W \mid S=s, A,Z, X] - \alpha_{0} - \alpha_{x} X}{\alpha_{u}} \right)\\
&\quad + \beta_{u,s} \left( \frac{\mathbb{E}[W \mid S=s, A,Z, X] - \alpha_{0} - \alpha_{x} X}{\alpha_{u}} \right) +  \beta_{x,s} X + \beta_{ax,s} AX\\
&= \beta_{0,s} - \frac{\beta_{u,s} \alpha_0}{\alpha_u} + \left( \beta_{a,s} - \frac{\beta_{au,s} \alpha_0}{\alpha_u} \right)A + \left( \beta_{ax,s} - \frac{\beta_{au,s} \alpha_x}{\alpha_u} \right)AX + \frac{\beta_{u,s}}{\alpha_u} \mathbb{E}[W \mid S = s, A,Z, X]\\
&\quad +\left( \beta_{x,s} - \frac{\beta_{u,s} \alpha_x}{\alpha_u} \right)X + \frac{\beta_{au,s}}{\alpha_u} A \mathbb{E}[W \mid S = s, A,Z, X]\\
&= \beta_{0,s}^* + \beta_{a,s}^*A + \beta_{ax,s}^* AX + \beta_{u,s}^* h_s(A,X,Z) + \beta_{x,s}^* X + \beta_{au,s}^* A h_s(A,X,Z)
\end{align*}
where $h_s(A,X,Z) = \mathbb{E}[W \mid S = s, A,X, Z]$ and
\begin{align*}
    &\beta_{0,s}^* = \beta_{0,s} - \frac{\beta_{u,s} \alpha_0}{\alpha_u}\\
    &\beta_{a,s}^* = \beta_{a,s} - \frac{\beta_{au,s} \alpha_0}{\alpha_u}\\
    &\beta_{ax,s}^* = \beta_{ax,s} - \frac{\beta_{au,s} \alpha_x}{\alpha_u}\\
    &\beta_{u,s}^* = \frac{\beta_{u,s}}{\alpha_u}\\
    &\beta_{x,s}^* = \beta_{x,s}  - \frac{\beta_{u,s} \alpha_x}{\alpha_u}\\
    &\beta_{au,s}^* = \frac{\beta_{au,s}}{\alpha_u}.
\end{align*}

\appsubsection{Proof of Result \ref{result:transportATE}}
We identify the untransported average treatment effect in the $S = s$ population as
\begin{align*}
&\mathbb{E}[\mu_1(s,X,U)-\mu_0(s,X,U)\mid S=s]\\
&= \mathbb{E}[\beta_{a,s}+\beta_{au,s}U+\beta_{ax,s}X\mid S=s]\\
&= \beta_{a,s}+\beta_{au,s}\mathbb{E}[U\mid S=s]
   +\beta_{ax,s}\mathbb{E}[X\mid S=s]\\
&= \beta_{a,s}
+\frac{\beta_{au,s}}{\alpha_u}
\{\mathbb{E}[W\mid S=s]-\alpha_0-\alpha_x\mathbb{E}[X\mid S=s]\}
+\beta_{ax,s}\mathbb{E}[X\mid S=s]\\
&= \beta_{a,s}^*
+\beta_{au,s}^*\mathbb{E}[W\mid S=s]
+\beta_{ax,s}^*\mathbb{E}[X\mid S=s].
\end{align*}
Here, the first equality follows from the additive structural model for the latent
CATE and the third equality follows by marginalizing the proxy structural model
for $W$, which gives
$\mathbb{E}[W\mid S=s]=\alpha_0+\alpha_u\mathbb{E}[U\mid S=s]
+\alpha_x\mathbb{E}[X\mid S=s]$.

Similarly, we identify the transported average treatment effect in the $S = s$ population as
\begin{align*}
&\mathbb{E}[\mu_1(1-s,X,U)-\mu_0(1-s,X,U)\mid S=s]\\
&= \mathbb{E}[\beta_{a,1-s}+\beta_{au,1-s}U+\beta_{ax,1-s}X\mid S=s]\\
&= \beta_{a,1-s}
+\beta_{au,1-s}\mathbb{E}[U\mid S=s]
+\beta_{ax,1-s}\mathbb{E}[X\mid S=s]\\
&= \beta_{a,1-s}
+\frac{\beta_{au,1-s}}{\alpha_u}
\{\mathbb{E}[W\mid S=s]-\alpha_0-\alpha_x\mathbb{E}[X\mid S=s]\}
+\beta_{ax,1-s}\mathbb{E}[X\mid S=s]\\
&= \beta_{a,1-s}^*
+\beta_{au,1-s}^*\mathbb{E}[W\mid S=s]
+\beta_{ax,1-s}^*\mathbb{E}[X\mid S=s].
\end{align*}

\appsubsection{Proof of Result \ref{result:CRR}}
We closely follow the steps for the proof of Result \ref{result:CATE}. Taking the expectation over $U$ conditional on $(A,X,Z,S=s)$ for the conditional mean of $W$, we again have
\begin{align*}
&\mathbb{E}[W \mid A, X, Z, S=s] = \alpha_0 + \alpha_u m_s(A,X,Z) + \alpha_x X
\end{align*}
where we define $m_s(A,X,Z) = \mathbb{E}[U \mid A,X,Z,S=s]$. Further, define $C_s(A,X) = \exp(\beta_{0,s}+\beta_{a,s}A + \beta_{ax,s} AX+\beta_{x,s}X)$. For the conditional outcome model, we have 
\begin{align*}
&\mathbb{E}[Y \mid A, X,Z, S=s]\\
&= \mathbb{E}[\mathbb{E}[Y \mid A, U, X,Z, S = s] \mid A, X, Z, S = s]\\
&= \mathbb{E}[\exp(\beta_{0,s} + \beta_{a,s}A+ \beta_{au,s} AU + \beta_{u,s} U + \beta_{ax,s} AX+\beta_{x,s}X) \mid A, X, Z, S=s]\\
&= \exp(\beta_{0,s}+\beta_{a,s}A + \beta_{ax,s} AX+\beta_{x,s}X) \mathbb{E}[\exp(\beta_{au,s} AU + \beta_{u,s} U) \mid A, X,Z, S=s]\\
&= C_s(A,X) \mathbb{E}[\exp(\{m_s(A,X,Z) + \epsilon \} \{ \beta_{au,s} A + \beta_{u,s}\}) \mid A, X,Z, S=s]\\
&= C_s(A,X)\exp(m_s(A,X,Z)\{\beta_{au,s} A + \beta_{u,s}\}) \mathbb{E}[\exp(\epsilon\{\beta_{au,s}A + \beta_{u,s}\})]\\
&= C_s(A,X)\exp(m_s(A,X,Z)\{\beta_{au,s} A + \beta_{u,s}\} + \log M_{\epsilon}(\beta_{au,s} A + \beta_{u,s}))\\
&= C_s(A,X)\exp(m_s(A,X,Z)\{\beta_{au,s} A + \beta_{u,s}\} + A \log \left( \frac{M_{\epsilon}(\beta_{au,s} + \beta_{u,s})}{M_{\epsilon}(\beta_{u,s})} \right) + \log M_{\epsilon} (\beta_{u,s}))\\
&= \exp(\widetilde{\beta}_{0,s} + \widetilde{\beta}_{a,s} A + \beta_{ax,s} AX+\beta_{x,s}X + \beta_{au,s} m_s(A,X,Z) A + \beta_{u,s} m_s(A,X,Z))
\end{align*}
where $\widetilde{\beta}_{0,s} = \beta_{0,s} + \log M_{\epsilon}(\beta_{u,s})$ and $\widetilde{\beta}_{a,s} = \beta_{a,s} + \log \left( \frac{M_{\epsilon}(\beta_{au,s} + \beta_{u,s})}{M_{\epsilon}(\beta_{u,s})} \right)$.
Note that the second-to-last equality holds because
\begin{align*}
\log M_{\epsilon}(\beta_{au,s}A + \beta_{u,s}) &= A\log M_{\epsilon}(\beta_{au,s} + \beta_{u,s}) + (1 - A) \log M_{\epsilon}(\beta_{u,s})\\
&= A \log \left( \frac{M_{\epsilon}(\beta_{au,s} + \beta_{u,s})}{M_{\epsilon}(\beta_{u,s})} \right) + \log M_{\epsilon} (\beta_{u,s})
\end{align*}
Hence, we have the following two regression equations:
\begin{align*}
\mathbb{E}[W \mid A, X, Z, S =s] &= \alpha_0 + \alpha_xX + \alpha_u m_s(A,X,Z)\\
\log(\mathbb{E}[Y \mid A, X, Z, S =s]) &= \widetilde{\beta}_{0,s} + \widetilde{\beta}_{a,s} A + \beta_{ax,s} AX+\beta_{x,s}X + \beta_{au,s} m_s(A,X,Z) A\\
&\qquad + \beta_{u,s} m_s(A,X,Z)
\end{align*}
As in the proof of Result \ref{result:CATE}, we solve for $m_s(A,X,Z)$ in the marginalized conditional mean for $W$ and then plug the resulting expression into the marginalized conditional mean for $Y$.
\begin{align*}
&\log(\mathbb{E}[Y \mid A, X,Z, S = s])\\
&= \widetilde{\beta}_{0,s} + \widetilde{\beta}_{a,s}A + \beta_{ax,s} AX+\beta_{x,s}X+\beta_{au,s}A \left( \frac{h_s(A,X,Z) - \alpha_{0} - \alpha_xX}{\alpha_u}\right)\\
&\qquad + \beta_{u,s} \left( \frac{h_s(A,X,Z) - \alpha_{0} - \alpha_xX}{\alpha_u} \right)\\
&= \widetilde{\beta}_{0,s} + \widetilde{\beta}_{a,s}A + \beta_{ax,s} AX+\beta_{x,s}X+\frac{\beta_{au,s}A h_s(A,X,Z)}{\alpha_u} - \frac{\beta_{au,s} \alpha_{0} A}{\alpha_u} - \frac{\beta_{au,s}A \alpha_xX}{\alpha_u}\\
&\qquad + \frac{\beta_{u,s} h_s(A,X,Z)}{\alpha_u} - \frac{\beta_{u,s}\alpha_{0}}{\alpha_u} - \frac{\beta_{u,s} \alpha_x X}{\alpha_u}\\
&= \left( \widetilde{\beta}_{0,s} - \frac{\beta_{u,s}\alpha_{0}}{\alpha_u} \right) + \left(\widetilde{\beta}_{a,s} - \frac{\beta_{au,s} \alpha_{0}}{\alpha_u} \right) A + \left(\beta_{ax,s} - \frac{\beta_{au,s} \alpha_x}{\alpha_u} \right) AX+ \left(\beta_{x,s}  - \frac{\beta_{u,s} \alpha_x}{\alpha_u}\right)X\\
&\qquad+ \frac{\beta_{u,s}h_s(A,X,Z)}{\alpha_u} + \frac{\beta_{au,s}}{\alpha_u}A h_s(A,X,Z)\\
&= \beta_{0,s}^* + \beta_{a,s}^* A + \beta_{ax,s}^* AX+\beta_{x,s}^*X+\beta_{u,s}^* h_s(A,X,Z) + \beta_{au,s}^* A h_s(A,X,Z)
\end{align*}
where
\begin{align*}
\beta_{0,s}^* &= \widetilde{\beta}_{0,s} - \frac{\beta_{u,s}\alpha_{0}}{\alpha_u}= \beta_{0,s} + \log M_{\epsilon}(\beta_{u,s}) - \frac{\beta_{u,s}\alpha_0}{\alpha_u}\\
\beta_{a,s}^* &= \widetilde{\beta}_{a,s} - \frac{\beta_{au,s} \alpha_{0}}{\alpha_u} = \beta_{a,s} + \log \left( \frac{M_{\epsilon}(\beta_{au,s} + \beta_{u,s})}{M_{\epsilon}(\beta_{u,s})} \right) - \frac{\beta_{au,s} \alpha_0}{\alpha_u}\\
\beta_{ax,s}^* &= \beta_{ax,s} - \frac{\beta_{au,s} \alpha_x}{\alpha_u}\\
\beta_{x,s}^* &= \beta_{x,s}  - \frac{\beta_{u,s} \alpha_x}{\alpha_u}\\
\beta_{u,s}^* &= \frac{\beta_{u,s}}{\alpha_u}\\
\beta_{au,s}^* &= \frac{\beta_{au,s}}{\alpha_u}\\
\end{align*}

\appsubsection{Proof of Result \ref{result:transportRR}}
First, consider the untransported marginal mean under treatment in the
$S=s$ population.
\begin{align*}
&\mathbb{E}[\mu_1(s,X,U)\mid S=s]\\
&=
\mathbb{E}\left[
\exp\{\beta_{0,s}+\beta_{a,s}
      +(\beta_{u,s}+\beta_{au,s})U
      +(\beta_{x,s}+\beta_{ax,s})X\}
\mid S=s
\right]\\
&=
\mathbb{E}\left[
\exp\{\beta_{0,s}+\beta_{a,s}
      +(\beta_{x,s}+\beta_{ax,s})X\}
\mathbb{E}\left[
\exp\{(\beta_{u,s}+\beta_{au,s})U\}
\mid A,X,Z,S=s
\right]
\mid S=s
\right].
\end{align*}
Here, the first equality follows from Assumption \ref{assump:CRRstructural} and the second equality follows by iterated expectations. Note that the inside expectation can be rewritten as
\begin{align*}
&\mathbb{E}\left[
\exp\{(\beta_{u,s}+\beta_{au,s})U\}
\mid A,X,Z,S=s
\right]\\
&=
\exp\{(\beta_{u,s}+\beta_{au,s})m_s(A,X,Z)\}
M_\epsilon(\beta_{u,s}+\beta_{au,s}),
\end{align*}
which follows from Assumption \ref{assump:CRRstructural}. Therefore
\begin{align*}
&\mathbb{E}[\mu_1(s,X,U)\mid S=s]\\
&=
\mathbb{E}\left[
\exp\left\{
\beta_{0,s}^*
+\beta_{a,s}^*
+(\beta_{x,s}^*+\beta_{ax,s}^*)X
+(\beta_{u,s}^*+\beta_{au,s}^*)h_s(A,X,Z)
\right\}
\mid S=s
\right],
\end{align*}
which follows from substituting $m_s(A,X,Z) = \{h_s(A,X,Z) - \alpha_0 - \alpha_xX\}/\alpha_u$ and collecting terms. For the transported marginal mean under treatment in the $S = s$ population, we can similarly show that
\begin{align*}
&\mathbb{E}\left[
\mu_0(s,X,U)
\frac{\mu_1(1-s,X,U)}{\mu_0(1-s,X,U)}
\mid S=s
\right]\\
&=
\mathbb{E}\left[
\exp\{\beta_{0,s}+\beta_{a,1-s}
      +(\beta_{u,s}+\beta_{au,1-s})U
      +(\beta_{x,s}+\beta_{ax,1-s})X\}
\mid S=s
\right]\\
&=\mathbb{E}\left[
\exp\{\beta_{0,s}+\beta_{a,1-s}
      +(\beta_{x,s}+\beta_{ax,1-s})X\}
\mathbb{E}\left[
\exp\{(\beta_{u,s}+\beta_{au,1-s})U\}
\mid A,X,Z,S=s
\right]
\mid S=s
\right].
\end{align*}
The inside expectation can be written as
\begin{align*}
&\mathbb{E}\left[
\exp\{(\beta_{u,s}+\beta_{au,1-s})U\}
\mid A,X,Z,S=s
\right]\\
&=
\exp\{(\beta_{u,s}+\beta_{au,1-s})m_s(A,X,Z)\}
M_\epsilon(\beta_{u,s}+\beta_{au,1-s}).
\end{align*}
Substituting $m_s(A,X,Z) = \{h_s(A,X,Z) - \alpha_0 - \alpha_xX\}/\alpha_u$ and collecting terms yields that the transported estimand is equal to
\begin{align*}
&\mathbb{E}\Big[
\exp\{
\beta_{0,s}+\beta_{a,1-s}
+\log M_\epsilon(\beta_{u,s}+\beta_{au,1-s})-\frac{(\beta_{u,s}+\beta_{au,1-s})\alpha_0}{\alpha_u}\\
&+\left(\beta_{x,s}+\beta_{ax,1-s}
-\frac{(\beta_{u,s}+\beta_{au,1-s})\alpha_x}{\alpha_u}\right)X +\frac{\beta_{u,s}+\beta_{au,1-s}}{\alpha_u}h_s(A,X,Z)
\}
\mid S=s
\Big].
\end{align*}
We want to show that this expression is equal to
\begin{align*}
\mathbb{E}\Big[
\exp\{
\beta_{0,s}^*
+\beta_{a,1-s}^*
+\beta_{x,s}^*X
+\beta_{ax,1-s}^*X+\beta_{u,s}^*h_s(A,X,Z)
+\beta_{au,1-s}^*h_s(A,X,Z)
\}
\mid S=s
\Big].
\end{align*}
Plugging in the starred expressions, note that
\begin{align*}
&\beta_{0,s}^*+\beta_{a,1-s}^*\\
&=
\beta_{0,s}+\beta_{a,1-s}
-\frac{(\beta_{u,s}+\beta_{au,1-s})\alpha_0}{\alpha_u}+
\log M_\epsilon(\beta_{u,s})
+
\log\left\{
\frac{M_\epsilon(\beta_{u,1-s}+\beta_{au,1-s})}
     {M_\epsilon(\beta_{u,1-s})}
\right\}\\
&=\beta_{0,s}+\beta_{a,1-s} - \frac{(\beta_{u,s}+\beta_{au,1-s})\alpha_0}{\alpha_u}+\log M_\epsilon(\beta_{u,s} + \beta_{au,1-s})
\end{align*}
where the last equality holds if $\beta_{u,s}=\beta_{u,1-s}$. Then the desired equality holds.

\appsubsection{Asymptotic Variance of the Two-Stage Estimator}
\label{app:asymp}

In this section, we derive the asymptotic distribution of the two stage estimator 
$\widehat{\theta} = (\widehat{\alpha}^\top,\widehat{\beta}^{\top})^{\top}$ using standard estimating equation theory.

Let $V_1$ denote the $n \times p_\alpha$ model matrix for the first-stage GLM. Denote the $i$th row of the model matrix as $V_{1i}^{\top}$, which contains an intercept and functions of $(A_i, X_i, Z_i, S_i)$. The first-stage GLM is
\begin{align*}
    \mu_{1i}= \mathbb{E}[W_i \mid A_i, X_i,Z_i,S_i]
    =  g_1^{-1}(V_{1i}^{\top} \alpha).
\end{align*}
Define fitted values from the first-stage GLM as $\widehat h_i = \mu_{1i}(\widehat\alpha)$ and the linear predictor as $\eta_{1i}(\alpha)= V_{1i}^{\top} \alpha$.

Let $V_2(\alpha)$ denote the $n \times p_\beta$ second-stage model matrix, whose $i$th row is denoted by $V_{2i}(\alpha)$, which contains an intercept and functions of 
$(A_i, X_i, Z_i, S_i,\eta_{1i}(\alpha))$. The second-stage GLM is
\begin{align*}
    \mu_{2i} = \mathbb{E}[Y_i \mid A_i, X_i, Z_i, S_i] = g_2^{-1} (V_{2i}(\alpha)^{\top} \beta)
\end{align*}

Define the stacked estimating equations
\begin{align*}
\Psi_i(\theta)
=
\begin{pmatrix}
\Psi_{1i}(\alpha) \\[4pt]
\Psi_{2i}(\alpha,\beta)
\end{pmatrix},
\qquad
\theta = (\alpha^{\top},\beta^{\top})^{\top},
\end{align*}
where
\begin{align*}
    \Psi_{1i}(\alpha) &= V_{1i}(W_i - \mu_{1i})\\
    \Psi_{2i}(\alpha,\beta)
    &= V_{2i}(\alpha)(Y_i - \mu_{2i}).
\end{align*}
The estimator $\widehat{\theta}$ satisfies
\begin{align*}
\frac{1}{n}\sum_{i=1}^n \Psi_i(\widehat\theta) = 0.
\end{align*}

Under standard regularity conditions for M-estimators, e.g., Theorem 5.21 of \citet{van2000asymptotic},
\begin{align*}
\sqrt{n}(\widehat\theta - \theta_0)
\xrightarrow{d}
\mathcal{N}(0,A^{-1} B (A^{-1})^{\top}),
\end{align*}
where
\begin{align*}
    A = \mathbb{E}\left[\frac{\partial \Psi_i(\theta_0)}{\partial \theta^{\top}}\right],
\qquad
B = \mathbb{E}\left[\Psi_i(\theta_0)\Psi_i(\theta_0)^{\top}\right].
\end{align*}
We now derive expressions for $B$, which can be written as
\begin{align*}
    B =
\begin{pmatrix}
B_{11} & B_{12} \\
B_{21} & B_{22}
\end{pmatrix},
\end{align*}
where
\begin{align*}
B_{11}
= \mathbb{E}\left[V_{1i}V_{1i}^{\top} (W_i - \mu_{1i})^2\right],
\qquad
B_{22}
= \mathbb{E}\left[V_{2i}(\alpha_0)V_{2i}(\alpha_0)^{\top} (Y_i - \mu_{2i})^2\right],
\end{align*}
\begin{align*}
B_{12}
= \mathbb{E}\left[
    V_{1i}(W_i - \mu_{1i})
    \cdot
    V_{2i}(\alpha_0)(Y_i - \mu_{2i})
\right],
\qquad
B_{21}=B_{12}^{\top}.
\end{align*}
The Jacobian $A$ can be written as
\begin{align*}
A =
\begin{pmatrix}
A_{11} & 0 \\
A_{21} & A_{22}
\end{pmatrix}.
\end{align*}
For a canonical link GLM,
\begin{align*}
\frac{\partial \Psi_{1i}}{\partial\alpha^{\top}}
=
- V_{1i} V_{1i}^\top \dot\mu_{1i},
\end{align*}
where 
$\dot\mu_{1i} = \partial\mu_{1i}/\partial\eta_{1i}$ equals 
$1$ when the identity link is used for the first-stage regression and $\mu_{1i}$ when the Poisson log link is used in the first-stage regression. Thus
\begin{align*}
A_{11} = -\mathbb{E}[V_{1i}V_{1i}^\top \dot\mu_{1i}].
\end{align*}
Similarly,
\begin{align*}
\frac{\partial \Psi_{2i}}{\partial\beta^{\top}}
=
- V_{2i}(\alpha_0) V_{2i}(\alpha_0)^{\top} \dot\mu_{2i},
\end{align*}
with $\dot\mu_{2i}$ defined analogously to $\dot\mu_{1i}$, except for the second-stage regression. Hence
\begin{align*}
A_{22} = -\mathbb{E}[V_{2i}(\alpha_0)V_{2i}(\alpha_0)^\top \dot\mu_{2i}].
\end{align*}
We now turn our attention to the cross block $A_{21} = \mathbb{E}\left[\partial \Psi_{2i}(\alpha,\beta)/\partial \alpha^\top\right]$. We have that
\begin{align*}
\frac{\partial \Psi_{2i}(\alpha,\beta)}{\partial \alpha^\top}
=
\frac{\partial V_{2i}(\alpha)}{\partial \alpha^\top}\{Y_i-\mu_{2i}(\alpha,\beta)\}-
V_{2i}(\alpha)\,\frac{\partial \mu_{2i}(\alpha,\beta)}{\partial \alpha^\top}.
\end{align*}
We first derive an expression for the derivative of $V_{2i}(\alpha)$. For ease of notation, let
\begin{align*}
\Gamma_i = \frac{\partial V_{2i}(\alpha)}{\partial \eta_{1i}(\alpha)}
\end{align*}
denote the $p_{\beta}$ vector whose $j$th entry is equal to 0 if the $j$th component of $V_{2i}(\alpha)$ does not involve $\eta_{1i}(\alpha)$, and otherwise equals the multiplier of $\eta_{1i}(\alpha)$ in the $j$th component of $V_{2i}(\alpha)$. Then
\begin{align*}
\frac{\partial V_{2i}(\alpha)}{\partial \alpha^\top}
=
\Gamma_i \frac{\partial \eta_{1i}(\alpha)}{\partial \alpha^\top} = \Gamma_i V_{1i}^{\top}
\end{align*}
Next, we derive the derivative of $\mu_{2i}(\alpha,\beta)$. We have
\begin{align*}
\frac{\partial \mu_{2i}(\alpha,\beta)}{\partial \alpha^\top} = \dot\mu_{2i} \frac{\partial\{ V_{2i} (\alpha)^{\top} \beta \}}{\partial \alpha^{\top}}= \dot \mu_{2i} \Gamma_i^\top\beta
V_{1i}^\top.
\end{align*}
Hence,
\begin{align*}
A_{21}
&=
\mathbb{E}[\Gamma_iV_{1i}^\top\{Y_i-\mu_{2i}\}
-
V_{2i}(\alpha_0)
\dot\mu_{2i}
\Gamma_i^\top\beta_0 V_{1i}^\top]\\
\end{align*}
The asymptotic distribution above motivates the following standard variance estimator. Let
\begin{align*}
    \widehat{A} = n^{-1} \sum_{i=1}^n \frac{\partial \Psi_i(\widehat{\theta})}{\partial \theta^{\top}},
\qquad
\widehat{B} = n^{-1} \sum_{i=1}^n \Psi_i(\widehat{\theta})\Psi_i(\widehat{\theta})^{\top}.
\end{align*}
Then
\begin{align*}
\widehat{\mathrm{Var}}(\widehat\theta)
=
\frac{1}{n}\,
\widehat A^{-1}
\widehat B
(\widehat A^{-1})^\top.
\end{align*}

\begin{figure}[ht]
\caption{An interaction DAG illustrating violations of conditional reconcilability, indicated by red arrows. $\Delta Y(a)$ denotes the conditional causal effect of $A$ on $Y$ on a given scale.}
\centering
\label{fig:DAGs_alt}
\includegraphics[width=.35\textwidth]{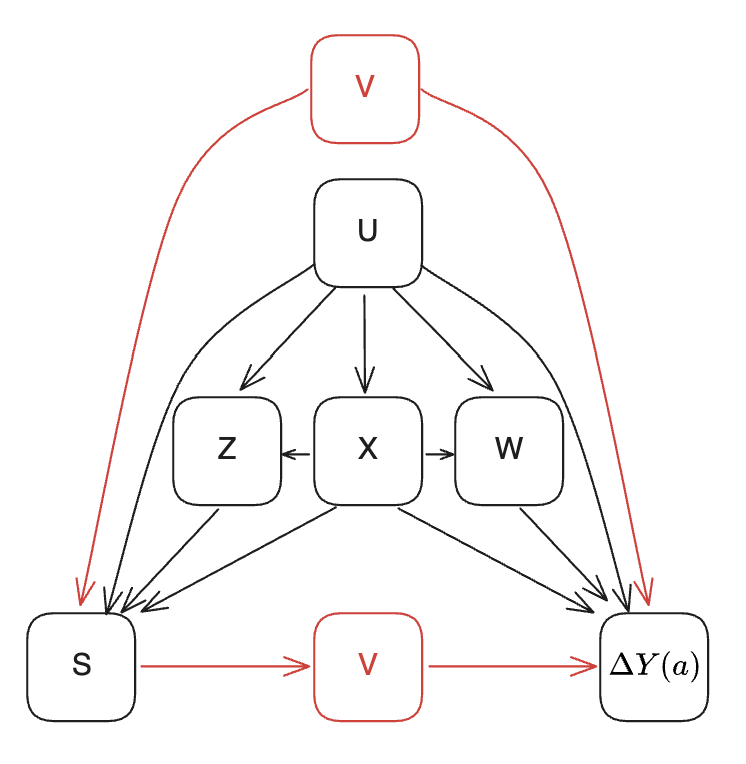}
\end{figure}

\appsection{Appendix C: Additional results}
\label{app:additional}
\label{app:appendixB}
\appsubsection{DAG illustrating violations of conditional reconcilability}
\label{app:DAGs_alternative}
Figure \ref{fig:DAGs_alt} illustrates one way that conditional reconcilability can fail. The omitted effect modifier $V$ need not be measured at or before study entry or treatment assignment. It may represent baseline population differences not captured by $(X,U)$ or differences induced by trial enrollment, including protocols, standards of care, or other trial-engagement effects \citep{dahabreh2019generalizing, ung2025generalizing}.

\appsubsection{Alternatives to Figure \ref{fig:DAGs}}
\label{app:altDAGs}
\begin{figure}[ht]
\caption{Alternative DAGs that satisfy Assumption \ref{assump:exchangeability} under the null hypothesis of conditional reconcilability. Red arrows indicate that these arrows have been flipped with respect to the original DAG in Figure \ref{fig:DAGs}.}
\centering
\label{fig:appendixDAGs}
\includegraphics[width=\textwidth]{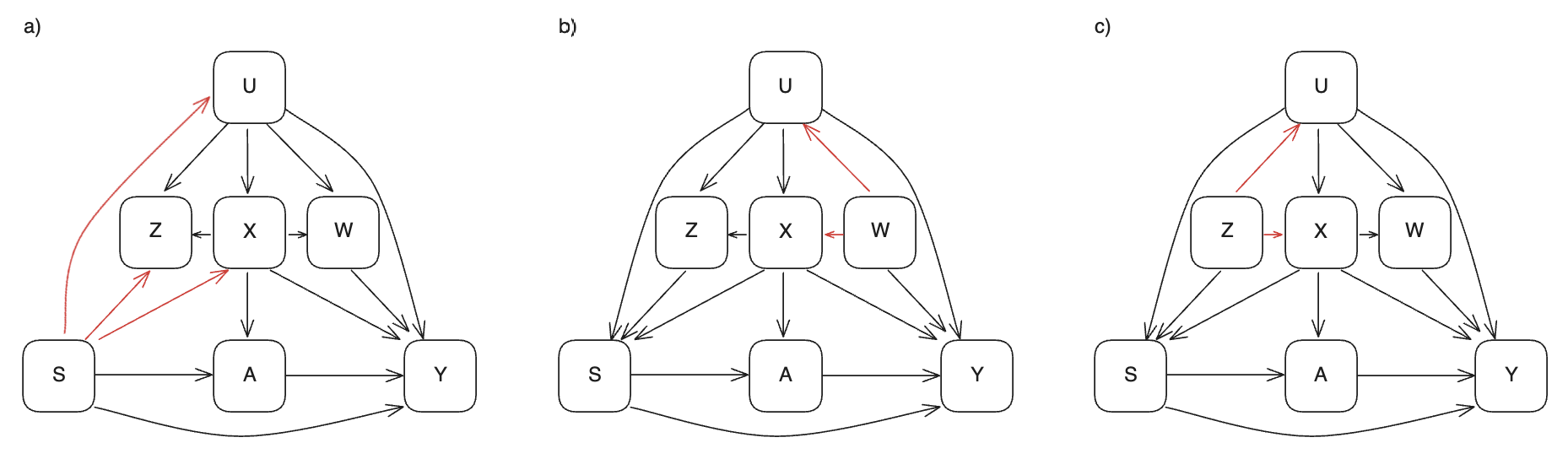}
\end{figure}

In Figure \ref{fig:appendixDAGs}, we depict additional (but not all) DAGs that satisfy Assumption \ref{assump:exchangeability} under the null hypothesis of conditional reconcilability. Figure \ref{fig:appendixDAGs}(a) illustrates a setting in which measured and unmeasured baseline covariates are downstream of trial participation $S$ \citep{su2025proximal}. In contrast to the primary DAG in Figure \ref{fig:DAGs}, which encodes differences in covariate distributions arising from selective enrollment of different populations into each trial, this alternative structure allows for trial participation itself to causally influence baseline covariates. This may arise if the two trials have different pre-randomization protocols, which in turn affect measured and/or unmeasured baseline covariates. Additional DAGs compatible with Assumption \ref{assump:exchangeability} in which $S$ is upstream of other covariates can be found in Figure S1 of \citet{su2025proximal}.

Figures \ref{fig:appendixDAGs}(b) and (c) depict alternative structures in which either the adjustment proxy $W$ or the reweighting proxy $Z$ is upstream of the unmeasured effect modifier $U$. These DAGs correspond to scenarios in which a measured baseline covariate causally influences the latent effect modifier, rather than only playing the role of a proxy. For example, measured behavioral factors (e.g., smoking status, alcohol use) may affect underlying biological pathways, which in turn may be effect modifiers.

\appsubsection{Robustness to working model misspecification and relaxation of parametric models}
\label{app:robustness}

In this section, we show that the proposed coefficient testing procedure in Section \ref{ss:hypothesisadditive} remains a valid test of the null hypothesis of conditional reconcilability on the additive scale, even if the first-stage regression is misspecified, under the following structural equation models:
\begin{assumption} \label{assump:CATEstructuralrelaxed}
    (Additive structural models). For $s \in \{0,1\}$,
\begin{align*}
    &\mathbb{E}[Y \mid A,U,X,Z,S=s]= m_s(U,X,Z)+ A(\beta_{a,s} + \beta_{au,s}U + \beta_{ax,s}X)\\
    &\mathbb{E}[W \mid A, U, X,Z, S = s] = \alpha_0 + \alpha_u U + \alpha_x X
\end{align*}
where $|\mathbb{E}[U \mid A, X, Z,S=s]| < \infty$ and $\alpha_u \neq 0$.
\end{assumption}
Assumption \ref{assump:CATEstructuralrelaxed} is a relaxed version of Assumption \ref{assump:CATEstructural} proposed in the main text. Namely, in the relaxed assumption, we do not require specification of the baseline outcome model $m_s(U,X,Z)$. Instead, we only require that the treatment effect is linear in $(U,X)$. In the following, we also work under the simplifying assumption that $A$ is marginally randomized and the $Z$ proxy satisfies certain conditions:
\begin{assumption} 
\label{assump:causalmarginal} (Causal consistency and randomization). Assume
    \begin{enumerate}[(i)]
        \item (Consistency) $Y(a) = Y$ whenever $A = a \in \{0,1\}$
        \item (Exchangeability) $\{Y(1),Y(0),U,Z,X\} \indep A \mid (S = s)$ for $s \in \{0,1\}$
    \end{enumerate}
\end{assumption}
The exchangeability assumption would be satisfied if $Z$ is measured at (or before) baseline. Otherwise, it would generally require that $A$ has no causal effect on $Z$. Under Assumptions \ref{assump:CATEstructuralrelaxed} and \ref{assump:causalmarginal}, the latent CATE can still be represented as
\begin{align*}
    &\mathbb{E}[Y(1) - Y(0) \mid U = u,X=x,S=s]\\
    &= \mathbb{E}[ \mathbb{E}[Y \mid A=1,U=u,X=x,S=s,Z] \mid U=u,X=x,S=s]\\
    &\qquad -\mathbb{E}[ \mathbb{E}[Y \mid A=0,U=u,X=x,S=s,Z] \mid U=u,X=x,S=s]\\
    &= \beta_{a,s} + \beta_{au,s} u +\beta_{ax,s}x
\end{align*}
and hence conditional reconcilability would still require that
\begin{align}
    \beta_{a,0} = \beta_{a,1}, \qquad \beta_{au,0} = \beta_{au,1}, \qquad \beta_{ax,0} = \beta_{ax,1}.
\end{align}
By following a proof similar to that of Result \ref{result:CATE}, it is not hard to show that under Assumptions \ref{assump:CATEstructuralrelaxed} and \ref{assump:causalmarginal}
\begin{align*}
    \mathbb{E}[Y \mid A, X, Z, S = s] = \widetilde{m}_s(X,Z) + A\{\beta_{a,s}^* + \beta_{ax,s}^* X + \beta_{au,s}^* h_s(X,Z) \},
\end{align*}
where $\widetilde{m}_s(X,Z) = \mathbb{E}[m_s(U,X,Z) \mid X,Z,S=s]$ and the starred coefficients are identical to the ones in the main text. We show the following result. For simplicity, we consider a version of the two-stage procedure where models are fit within each stratum $S = s$ rather than with interactions. However, as long as all interactions with $S$ and the other terms are fit, the two approaches are equivalent.
\begin{result}[Robustness to first-stage working-model misspecification]
\label{result:robust_working_model}
Suppose Assumptions \ref{assump:CATEstructuralrelaxed} and
\ref{assump:causalmarginal} hold. Let
$b(X,Z)$ denote the vector of basis functions used in the first-stage
working regression, with $(1,X^\top,Z^\top)^\top$ contained in the linear span of
$b(X,Z)$. For each $s \in \{0,1\}$, define $\widetilde h_s(X,Z)$ as the
population least-squares projection of $W$ onto $b(X,Z)$ among individuals
with $S=s$:
\begin{align*}
    \widetilde h_s(X,Z)=b(X,Z)^\top\widetilde\gamma_s,\qquad
    \mathbb{E}\!\left[
    b(X,Z)\{W-b(X,Z)^\top\widetilde\gamma_s\}\mid S=s
    \right]=0.
\end{align*}
The first-stage working model need not be correctly specified, so
$\widetilde h_s(X,Z)$ need not equal
$h_s(X,Z)=\mathbb{E}(W\mid X,Z,S=s)$.

For each $s$, let $V_s(X,Z)$ be a vector of baseline regressors such that
each component of $V_s(X,Z)$ lies in the linear span of $b(X,Z)$, and such
that the linear span of $V_s(X,Z)$ contains
$(1,X^\top,\widetilde h_s(X,Z))^\top$. Define
\[
    \widetilde V_s(X,Z,A)
    =
    \{V_s(X,Z)^\top,\ A,\ A X^\top,\ A\widetilde h_s(X,Z)\}^\top.
\]
Let $\beta_{v,s}^*$ denote the population least-squares coefficient from
projecting $\widetilde m_s(X,Z)$ onto $V_s(X,Z)$ among individuals with
$S=s$, i.e.,
\[
    \mathbb{E}\!\left[
    V_s(X,Z)\{\widetilde m_s(X,Z)-V_s(X,Z)^\top\beta_{v,s}^*\}
    \mid S=s
    \right]=0.
\]
Then
\begin{align}
\label{eq:misspecifiedEE}
\mathbb{E}\!\left[
\widetilde V_s
\left\{
Y
- V_s^\top\beta_{v,s}^*
- \beta_{a,s}^*A
- A X^\top\beta_{ax,s}^*
- \beta_{au,s}^* A\widetilde h_s
\right\}
\mid S=s
\right]
=0,
\end{align}
where
\begin{align*}
    \beta_{a,s}^*
    &= \beta_{a,s}-\frac{\beta_{au,s}\alpha_0}{\alpha_u},\\
    \beta_{ax,s}^*
    &= \beta_{ax,s}-\frac{\beta_{au,s}\alpha_x}{\alpha_u},\\
    \beta_{au,s}^*
    &= \frac{\beta_{au,s}}{\alpha_u}.
\end{align*}
\end{result}
Under mild regularity conditions ensuring that the solutions to the population estimating equation in (\ref{eq:misspecifiedEE}) are unique, the above result suggests that solutions to the proposed two-stage procedure correspond to the desired $(\beta_{a,s}^*, \beta_{au,s}^*, \beta_{ax,s}^*)$. This holds even if one were to misspecify the first-stage regression such that $\mathbb{E}[W \mid X=x,Z=z,S=s] \neq \widetilde{h}_s(x,z)$ and similarly if one misspecifies the baseline outcome model such that $\mathbb{E}[Y \mid A=0,X=x,Z=z,S=s] \neq v_s^{\top} \beta_{v,s}^*$. We provide a proof of the above result, which draws some parallels to proofs of the robustness of two-stage least squares to first-stage misspecification in the context of instrumental variable analyses.
\begin{proof}
    We can rewrite the left-hand side of Equation \ref{eq:misspecifiedEE} as
\begin{align*}
&\mathbb{E}[\widetilde V_s (Y - V_s^{\top} \beta_{v,s}^* - \widetilde m_s + \widetilde m_s - \beta_{a,s}^*A- \beta_{ax,s}^* AX - \beta_{au,s}^*A \widetilde h_s - \beta_{au,s}^*Ah_s + \beta_{au,s}^*Ah_s) \mid S = s]\\
&= \mathbb{E}[\widetilde V_s (Y  - \widetilde m_s - \beta_{a,s}^*A- \beta_{ax,s}^* AX  - \beta_{au,s}^*Ah_s ) \mid S = s]\\
&\quad + \mathbb{E}[ \widetilde{V}_s(\widetilde m_s - V_s^{\top} \beta_{v,s}^*)\mid S = s] + \beta_{au,s}^* \mathbb{E}[\widetilde V_s A(h_s - \widetilde h_s) \mid S = s]
\end{align*}
The first term will be 0 because $\widetilde V_s$ is a function of $(A,X,Z)$, and the residual in the expectation is centered at the true conditional mean of $Y$ (under Assumptions \ref{assump:CATEstructuralrelaxed} and \ref{assump:causalmarginal}). The second term will be 0 if
\begin{align*}
&\mathbb{E}[\widetilde m_s - V_s^{\top} \beta_{v,s}^* \mid S = s] = 0,\\
&\mathbb{E}[V_s(\widetilde m_s - V_s^{\top} \beta_{v,s}^*) \mid S = s] = 0,\\
&\mathbb{E}[A(\widetilde m_s - V_s^{\top} \beta_{v,s}^*) \mid S = s] = 0,\\
&\mathbb{E}[AX(\widetilde m_s - V_s^{\top} \beta_{v,s}^*) \mid S = s] = 0,\\
&\mathbb{E}[A \widetilde h_s(\widetilde m_s - V_s^{\top} \beta_{v,s}^*) \mid S = s] = 0.
\end{align*}
The first four terms can be shown to be 0 if the basis for $V_s$ includes $(1, X)$ and under marginal randomization of $A$, which lets us factor out $\mathbb{E}[A \mid S = s]$ from the above expressions. The fifth term will be 0 if $\widetilde h_s$ lies in the subspace spanned by the basis of $V_s$. Given that we assume the linear span of $V_s$ contains $(1,X,\widetilde h_s)$, the above will be satisfied.

Lastly, we analyze $\beta_{au,s}^* \mathbb{E}[\widetilde V_s A(h_s - \widetilde h_s) \mid S = s]$. Note that $\widetilde V_s A = (V_sA, A, AX, A \widetilde h_s)$. Hence, the relevant moment equations are:
\begin{align*}
&\mathbb{E}[A(h_s - \widetilde h_s) \mid S = s]=0\\
&\mathbb{E}[V_sA(h_s - \widetilde h_s) \mid S = s]=0\\
&\mathbb{E}[AX(h_s - \widetilde h_s) \mid S = s]=0\\
&\mathbb{E}[A \widetilde h_s(h_s - \widetilde h_s) \mid S = s]=0
\end{align*}
By marginal randomization of $A$, we can factor out an $\mathbb{E}[A \mid S = s]$. Then, all the terms can be shown to be equal to 0 as long as $(1,V_s,X,\widetilde h_s)$ lies in the span of the basis of the first-stage regression $b(X,Z)$, which holds by assumption.
\end{proof}
In general, other modifications to the proposed assumptions would cause the procedure to lose robustness guarantees. For example, if the model for $W$ were specified with a log link, such as in Appendix \ref{app:logproxy}, then we generally lose robustness to misspecification of the first-stage regression but can maintain robustness to misspecification of the baseline outcome model, so long as we include $\log h_s$ rather than $h_s$ in $V_s$. Intuitively, the reason that we would lose this robustness is because the final term would be a residual term of the form $\mathbb{E}[\widetilde V A(\log h_s - \log \widetilde h_s)]$, which would not necessarily be equal to 0 under population least squares. Likewise, if the model for $Y$ is specified with a log link (even if the model for $W$ is specified with an identity link), we would likely lose both forms of robustness, for similar reasons. While the above proofs make the assumption of marginal randomization, this assumption can be relaxed to conditional randomization (with $Z$ included in the conditional randomization statement); however, this will generally require the first-stage regression to include functions of the propensity score $\mathbb{E}[A \mid X] = \pi_s(X)$ in the basis. It will also require existing interaction terms with $A$ in the second-stage regression to be fitted as centered interaction terms $(A - \pi_s(X))$.

\appsubsection{Log model for the adjustment proxy}
\label{app:logproxy}
In this section, we consider alternatives to the structural equation model for $W$ presented in Assumptions \ref{assump:CATEstructural} and \ref{assump:CRRstructural}. In particular, consider
\begin{align*}
    \log (\mathbb{E}(W \mid A,U,X,Z,S=s)) = \alpha_0 + \alpha_u U + \alpha_x X
\end{align*}
which specifies a structural model for the conditional mean of $W$ on the log scale. We now derive results analogous to Results \ref{result:CATE} and \ref{result:CRR}. Taking the expectation over $U$ conditional on $(A,X,Z,S=s)$, we have
\begin{align*}
&\mathbb{E}[W \mid A, X, Z, S=s]\\
&= \mathbb{E}[\mathbb{E}[W \mid A, U, X, Z, S = s] \mid A, X, Z, S = s]\\
&= \mathbb{E}[\exp(\alpha_0 + \alpha_{u} U + \alpha_{x} X) \mid A, X, Z, S=s]\\
&= \exp(\alpha_0 + \alpha_xX) \mathbb{E}[\exp(\alpha_u U) \mid A, X, Z, S=s]\\
&= \exp(\alpha_0 + \alpha_x X) \mathbb{E}[\exp(\alpha_u (m_s(A,X,Z) + \epsilon)) \mid A, X, Z, S=s]\\
&= \exp(\alpha_0 + \alpha_x X+\alpha_u m_s(A,X,Z)) \mathbb{E}[\exp(\alpha_u \epsilon)]\\
&= \exp(\alpha_0 + \alpha_x X + \alpha_u m_s(A,X,Z) + \log M_{\epsilon}(\alpha_u))\\
&= \exp(\widetilde{\alpha}_{0} + \alpha_xX +\alpha_u m_s(A,X,Z))
\end{align*}
where $\widetilde{\alpha}_{0} = \alpha_0 + \log M_{\epsilon}(\alpha_u)$ and $m_s(A,X,Z) = \mathbb{E}[U \mid A,X,Z,S=s]$. We first consider the additive scale, i.e., if the model for the conditional mean of $Y$ is specified according to Assumption \ref{assump:CATEstructural}. From the proof of Result \ref{result:CATE}, we have that
\begin{align*}
    &\mathbb{E}[Y \mid A,X,Z,S=s]\\
    &= \beta_{0,s} + \beta_{a,s} A + \beta_{au,s} A m_s(A,X,Z) + \beta_{u,s} m_s(A,X,Z) + \beta_{x,s} X + \beta_{ax,s} AX\\
\end{align*}
Solving for $m_s(A,X,Z)$ in the marginalized conditional mean for $W$ and then plugging the resulting expression into the marginalized conditional mean for $Y$ yields
\begin{align*}
&\mathbb{E}[Y \mid A, X, Z, S = s]\\
&= \beta_{0,s} + \beta_{a,s} A + \beta_{au,s} A \left(\frac{\log h_s(A,X,Z) - \widetilde{\alpha}_{0} - \alpha_{x} X}{\alpha_{u}} \right)\\
&\quad + \beta_{u,s} \left( \frac{\log h_s(A,X,Z) - \widetilde{\alpha}_0 - \alpha_{x} X}{\alpha_{u}} \right) +  \beta_{x,s} X + \beta_{ax,s} AX\\
&= \beta_{0,s} - \frac{\beta_{u,s} \widetilde{\alpha}_0}{\alpha_u} + \left( \beta_{a,s} - \frac{\beta_{au,s} \widetilde{\alpha}_0}{\alpha_u} \right)A + \left( \beta_{ax,s} - \frac{\beta_{au,s} \alpha_x}{\alpha_u} \right)AX + \frac{\beta_{u,s}}{\alpha_u} \log h_s(A,X,Z)\\
&\quad +\left( \beta_{x,s} - \frac{\beta_{u,s} \alpha_x}{\alpha_u} \right)X + \frac{\beta_{au,s}}{\alpha_u} A \log h_s(A,X,Z)\\
&= \beta_{0,s}^* + \beta_{a,s}^*A + \beta_{ax,s}^* AX + \beta_{u,s}^* \log h_s(A,X,Z) + \beta_{x,s}^* X + \beta_{au,s}^* A \log h_s(A,X,Z)
\end{align*}
where $h_s(A,X,Z) = \mathbb{E}[W \mid A, X, Z,S=s]$ and
\begin{align*}
    &\beta_{0,s}^* = \beta_{0,s} - \frac{\beta_{u,s} \widetilde{\alpha}_0}{\alpha_u}\\
    &\beta_{a,s}^* = \beta_{a,s} - \frac{\beta_{au,s} \widetilde{\alpha}_0}{\alpha_u}\\
    &\beta_{ax,s}^* = \beta_{ax,s} - \frac{\beta_{au,s} \alpha_x}{\alpha_u}\\
    &\beta_{u,s}^* = \frac{\beta_{u,s}}{\alpha_u}\\
    &\beta_{x,s}^* = \beta_{x,s}  - \frac{\beta_{u,s} \alpha_x}{\alpha_u}\\
    &\beta_{au,s}^* = \frac{\beta_{au,s}}{\alpha_u}.
\end{align*}
Next, we consider the case when the model for the conditional mean of $Y$ is according to Assumption \ref{assump:CRRstructural}. From the proof of Result \ref{result:CRR}, we have that
\begin{align*}
&\mathbb{E}[Y \mid A, X,Z, S=s]\\
&= \exp(\widetilde{\beta}_{0,s} + \widetilde{\beta}_{a,s} A + \beta_{ax,s} AX+\beta_{x,s}X + \beta_{au,s} m_s(A,X,Z) A + \beta_{u,s} m_s(A,X,Z))
\end{align*}
where $\widetilde{\beta}_{0,s} = \beta_{0,s} + \log M_{\epsilon}(\beta_{u,s})$ and $\widetilde{\beta}_{a,s} = \beta_{a,s} + \log \left( \frac{M_{\epsilon}(\beta_{au,s} + \beta_{u,s})}{M_{\epsilon}(\beta_{u,s})} \right)$. Again, plugging in the expression for $m_s(A,X,Z)$ from the marginalized conditional mean for $W$ yields
\begin{align*}
&\log(\mathbb{E}[Y \mid A, X,Z, S = s])\\
&= \widetilde{\beta}_{0,s} + \widetilde{\beta}_{a,s}A + \beta_{ax,s} AX+\beta_{x,s}X+\beta_{au,s}A \left( \frac{\log h_s(A,X,Z) - \widetilde{\alpha}_{0} - \alpha_xX}{\alpha_u}\right)\\
&\qquad+ \beta_{u,s} \left( \frac{\log h_s(A,X,Z) - \widetilde{\alpha}_{0} - \alpha_xX}{\alpha_u} \right)\\
&= \widetilde{\beta}_{0,s} + \widetilde{\beta}_{a,s}A + \beta_{ax,s} AX+\beta_{x,s}X+\frac{\beta_{au,s}A \log h_s(A,X,Z)}{\alpha_u} - \frac{\beta_{au,s} \widetilde{\alpha}_{0} A}{\alpha_u} - \frac{\beta_{au,s}A \alpha_xX}{\alpha_u}\\
&\qquad+ \frac{\beta_{u,s} \log h_s(A,X,Z)}{\alpha_u} - \frac{\beta_{u,s}\widetilde{\alpha}_{0}}{\alpha_u} - \frac{\beta_{u,s} \alpha_x X}{\alpha_u}\\
&= \left( \widetilde{\beta}_{0,s} - \frac{\beta_{u,s}\widetilde{\alpha}_{0}}{\alpha_u} \right) + \left(\widetilde{\beta}_{a,s} - \frac{\beta_{au,s} \widetilde{\alpha}_{0}}{\alpha_u} \right) A + \left(\beta_{ax,s} - \frac{\beta_{au,s} \alpha_x}{\alpha_u} \right) AX+ \left(\beta_{x,s}  - \frac{\beta_{u,s} \alpha_x}{\alpha_u}\right)X\\
&\qquad + \frac{\beta_{u,s}\log h_s(A,X,Z)}{\alpha_u} + \frac{\beta_{au,s}}{\alpha_u}A \log h_s(A,X,Z)\\
&= \beta_{0,s}^* + \beta_{a,s}^* A + \beta_{ax,s}^* AX+\beta_{x,s}^*X+\beta_{u,s}^* \log h_s(A,X,Z) + \beta_{au,s}^* A \log h_s(A,X,Z)
\end{align*}
where
\begin{align*}
\beta_{0,s}^* &= \widetilde{\beta}_{0,s} - \frac{\beta_{u,s}\widetilde{\alpha}_{0}}{\alpha_u}= \beta_{0,s} + \log M_{\epsilon}(\beta_{u,s}) - \frac{\beta_{u,s}\alpha_0}{\alpha_u}-\frac{\beta_{u,s}\log M_{\epsilon}(\alpha_u)}{\alpha_u}\\
\beta_{a,s}^* &= \widetilde{\beta}_{a,s} - \frac{\beta_{au,s} \widetilde{\alpha}_{0}}{\alpha_u} = \beta_{a,s} + \log \left( \frac{M_{\epsilon}(\beta_{au,s} + \beta_{u,s})}{M_{\epsilon}(\beta_{u,s})} \right) - \frac{\beta_{au,s} \alpha_0}{\alpha_u} - \frac{\beta_{au,s} \log M_{\epsilon}(\alpha_u)}{\alpha_u}\\
\beta_{ax,s}^* &= \beta_{ax,s} - \frac{\beta_{au,s} \alpha_x}{\alpha_u}\\
\beta_{x,s}^* &= \beta_{x,s}  - \frac{\beta_{u,s} \alpha_x}{\alpha_u}\\
\beta_{u,s}^* &= \frac{\beta_{u,s}}{\alpha_u}\\
\beta_{au,s}^* &= \frac{\beta_{au,s}}{\alpha_u}\\
\end{align*}

\appsubsection{Simulations}
\label{app:simulations}

We evaluate the finite-sample performance of the proposed NHST tests in Section \ref{s:hypothesis} for the null hypothesis of conditional reconcilability on both the additive and multiplicative scales. For each observation, we generated measured and unmeasured baseline covariates, proxy variables, trial membership, and treatment assignment as
\begin{align*}
&U \sim \mathcal{N}(0, \sigma_U^2), \qquad X \mid U\sim \mathcal{N}_3(\gamma^{\top} (1,U)^{\top}, (\sigma_X^2 - \rho_X) I_{3 \times 3} + \rho_X\mathbf{1}_{3 \times 3}),\\
&Z \mid X,U\sim \mathcal{N}(\eta^{\top} (1,U,X^{\top})^{\top}, \sigma_Z^2), \qquad W \mid X,U\sim \mathcal{N}(\alpha^{\top} (1,U,X^{\top})^{\top}, \sigma_W^2),\\
&S \mid X,U,Z\sim \text{Bernoulli}(\text{expit}\{\kappa^{\top}(1,U,Z,X^{\top})^{\top}\}), \qquad A \sim \text{Bernoulli}(0.5)\\
\end{align*}
To assess reconcilability on the additive scale, outcomes were generated from
\begin{align*}
    Y \mid U,X,Z,W,A,S=s\sim \mathcal{N}(\beta_s^{\top}(1,A,U,AU,X^{\top},AX^{\top},W)^{\top}, \sigma_Y^2=1),
\end{align*}
whereas on the multiplicative scale, we generated outcomes from
\begin{align*}
    Y \mid U,X,Z,W,A,S=s \sim \text{Poisson}(\exp\{ \beta_s^{\top}(1,A,U,AU,X^{\top},AX^{\top},W)^{\top}\}).
\end{align*}
Across both scales, we fixed
\begin{align*}
    &\gamma=\begin{pmatrix}
    0.4 & 0.2 & -0.2 \\
    -0.2 & 0.2 & 0.4
\end{pmatrix}, \qquad \eta=(-0.3, -0.5,0.2,0.4,0.4)^{\top},\\
    &\alpha=(-0.2,0.5,0.4,0.5,0.6)^{\top}, \qquad \kappa=(0.6,0.5,0.6,0.4,0.3,-0.4)^{\top},\\
    &\beta_0=(0.3,0.4,0.5,0.3,-0.3,-0.2,-0.3,0.5,-0.4,0.2,-0.2)^{\top}\\
    &\sigma_U^2= \sigma_X^2 =\sigma_Z^2 = \sigma_W^2 = 1, \qquad \rho_X = 0.2.
\end{align*}
Under the null hypothesis for the additive scale, we set
\begin{align*}
    \beta_1=(0.3, 0.4, 0.7, 0.3, 0.2, -0.5, -0.1,0.5, -0.4, 0.2, 0.2)^{\top}.
\end{align*}
For the multiplicative scale, we set
\begin{align*}
    \beta_1=(0.3, 0.4, 0.5, 0.3, 0.2, -0.5, -0.1,0.5, -0.4, 0.2, 0.2)^{\top}
\end{align*}
so that the coefficient on $U$ in the outcome model is the same across $s \in \{0,1\}$, matching the additional assumption required for the transportability result in Result \ref{result:transportRR} to hold. In both settings, although $\beta_1 \neq \beta_0$, the latent conditional effects are identical across trials under the respective additive and multiplicative null hypotheses. In particular, for $s \in \{0,1\}$,
\begin{align*}
    &\mu_1(s,x,u)-\mu_0(s,x,u)= 0.4 + 0.3u + (0.5,-0.4,0.2) x\\
    &\mu_1(s,x,u)/\mu_0(s,x,u)= \exp\{0.4 + 0.3u + (0.5,-0.4,0.2) x\}
\end{align*}
Under the alternative hypothesis, we held $\beta_0$ fixed and chose $\beta_1$ such that the latent CATE/CRR differed across trials. For the additive scale, we set
\begin{align*}
    &\beta_1=(0.3, 0.2, 0.7, 0.6, 0.2, -0.5, -0.1,0.8, -0.2, 0, 0.2)^{\top}\\
    &\mu_1(1,x,u)-\mu_0(1,x,u)=0.2 + 0.6u + (0.8,-0.2,0) x
\end{align*}
and for the multiplicative scale, we set
\begin{align*}
    &\beta_1=(0.3, 0.3, 0.5, 0.5, 0.2, -0.5, -0.1,0.7, -0.2, 0, 0.2)^{\top}\\
    &\mu_1(1,x,u)/\mu_0(1,x,u)=\exp\{0.3 + 0.5u + (0.7,-0.2,0) x\}.
\end{align*}
For each $n \in \{1000,2000,3000,4000\}$, we generated 500 simulated datasets and evaluated the empirical Type I error and power for the transportability and coefficient-testing procedures, benchmarking against an oracle procedure with direct access to $U$. Transportability estimators targeted $\mathbb{E}[Y(1) - Y(0) \mid S = s]$ for $s \in \{0,1\}$ on the additive scale and $\mathbb{E}[Y(1) \mid S = s]$ for $s \in \{0,1\}$ on the multiplicative scale. For all methods, the first-stage regression for $h_S(X,Z)$ was fit using a linear regression of $W$ on $(X^{\top},Z,S)^{\top}$, including all main effects and interactions. We did not include $A$ in the first-stage regression, as $A$ was fully randomized. The second-stage regression was then fit according to the procedures in Section \ref{s:hypothesis}. Standard errors for all test statistics were computed via the nonparametric bootstrap. For the test statistic for the coefficient-testing approach, we additionally evaluated the analytic variance estimator based on the asymptotic variance in Appendix \ref{app:asymp}.

Figure \ref{fig:nhst} summarizes the results. Under the null, all procedures controlled Type I error near the nominal level on both the additive and multiplicative scales. Under the alternative, the oracle test was most powerful, followed by the coefficient testing method, with transportability procedures exhibiting the lowest power. Though we do not provide theoretical guarantees that the coefficient testing method will uniformly be more powerful than the transportability procedures, we expect that similar empirical results would hold for a broad class of alternatives. Lastly, the close agreement in rejection rates between the bootstrap and analytic variance implementations of the coefficient test suggests that the asymptotic variance approximation is performing well in finite samples.

\clearpage
\vspace*{\fill}
\begin{figure}[H]
  \centering
  \includegraphics[width=\linewidth]{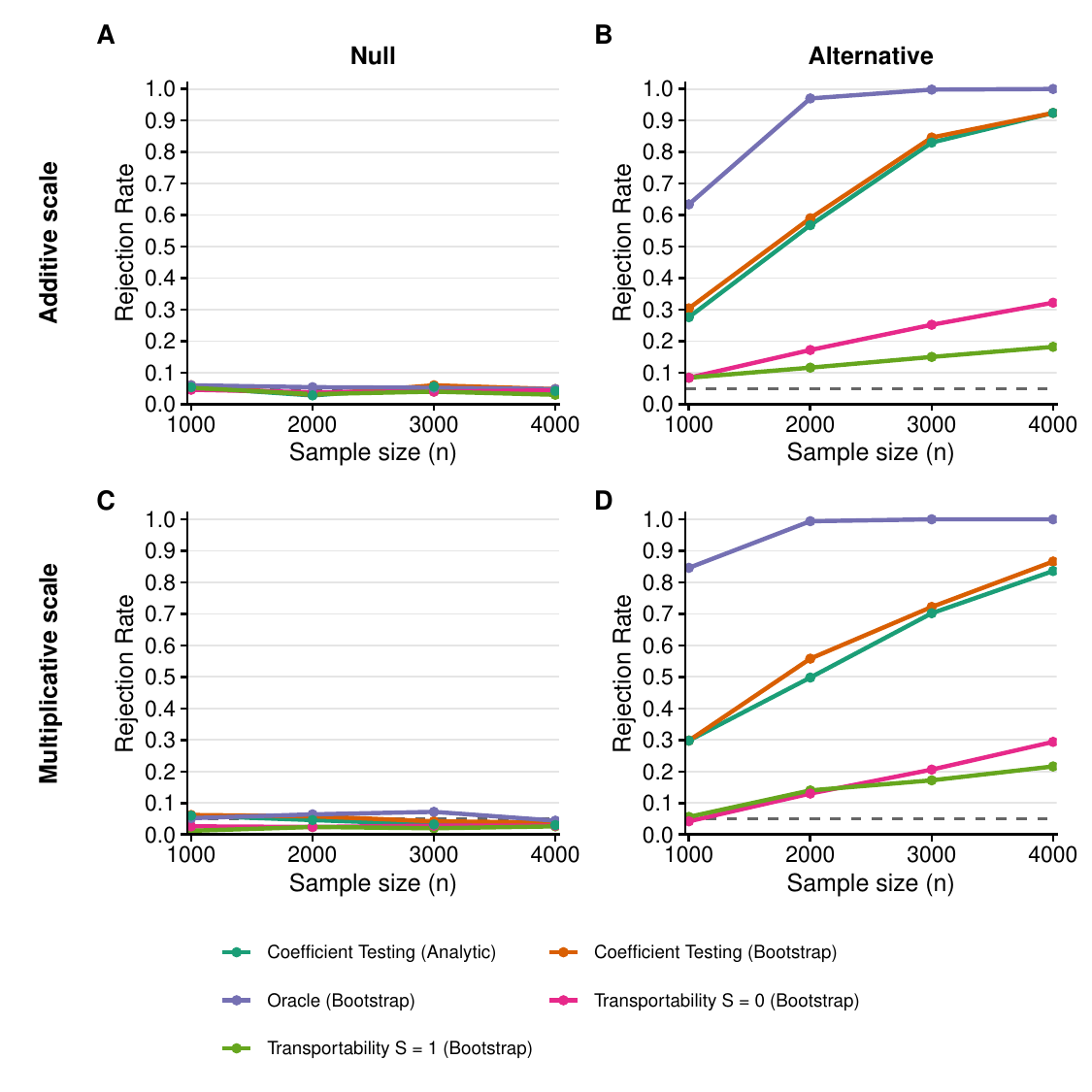}
  \caption{
    Rejection rates under the null hypothesis (panels A, C) and the alternative hypothesis (panels B, D) for the proposed tests. The top row corresponds to the additive scale and the bottom row to the multiplicative scale. Rejection is defined as a test with $p \leq 0.05$.
  }
  \label{fig:nhst}
\end{figure}
\vspace*{\fill}
\clearpage

\appsubsection{Distribution of baseline covariates in Meis and PROLONG}
\label{app:databaseline}

A complete summary of measured baseline characteristics is provided in Table \ref{table1}.

\begin{table}[t] 
\centering
\caption{\label{table1}Distribution of baseline covariates by treatment assignment and trial}
\centering
\resizebox{\ifdim\width>\linewidth\linewidth\else\width\fi}{!}{
\begin{tabular}[t]{>{\raggedright\arraybackslash}p{16em}>{\raggedright\arraybackslash}p{6em}>{\raggedright\arraybackslash}p{6em}>{\raggedright\arraybackslash}p{6em}>{\raggedright\arraybackslash}p{6em}}
\toprule
\multicolumn{1}{c}{ } & \multicolumn{2}{c}{PROLONG} & \multicolumn{2}{c}{Meis} \\
\cmidrule(l{3pt}r{3pt}){2-3} \cmidrule(l{3pt}r{3pt}){4-5}
  & Placebo (n=568) & 17OHP-C (n=1108) & Placebo (n=149) & 17OHP-C (n=294)\\
\midrule
Age (years), mean (SD) & 29.9 (5.23) & 29.9 (5.15) & 26.5 (5.42) & 26.1 (5.70)\\
\addlinespace
Race &  &  &  & \\
\hspace{1em}Non-Hispanic Black & 40 (7.0\%) & 70 (6.3\%) & 89 (59.7\%) & 177 (60.2\%)\\
\hspace{1em}Non-Hispanic White & 445 (78.3\%) & 895 (80.8\%) & 34 (22.8\%) & 74 (25.2\%)\\
\hspace{1em}Hispanic, Asian, or Other & 83 (14.6\%) & 143 (12.9\%) & 26 (17.4\%) & 43 (14.6\%)\\
\addlinespace
Gestational age at qualifying prior PTB in weeks, mean (SD) & 31.6 (4.13) & 31.3 (4.32) & 31.3 (4.21) & 30.5 (4.62)\\
\addlinespace
Number of prior PTBs, mean (SD) & 1.14 (0.407) & 1.18 (0.543) & 1.61 (0.913) & 1.40 (0.726)\\
\addlinespace
Married or living with partner & 512 (90.1\%) & 993 (89.6\%) & 69 (46.3\%) & 149 (50.7\%)\\
\addlinespace
Pre-pregnancy BMI, mean (SD) & 24.7 (8.70) & 24.3 (7.09) & 26.0 (6.96) & 27.0 (7.94)\\
\addlinespace
Years of education, mean (SD) & 13.0 (2.37) & 13.1 (2.37) & 12.0 (2.34) & 11.8 (2.25)\\
\addlinespace
Smoking during pregnancy & 40 (7.0\%) & 90 (8.1\%) & 30 (20.1\%) & 65 (22.1\%)\\
\addlinespace
Alcohol use during pregnancy & 17 (3.0\%) & 23 (2.1\%) & 10 (6.7\%) & 26 (8.8\%)\\
\addlinespace
Substance use during pregnancy & 8 (1.4\%) & 14 (1.3\%) & 4 (2.7\%) & 11 (3.7\%)\\
\bottomrule
\end{tabular}}
\end{table}

\appsubsection{Secondary data analyses on the additive scale}
\label{app:data}
In this section, we consider secondary analyses of the Meis and PROLONG data on the additive scale using different proxy variables. These analyses are intended to be exploratory and should not be regarded as definitive evidence regarding reconcilability. We first consider the following analyses:
\begin{itemize}
    \item Analysis 1b: W is the gestational age at qualifying prior PTB, Z is the number of prior PTBs
    \item Analysis 2b: W is pre-pregnancy BMI, Z is gestational age at qualifying prior PTB
    \item Analysis 3a: W is number of prior PTBs, Z is pre-pregnancy BMI
    \item Analysis 3b: W is pre-pregnancy BMI, Z is number of prior PTBs
    \item Analysis 4: W is number of prior PTBs, X is maternal age, Z comprises all remaining baseline covariates
    \item Analysis 5: W is gestational age at qualifying prior PTB, X is maternal age, Z comprises all remaining baseline covariates
    \item Analysis 6: W is pre-pregnancy BMI, X is maternal age, Z comprises all remaining baseline covariates
\end{itemize}
Note that in the above, Analyses 1b and 2b correspond to the analyses in the main text, except with the roles of the adjustment and reweighting proxies swapped (as in Analysis 3a vs. 3b). Analyses 4--6 were motivated from the perspective that it is unlikely for many of the measured covariates to be direct effect modifiers of the treatment effect in the sense of $(X,U,W)$ in Figure \ref{fig:DAGs}(b).

We first report measures of proxy relevance for the different analyses in Table \ref{tab:relevance_cont}. We consider estimates, naive standard errors, and p-values for coefficients in the second-stage regression for terms containing $\widehat{h}_S(X,Z)$. We also report the p-value from a joint F-test for the reported coefficients. Strongly significant coefficient estimates would be indicative of stronger proxies. Though proxies seem to be informative in some analyses (Analyses 1b, 2b, 3a, 4, 5, 6), in other analyses they seem to be weaker (Analyses 1, 2, 3b).

Next, we report the additional results for testing the null hypothesis of conditional and marginal reconcilability on the additive scale for the Meis and PROLONG trials for all of the above analyses in Table \ref{tab:table2_appendix}. These results are intended to be exploratory. Across most analyses, the proposed coefficient testing procedure has a lower p-value than the transportability method, which is consistent with it having greater power when conditional reconcilability is the null hypothesis of interest. Note that in some analyses (e.g., Analysis 3b), this trend does not hold, as the theory does not predict the coefficient testing method to universally dominate the transportability method in terms of power across all alternatives.

Lastly, we report additional results for equivalence testing on the additive scale in Table \ref{tab:table3_appendix} for marginal reconcilability. The reported LEADs across all analyses are likely higher than one would consider necessary to claim equivalence at clinically insignificant margins. While reconciliation proportions in the PROLONG direction in some analyses are moderately high, the mean reconciliation proportion across the analyses is low.

\begin{table}[!h]
\centering
\caption{\label{tab:relevance_cont}Supplementary measures of proxy relevance for different analyses using naive standard errors (SEs) reported directly from the second-stage regression on the additive scale. Joint-test p-values are from F-tests of the null that all proxy-relevance coefficients are zero.}
\centering
\resizebox{\ifdim\width>\linewidth\linewidth\else\width\fi}{!}{
\begin{tabular}[t]{>{\raggedright\arraybackslash}p{12em}>{\raggedright\arraybackslash}p{11em}>{\raggedright\arraybackslash}p{8em}>{\raggedright\arraybackslash}p{8em}>{\raggedright\arraybackslash}p{8em}}
\toprule
 & Coefficients & Estimate & SE & p-value\\
\midrule
 & $\widehat{\beta}_{u,0}^*$ & -3.718 & 2.625 & 0.157\\

 & $\widehat{\beta}_{u,1}^* - \widehat{\beta}_{u,0}^*$ & 5.603 & 3.278 & 0.088\\

 & $\widehat{\beta}_{au,0}^*$ & -1.501 & 4.018 & 0.709\\

 & $\widehat{\beta}_{au,1}^* - \widehat{\beta}_{au,0}^*$ & -0.859 & 4.799 & 0.858\\

\multirow[t]{-5}{12em}{\raggedright\arraybackslash Analysis 1: W is number of prior PTBs, Z is gestational age at prior PTB} & Joint test & -- & -- & 0.204\\
\cmidrule{1-5}
 & $\widehat{\beta}_{u,0}^*$ & -0.585 & 0.335 & 0.081\\

 & $\widehat{\beta}_{u,1}^* - \widehat{\beta}_{u,0}^*$ & -0.305 & 0.404 & 0.450\\

 & $\widehat{\beta}_{au,0}^*$ & -0.042 & 0.559 & 0.940\\

 & $\widehat{\beta}_{au,1}^* - \widehat{\beta}_{au,0}^*$ & 0.563 & 0.757 & 0.457\\

\multirow[t]{-5}{12em}{\raggedright\arraybackslash Analysis 1b: W is gestational age at prior PTB, Z is number of prior PTBs} & Joint test & -- & -- & 0.000\\
\cmidrule{1-5}
 & $\widehat{\beta}_{u,0}^*$ & 0.084 & 0.295 & 0.775\\

 & $\widehat{\beta}_{u,1}^* - \widehat{\beta}_{u,0}^*$ & 0.236 & 0.367 & 0.520\\

 & $\widehat{\beta}_{au,0}^*$ & -0.699 & 0.566 & 0.217\\

 & $\widehat{\beta}_{au,1}^* - \widehat{\beta}_{au,0}^*$ & 0.547 & 0.686 & 0.425\\

\multirow[t]{-5}{12em}{\raggedright\arraybackslash Analysis 2: W is gestational age at prior PTB, Z is pre-pregnancy BMI} & Joint test & -- & -- & 0.389\\
\cmidrule{1-5}
 & $\widehat{\beta}_{u,0}^*$ & 0.067 & 0.265 & 0.801\\

 & $\widehat{\beta}_{u,1}^* - \widehat{\beta}_{u,0}^*$ & -0.532 & 0.328 & 0.105\\

 & $\widehat{\beta}_{au,0}^*$ & -0.474 & 0.386 & 0.219\\

 & $\widehat{\beta}_{au,1}^* - \widehat{\beta}_{au,0}^*$ & 0.289 & 0.466 & 0.535\\

\multirow[t]{-5}{12em}{\raggedright\arraybackslash Analysis 2b: W is pre-pregnancy BMI, Z is gestational age at prior PTB} & Joint test & -- & -- & 0.000\\
\cmidrule{1-5}
 & $\widehat{\beta}_{u,0}^*$ & -0.132 & 2.964 & 0.964\\

 & $\widehat{\beta}_{u,1}^* - \widehat{\beta}_{u,0}^*$ & 0.544 & 3.704 & 0.883\\

 & $\widehat{\beta}_{au,0}^*$ & -3.677 & 3.778 & 0.330\\

 & $\widehat{\beta}_{au,1}^* - \widehat{\beta}_{au,0}^*$ & 3.968 & 4.608 & 0.389\\

\multirow[t]{-5}{12em}{\raggedright\arraybackslash Analysis 3a: W is number of prior PTBs, Z is pre-pregnancy BMI} & Joint test & -- & -- & 0.571\\
\cmidrule{1-5}
 & $\widehat{\beta}_{u,0}^*$ & 0.281 & 0.372 & 0.450\\

 & $\widehat{\beta}_{u,1}^* - \widehat{\beta}_{u,0}^*$ & -0.245 & 0.416 & 0.556\\

 & $\widehat{\beta}_{au,0}^*$ & 0.495 & 0.428 & 0.248\\

 & $\widehat{\beta}_{au,1}^* - \widehat{\beta}_{au,0}^*$ & -0.354 & 0.505 & 0.483\\

\multirow[t]{-5}{12em}{\raggedright\arraybackslash Analysis 3b: W is pre-pregnancy BMI, Z is number of prior PTBs} & Joint test & -- & -- & 0.005\\
\cmidrule{1-5}
 & $\widehat{\beta}_{u,0}^*$ & -3.733 & 1.240 & 0.003\\

 & $\widehat{\beta}_{u,1}^* - \widehat{\beta}_{u,0}^*$ & 0.547 & 1.497 & 0.715\\

 & $\widehat{\beta}_{au,0}^*$ & -0.061 & 2.127 & 0.977\\

 & $\widehat{\beta}_{au,1}^* - \widehat{\beta}_{au,0}^*$ & 3.058 & 2.544 & 0.230\\

\multirow[t]{-5}{12em}{\raggedright\arraybackslash Analysis 4: W is number of prior PTBs, X is maternal age, Z is all other covariates} & Joint test & -- & -- & 0.000\\
\cmidrule{1-5}
 & $\widehat{\beta}_{u,0}^*$ & -0.237 & 0.281 & 0.400\\

 & $\widehat{\beta}_{u,1}^* - \widehat{\beta}_{u,0}^*$ & -0.257 & 0.334 & 0.442\\

 & $\widehat{\beta}_{au,0}^*$ & -0.935 & 0.469 & 0.046\\

 & $\widehat{\beta}_{au,1}^* - \widehat{\beta}_{au,0}^*$ & 1.587 & 0.562 & 0.005\\

\multirow[t]{-5}{12em}{\raggedright\arraybackslash Analysis 5: W is gestational age at prior PTB, X is maternal age, Z is all other covariates} & Joint test & -- & -- & 0.001\\
\cmidrule{1-5}
 & $\widehat{\beta}_{u,0}^*$ & -0.185 & 0.086 & 0.033\\

 & $\widehat{\beta}_{u,1}^* - \widehat{\beta}_{u,0}^*$ & -0.049 & 0.107 & 0.649\\

 & $\widehat{\beta}_{au,0}^*$ & 0.496 & 0.171 & 0.004\\

 & $\widehat{\beta}_{au,1}^* - \widehat{\beta}_{au,0}^*$ & -0.366 & 0.209 & 0.080\\

\multirow[t]{-5}{12em}{\raggedright\arraybackslash Analysis 6: W is pre-pregnancy BMI, X is maternal age, Z is all other covariates} & Joint test & -- & -- & 0.000\\
\bottomrule
\end{tabular}}
\end{table}

\begin{table}[!h]
\centering
\caption{\label{tab:table2_appendix}Supplementary results for testing the null hypothesis of conditional and marginal reconcilability on the additive scale for the Meis and PROLONG trials. $\widehat{\phi}_{ATE,s}$ and $\widehat{\theta}_{ATE,s}$ denote the transported and untransported estimates for the ATE in trial $s$, respectively, as defined in Section \ref{ss:hypothesisadditive}. $\widehat{\theta}_{ATE,s}$ was taken to be the standard difference-in-means estimator. 95\% confidence intervals and p-values were computed using the nonparametric bootstrap with 5,000 resamples. All estimates are in weeks (gestational age at delivery).}
\centering
\resizebox{\ifdim\width>\linewidth\linewidth\else\width\fi}{!}{
\begin{tabular}[t]{>{\raggedright\arraybackslash}p{14em}>{\raggedright\arraybackslash}p{4em}>{\raggedright\arraybackslash}p{6em}>{\raggedright\arraybackslash}p{8em}>{\raggedright\arraybackslash}p{8em}>{\raggedright\arraybackslash}p{8em}>{\raggedright\arraybackslash}p{4em}}
\toprule
 & Method & Target & $\widehat{\phi}_{ATE,s}$ & $\widehat{\theta}_{ATE,s}$ & $\widehat{\phi}_{ATE,s} - \widehat{\theta}_{ATE,s}$ & p-value\\
\midrule
 &  & Meis & -0.09 (-1.32, 1.13) & 1.09 (0.15, 2.03) & -1.19 (-2.72, 0.35) & 0.129\\

 & \multirow[t]{-2}{4em}{\raggedright\arraybackslash Transport} & PROLONG & 1.34 (-0.44, 3.13) & 0.10 (-0.24, 0.44) & 1.24 (-0.58, 3.06) & 0.180\\

\multirow[t]{-3}{14em}{\raggedright\arraybackslash Analysis 1b: W is gestational age at prior PTB, Z is number of prior PTBs} & Coefficient & -- & -- & -- & -- & 0.125\\

\cmidrule{1-7}
 &  & Meis & 0.23 (-1.48, 1.94) & 1.09 (0.15, 2.03) & -0.86 (-2.80, 1.08) & 0.385\\

 & \multirow[t]{-2}{4em}{\raggedright\arraybackslash Transport} & PROLONG & 0.73 (-0.93, 2.38) & 0.10 (-0.24, 0.44) & 0.63 (-1.07, 2.32) & 0.468\\

\multirow[t]{-3}{14em}{\raggedright\arraybackslash Analysis 2b: W is pre-pregnancy BMI, Z is gestational age at prior PTB} & Coefficient & -- & -- & -- & -- & 0.065\\

\cmidrule{1-7}
 &  & Meis & -0.09 (-2.54, 2.35) & 1.09 (0.15, 2.03) & -1.18 (-3.82, 1.45) & 0.379\\

 & \multirow[t]{-2}{4em}{\raggedright\arraybackslash Transport} & PROLONG & 0.13 (-2.44, 2.69) & 0.10 (-0.24, 0.44) & 0.03 (-2.56, 2.61) & 0.985\\

\multirow[t]{-3}{14em}{\raggedright\arraybackslash Analysis 3a: W is number of prior PTBs, Z is pre-pregnancy BMI} & Coefficient & -- & -- & -- & -- & 0.241\\

\cmidrule{1-7}
 &  & Meis & 0.06 (-1.58, 1.69) & 1.09 (0.15, 2.03) & -1.03 (-2.91, 0.84) & 0.281\\

 & \multirow[t]{-2}{4em}{\raggedright\arraybackslash Transport} & PROLONG & 1.56 (-0.30, 3.42) & 0.10 (-0.24, 0.44) & 1.46 (-0.43, 3.35) & 0.131\\

\multirow[t]{-3}{14em}{\raggedright\arraybackslash Analysis 3b: W is pre-pregnancy BMI, Z is number of prior PTBs} & Coefficient & -- & -- & -- & -- & 0.451\\

\cmidrule{1-7}
 &  & Meis & 0.28 (-0.91, 1.46) & 1.09 (0.15, 2.03) & -0.81 (-2.32, 0.70) & 0.291\\

 & \multirow[t]{-2}{4em}{\raggedright\arraybackslash Transport} & PROLONG & -0.92 (-3.01, 1.17) & 0.10 (-0.24, 0.44) & -1.02 (-3.14, 1.11) & 0.348\\

\multirow[t]{-3}{14em}{\raggedright\arraybackslash Analysis 4: W is number of prior PTBs, X is maternal age, Z is all other covariates} & Coefficient & -- & -- & -- & -- & 0.091\\

\cmidrule{1-7}
 &  & Meis & 0.32 (-0.32, 0.95) & 1.09 (0.15, 2.03) & -0.78 (-1.92, 0.37) & 0.183\\

 & \multirow[t]{-2}{4em}{\raggedright\arraybackslash Transport} & PROLONG & 0.99 (-0.36, 2.34) & 0.10 (-0.24, 0.44) & 0.89 (-0.51, 2.28) & 0.212\\

\multirow[t]{-3}{14em}{\raggedright\arraybackslash Analysis 5: W is gestational age at prior PTB, X is maternal age, Z is all other covariates} & Coefficient & -- & -- & -- & -- & 0.032\\

\cmidrule{1-7}
 &  & Meis & -0.01 (-0.93, 0.91) & 1.09 (0.15, 2.03) & -1.10 (-2.43, 0.23) & 0.104\\

 & \multirow[t]{-2}{4em}{\raggedright\arraybackslash Transport} & PROLONG & 1.35 (-0.19, 2.89) & 0.10 (-0.24, 0.44) & 1.25 (-0.34, 2.83) & 0.123\\

\multirow[t]{-3}{14em}{\raggedright\arraybackslash Analysis 6: W is pre-pregnancy BMI, X is maternal age, Z is all other covariates} & Coefficient & -- & -- & -- & -- & 0.026\\
\bottomrule
\end{tabular}}
\end{table}

\begin{table}[!h]
\centering
\caption{\label{tab:table3_appendix}Supplementary results for equivalence testing for marginal reconcilability on the additive scale for the Meis and PROLONG trials. Estimate denotes the plug-in estimate of the quantity defining the corresponding equivalence null hypothesis on the additive scale (Section \ref{s:equiv}) and is presented with a 90\% confidence interval computed using the nonparametric bootstrap. 95\% confidence intervals for the reconciliation proportion (RP) were also computed using the nonparametric bootstrap. Reported p-values correspond to equivalence tests with equivalence margins of 1 week (gestational age at delivery) in both populations.}
\centering
\resizebox{\ifdim\width>\linewidth\linewidth\else\width\fi}{!}{
\begin{tabular}[t]{>{\raggedright\arraybackslash}p{14em}>{\raggedright\arraybackslash}p{8em}>{\raggedright\arraybackslash}p{6em}>{\raggedright\arraybackslash}p{10em}>{\raggedright\arraybackslash}p{6em}>{\raggedright\arraybackslash}p{6em}>{\raggedright\arraybackslash}p{10em}}
\toprule
Analysis & Method & Hypothesis & Estimate (CI) & p-value & LEAD & RP (CI)\\
\midrule
 &  & Meis & 1.19 (-0.10, 2.47) & 0.594 & 2.47 & -0.20 (-5.30, 0.77)\\

 &  & PROLONG & -1.24 (-2.77, 0.28) & 0.603 & 2.77 & -0.25 (-8.78, 0.84)\\

 &  & Union & -- & 0.603 & 2.77 & --\\

 & \multirow{-4}{8em}{\raggedright\arraybackslash Proximal} & Mean & 1.21 (0.24, 2.19) & 0.641 & 2.19 & -0.22 (-3.42, 0.66)\\
\cmidrule{2-7}
 &  & Meis & 1.34 (0.08, 2.59) & 0.671 & 2.59 & -0.35 (-5.31, 0.71)\\

 &  & PROLONG & -1.10 (-2.48, 0.28) & 0.547 & 2.48 & -0.11 (-6.66, 0.84)\\

 &  & Union & -- & 0.671 & 2.59 & --\\

\multirow{-8}{14em}[0.5\dimexpr\aboverulesep+\belowrulesep+\cmidrulewidth]{\raggedright\arraybackslash Analysis 1b: W is gestational age at prior PTB, Z is number of prior PTBs} & \multirow{-4}{8em}{\raggedright\arraybackslash Non-proximal} & Mean & 1.22 (0.24, 2.19) & 0.643 & 2.19 & -0.23 (-3.07, 0.63)\\
\cmidrule{1-7}
 &  & Meis & 0.86 (-0.77, 2.49) & 0.445 & 2.49 & 0.13 (-5.27, 0.88)\\

 &  & PROLONG & -0.63 (-2.04, 0.79) & 0.332 & 2.04 & 0.37 (-6.45, 0.95)\\

 &  & Union & -- & 0.445 & 2.49 & --\\

 & \multirow{-4}{8em}{\raggedright\arraybackslash Proximal} & Mean & 0.74 (-0.10, 1.59) & 0.309 & 1.59 & 0.25 (-2.40, 0.83)\\
\cmidrule{2-7}
 &  & Meis & 1.40 (0.16, 2.64) & 0.701 & 2.64 & -0.41 (-4.78, 0.66)\\

 &  & PROLONG & -0.25 (-1.52, 1.02) & 0.166 & 1.52 & 0.75 (-2.29, 0.98)\\

 &  & Union & -- & 0.701 & 2.64 & --\\

\multirow{-8}{14em}[0.5\dimexpr\aboverulesep+\belowrulesep+\cmidrulewidth]{\raggedright\arraybackslash Analysis 2b: W is pre-pregnancy BMI, Z is gestational age at prior PTB} & \multirow{-4}{8em}{\raggedright\arraybackslash Non-proximal} & Mean & 0.83 (0.07, 1.58) & 0.352 & 1.58 & 0.17 (-2.06, 0.77)\\
\cmidrule{1-7}
 &  & Meis & 1.18 (-1.03, 3.40) & 0.555 & 3.40 & -0.19 (-8.90, 0.86)\\

 &  & PROLONG & -0.03 (-2.20, 2.15) & 0.230 & 2.20 & 0.97 (0.67, 1.00)\\

 &  & Union & -- & 0.555 & 3.40 & --\\

 & \multirow{-4}{8em}{\raggedright\arraybackslash Proximal} & Mean & 0.60 (-0.51, 1.72) & 0.280 & 1.72 & 0.39 (-2.27, 0.89)\\
\cmidrule{2-7}
 &  & Meis & 1.30 (0.05, 2.55) & 0.654 & 2.55 & -0.31 (-5.23, 0.72)\\

 &  & PROLONG & -1.05 (-2.40, 0.30) & 0.523 & 2.40 & -0.06 (-6.74, 0.86)\\

 &  & Union & -- & 0.654 & 2.55 & --\\

\multirow{-8}{14em}[0.5\dimexpr\aboverulesep+\belowrulesep+\cmidrulewidth]{\raggedright\arraybackslash Analysis 3a: W is number of prior PTBs, Z is pre-pregnancy BMI} & \multirow{-4}{8em}{\raggedright\arraybackslash Non-proximal} & Mean & 1.17 (0.22, 2.13) & 0.618 & 2.13 & -0.18 (-2.91, 0.64)\\
\cmidrule{1-7}
 &  & Meis & 1.03 (-0.54, 2.61) & 0.513 & 2.61 & -0.04 (-5.64, 0.84)\\

 &  & PROLONG & -1.46 (-3.05, 0.13) & 0.683 & 3.05 & -0.47 (-9.82, 0.80)\\

 &  & Union & -- & 0.683 & 3.05 & --\\

 & \multirow{-4}{8em}{\raggedright\arraybackslash Proximal} & Mean & 1.25 (0.14, 2.35) & 0.642 & 2.35 & -0.26 (-3.84, 0.67)\\
\cmidrule{2-7}
 &  & Meis & 1.30 (0.05, 2.55) & 0.654 & 2.55 & -0.31 (-5.23, 0.72)\\

 &  & PROLONG & -1.05 (-2.40, 0.30) & 0.523 & 2.40 & -0.06 (-6.74, 0.86)\\

 &  & Union & -- & 0.654 & 2.55 & --\\

\multirow{-8}{14em}[0.5\dimexpr\aboverulesep+\belowrulesep+\cmidrulewidth]{\raggedright\arraybackslash Analysis 3b: W is pre-pregnancy BMI, Z is number of prior PTBs} & \multirow{-4}{8em}{\raggedright\arraybackslash Non-proximal} & Mean & 1.17 (0.22, 2.13) & 0.618 & 2.13 & -0.18 (-2.91, 0.64)\\
\cmidrule{1-7}
 &  & Meis & 0.81 (-0.45, 2.08) & 0.405 & 2.08 & 0.18 (-4.98, 0.89)\\

 &  & PROLONG & 1.02 (-0.77, 2.80) & 0.506 & 2.80 & -0.03 (-15.90, 0.94)\\

 &  & Union & -- & 0.506 & 2.80 & --\\

 & \multirow{-4}{8em}{\raggedright\arraybackslash Proximal} & Mean & 0.92 (0.13, 1.70) & 0.430 & 1.70 & 0.08 (-4.19, 0.84)\\
\cmidrule{2-7}
 &  & Meis & 0.95 (0.07, 1.83) & 0.463 & 1.83 & 0.04 (-1.11, 0.56)\\

 &  & PROLONG & -0.30 (-1.22, 0.62) & 0.105 & 1.22 & 0.70 (-2.25, 0.97)\\

 &  & Union & -- & 0.463 & 1.83 & --\\

\multirow{-8}{14em}[0.5\dimexpr\aboverulesep+\belowrulesep+\cmidrulewidth]{\raggedright\arraybackslash Analysis 4: W is number of prior PTBs, X is maternal age, Z is all other covariates} & \multirow{-4}{8em}{\raggedright\arraybackslash Non-proximal} & Mean & 0.63 (-0.02, 1.27) & 0.170 & 1.27 & 0.37 (-0.79, 0.78)\\
\cmidrule{1-7}
 &  & Meis & 0.78 (-0.18, 1.74) & 0.350 & 1.74 & 0.22 (-1.81, 0.78)\\

 &  & PROLONG & -0.89 (-2.06, 0.28) & 0.438 & 2.06 & 0.10 (-5.92, 0.88)\\

 &  & Union & -- & 0.438 & 2.06 & --\\

 & \multirow{-4}{8em}{\raggedright\arraybackslash Proximal} & Mean & 0.83 (0.04, 1.63) & 0.364 & 1.63 & 0.16 (-1.60, 0.73)\\
\cmidrule{2-7}
 &  & Meis & 0.95 (0.07, 1.83) & 0.463 & 1.83 & 0.04 (-1.11, 0.56)\\

 &  & PROLONG & -0.30 (-1.22, 0.62) & 0.105 & 1.22 & 0.70 (-2.25, 0.97)\\

 &  & Union & -- & 0.463 & 1.83 & --\\

\multirow{-8}{14em}[0.5\dimexpr\aboverulesep+\belowrulesep+\cmidrulewidth]{\raggedright\arraybackslash Analysis 5: W is gestational age at prior PTB, X is maternal age, Z is all other covariates} & \multirow{-4}{8em}{\raggedright\arraybackslash Non-proximal} & Mean & 0.63 (-0.02, 1.27) & 0.170 & 1.27 & 0.37 (-0.79, 0.78)\\
\cmidrule{1-7}
 &  & Meis & 1.10 (-0.01, 2.22) & 0.559 & 2.22 & -0.11 (-3.49, 0.73)\\

 &  & PROLONG & -1.25 (-2.58, 0.08) & 0.620 & 2.58 & -0.26 (-7.14, 0.81)\\

 &  & Union & -- & 0.620 & 2.58 & --\\

 & \multirow{-4}{8em}{\raggedright\arraybackslash Proximal} & Mean & 1.17 (0.26, 2.09) & 0.623 & 2.09 & -0.18 (-2.81, 0.63)\\
\cmidrule{2-7}
 &  & Meis & 0.95 (0.07, 1.83) & 0.463 & 1.83 & 0.04 (-1.11, 0.56)\\

 &  & PROLONG & -0.30 (-1.22, 0.62) & 0.105 & 1.22 & 0.70 (-2.25, 0.97)\\

 &  & Union & -- & 0.463 & 1.83 & --\\

\multirow{-8}{14em}[0.5\dimexpr\aboverulesep+\belowrulesep+\cmidrulewidth]{\raggedright\arraybackslash Analysis 6: W is pre-pregnancy BMI, X is maternal age, Z is all other covariates} & \multirow{-4}{8em}{\raggedright\arraybackslash Non-proximal} & Mean & 0.63 (-0.02, 1.27) & 0.170 & 1.27 & 0.37 (-0.79, 0.78)\\
\bottomrule
\end{tabular}}
\end{table}

\appsubsection{Secondary data analyses on the multiplicative scale}
\label{app:mult}

In this section, we consider secondary analyses of the Meis and PROLONG data on the multiplicative scale. These analyses are exploratory and should not be interpreted as definitive evidence regarding reconcilability. Rather, they illustrate how our methods can be applied and interpreted on the multiplicative scale while highlighting several practical challenges that arise in this setting compared to the additive scale setting.

We consider the same set of analyses as on the additive scale. However, for most analyses, we categorize pre-pregnancy BMI into the following groups: $<$18.5, 18.5--25, 25--30, 30--35, and $\geq 35$ kg/m$^2$. This modification was motivated by the sensitivity of the log-link outcome model used for the multiplicative-scale analyses to a small number of influential BMI outliers. We retain pre-pregnancy BMI as a continuous variable in analyses where it serves as the adjustment proxy $W$ (Analyses 2b, 3b, and 6), as the current methodology would require further extension to accommodate a categorical adjustment proxy.

We report the results of the proximal tests of the null hypotheses of conditional and marginal reconcilability on the multiplicative scale in Table \ref{tab:nhst_mult}. Analyses 1--3 exhibit large standard errors for the transportability estimator $\widehat{\phi}_{RR,s}$ due to the log-link outcome model, highlighting a practical challenge of conducting reconciliation analyses on the multiplicative scale. Consequently, in Analyses 1--3, the transportability tests reject neither marginal nor conditional reconcilability, and the coefficient tests do not reject conditional reconcilability. $\widehat{\phi}_{RR,s}$ does have substantially narrower confidence intervals in Analyses 4--6. In particular, for Analyses 5--6, the transportability tests reject the null hypotheses of both marginal and conditional reconcilability in both transport directions. While this suggests the presence of additional omitted effect modifiers on the multiplicative scale beyond those included in $X$ and/or proxied by $(Z,W)$, those effect modifiers need not necessarily be conditionally imbalanced between the two trials. As discussed in Section \ref{ss:hypothesismultiplicative}, a lack of reconciliation may also arise from differences in the conditional baseline risks between the trials, a key distinction from the additive scale.

Unlike on the additive scale, the p-values from the coefficient test are not generally smaller than those from the transportability tests. This is unsurprising because, on the multiplicative scale, identification of the transported marginal estimand under the proposed structural models additionally requires the assumption that $\beta_{u,0}=\beta_{u,1}$. Although conditional reconcilability continues to imply marginal reconcilability at the causal estimand level, Result \ref{result:transportRR} identifies the transported marginal estimand only under this additional assumption. Consequently, the transportability procedure relies on a stronger set of assumptions than the coefficient test, and therefore one should not directly compare the power of the two procedures.

Next, we report the results of the proximal equivalence tests on the multiplicative scale in Table \ref{tab:equiv_mult}. We consider an equivalence margin of 0.1 on the log scale, corresponding approximately to a 10\% difference in the relative risks between the transported and untransported estimands. None of the equivalence tests are statistically significant, indicating insufficient evidence to conclude marginal reconcilability at this margin. The LEADs are large across all analyses, particularly for Analyses 1--3, reflecting the high variability of the transportability estimator. Even for Analyses 4--6, the LEADs are approximately 0.7 on the log scale, implying that the ratio of the transported and untransported relative risks would need to lie roughly between 0.5 and 2.0 to establish marginal reconcilability. Such a margin would likely be considered too large to be clinically insignificant. Across most analyses, the large LEADs are accompanied by small estimated reconciliation proportions.

Lastly, we report measures of proxy relevance on the multiplicative scale in Table \ref{tab:relevance_mult}. As on the additive scale, larger and more statistically significant coefficient estimates are indicative of stronger proxies. The proxies appear reasonably informative in several analyses (Analyses 1b, 2, 4, 5, 6), whereas they appear substantially weaker in others (Analyses 1, 2b, 3a, 3b).

\begin{table}[!h]
\centering
\caption{\label{tab:nhst_mult}Supplementary results for testing the null hypothesis of conditional and marginal reconcilability on the multiplicative scale for the Meis and PROLONG trials using the binary outcome of delivery before 37 weeks of gestation. $\widehat{\phi}_{RR,s}$ and $\widehat{\theta}_{RR,s}$ denote the transported and untransported estimates for the RR in trial $s$, respectively. $\widehat{\theta}_{RR,s}$ was taken to be the standard ratio-of-means estimator. 95\% confidence intervals and p-values were computed using the nonparametric bootstrap with 5,000 resamples.}
\centering
\resizebox{\ifdim\width>\linewidth\linewidth\else\width\fi}{!}{
\begin{tabular}[t]{>{\raggedright\arraybackslash}p{14em}>{\raggedright\arraybackslash}p{4em}>{\raggedright\arraybackslash}p{6em}>{\raggedright\arraybackslash}p{10em}>{\raggedright\arraybackslash}p{8em}>{\raggedright\arraybackslash}p{10em}>{\raggedright\arraybackslash}p{4em}}
\toprule
 & Method & Target & $\widehat{\phi}_{RR,s}$ & $\widehat{\theta}_{RR,s}$ & $\log(\widehat{\phi}_{RR,s}/\widehat{\theta}_{RR,s})$ & p-value\\
\midrule
 &  & Meis & 0.93 (0.02, 50.45) & 0.66 (0.53, 0.81) & 0.35 (-3.65, 4.35) & 0.865\\

 & \multirow[t]{-2}{4em}{\raggedright\arraybackslash Transport} & PROLONG & 0.70 (0.00, 17624.49) & 1.06 (0.88, 1.28) & -0.41 (-10.55, 9.72) & 0.936\\

\multirow[t]{-3}{14em}{\raggedright\arraybackslash Analysis 1: W is number of prior PTBs, Z is gestational age at prior PTB} & Coefficient & -- & -- & -- & -- & 0.968\\

\cmidrule{1-7}
 &  & Meis & 1.11 (0.02, 56.07) & 0.66 (0.53, 0.81) & 0.53 (-3.39, 4.45) & 0.791\\

 & \multirow[t]{-2}{4em}{\raggedright\arraybackslash Transport} & PROLONG & 0.77 (0.00, 19995.16) & 1.06 (0.88, 1.28) & -0.32 (-10.49, 9.85) & 0.950\\

\multirow[t]{-3}{14em}{\raggedright\arraybackslash Analysis 1b: W is gestational age at prior PTB, Z is number of prior PTBs} & Coefficient & -- & -- & -- & -- & 0.991\\

\cmidrule{1-7}
 &  & Meis & 1.04 (0.02, 49.35) & 0.66 (0.53, 0.81) & 0.46 (-3.40, 4.32) & 0.814\\

 & \multirow[t]{-2}{4em}{\raggedright\arraybackslash Transport} & PROLONG & 0.87 (0.00, 10339.94) & 1.06 (0.88, 1.28) & -0.20 (-9.59, 9.19) & 0.967\\

\multirow[t]{-3}{14em}{\raggedright\arraybackslash Analysis 2: W is gestational age at prior PTB, Z is pre-pregnancy BMI} & Coefficient & -- & -- & -- & -- & 0.740\\

\cmidrule{1-7}
 &  & Meis & 0.97 (0.02, 39.78) & 0.66 (0.53, 0.81) & 0.40 (-3.32, 4.11) & 0.835\\

 & \multirow[t]{-2}{4em}{\raggedright\arraybackslash Transport} & PROLONG & 0.82 (0.00, 8622.57) & 1.06 (0.88, 1.28) & -0.25 (-9.51, 9.01) & 0.958\\

\multirow[t]{-3}{14em}{\raggedright\arraybackslash Analysis 2b: W is pre-pregnancy BMI, Z is gestational age at prior PTB} & Coefficient & -- & -- & -- & -- & 0.908\\

\cmidrule{1-7}
 &  & Meis & 0.72 (0.01, 45.23) & 0.66 (0.53, 0.81) & 0.09 (-4.05, 4.24) & 0.965\\

 & \multirow[t]{-2}{4em}{\raggedright\arraybackslash Transport} & PROLONG & 1.31 (0.00, 6981.44) & 1.06 (0.88, 1.28) & 0.21 (-8.37, 8.80) & 0.961\\

\multirow[t]{-3}{14em}{\raggedright\arraybackslash Analysis 3a: W is number of prior PTBs, Z is pre-pregnancy BMI} & Coefficient & -- & -- & -- & -- & 0.898\\

\cmidrule{1-7}
 &  & Meis & 1.01 (0.02, 54.96) & 0.66 (0.53, 0.81) & 0.43 (-3.57, 4.43) & 0.833\\

 & \multirow[t]{-2}{4em}{\raggedright\arraybackslash Transport} & PROLONG & 0.76 (0.00, 3808.24) & 1.06 (0.88, 1.28) & -0.33 (-8.85, 8.19) & 0.940\\

\multirow[t]{-3}{14em}{\raggedright\arraybackslash Analysis 3b: W is pre-pregnancy BMI, Z is number of prior PTBs} & Coefficient & -- & -- & -- & -- & 0.880\\

\cmidrule{1-7}
 &  & Meis & 0.96 (0.64, 1.45) & 0.66 (0.53, 0.81) & 0.38 (-0.07, 0.84) & 0.097\\

 & \multirow[t]{-2}{4em}{\raggedright\arraybackslash Transport} & PROLONG & 0.92 (0.55, 1.55) & 1.06 (0.88, 1.28) & -0.14 (-0.69, 0.41) & 0.627\\

\multirow[t]{-3}{14em}{\raggedright\arraybackslash Analysis 4: W is number of prior PTBs, X is maternal age, Z is all other covariates} & Coefficient & -- & -- & -- & -- & 0.180\\

\cmidrule{1-7}
 &  & Meis & 1.02 (0.74, 1.41) & 0.66 (0.53, 0.81) & 0.44 (0.06, 0.82) & 0.024\\

 & \multirow[t]{-2}{4em}{\raggedright\arraybackslash Transport} & PROLONG & 0.70 (0.51, 0.97) & 1.06 (0.88, 1.28) & -0.41 (-0.78, -0.04) & 0.030\\

\multirow[t]{-3}{14em}{\raggedright\arraybackslash Analysis 5: W is gestational age at prior PTB, X is maternal age, Z is all other covariates} & Coefficient & -- & -- & -- & -- & 0.079\\

\cmidrule{1-7}
 &  & Meis & 1.09 (0.81, 1.45) & 0.66 (0.53, 0.81) & 0.51 (0.15, 0.86) & 0.006\\

 & \multirow[t]{-2}{4em}{\raggedright\arraybackslash Transport} & PROLONG & 0.70 (0.51, 0.96) & 1.06 (0.88, 1.28) & -0.41 (-0.78, -0.04) & 0.028\\

\multirow[t]{-3}{14em}{\raggedright\arraybackslash Analysis 6: W is pre-pregnancy BMI, X is maternal age, Z is all other covariates} & Coefficient & -- & -- & -- & -- & 0.154\\
\bottomrule
\end{tabular}}
\end{table}

\begin{table}[!h]
\centering
\caption{\label{tab:equiv_mult}Supplementary results for equivalence testing for marginal reconcilability on the multiplicative scale for the Meis and PROLONG trials using the binary outcome of delivery before 37 weeks of gestation. Estimate denotes the plug-in estimate of the quantity defining the corresponding proximal equivalence null hypothesis on the multiplicative scale (Section \ref{s:equiv}) and is presented with a 90\% confidence interval computed using the nonparametric bootstrap. 95\% confidence intervals for the reconciliation proportion (RP) were also computed using the nonparametric bootstrap. Reported p-values correspond to equivalence tests with equivalence margins of 0.1 (on the log scale). The LEAD is also reported on the log scale.}
\centering
\resizebox{\ifdim\width>\linewidth\linewidth\else\width\fi}{!}{
\begin{tabular}[t]{>{\raggedright\arraybackslash}p{16em}>{\raggedright\arraybackslash}p{5em}>{\raggedright\arraybackslash}p{9em}>{\raggedright\arraybackslash}p{5em}>{\raggedright\arraybackslash}p{5em}>{\raggedright\arraybackslash}p{9em}}
\toprule
Analysis & Hypothesis & Estimate (CI) & p-value & LEAD & RP (CI)\\
\midrule
 & Meis & -0.35 (-3.70, 3.01) & 0.548 & 3.70 & 0.28 (-3.60, 0.89)\\

 & PROLONG & 0.41 (-8.09, 8.92) & 0.524 & 8.92 & 0.14 (-41.08, 0.98)\\

 & Union & -- & 0.548 & 8.92 & --\\

\multirow{-4}{16em}{\raggedright\arraybackslash Analysis 1: W is number of prior PTBs, Z is gestational age at prior PTB} & Mean & 0.38 (-3.93, 4.69) & 0.543 & 4.69 & 0.21 (-9.43, 0.94)\\
\cmidrule{1-6}
 & Meis & -0.53 (-3.82, 2.76) & 0.585 & 3.82 & -0.11 (-3.28, 0.71)\\

 & PROLONG & 0.32 (-8.21, 8.86) & 0.517 & 8.86 & 0.33 (-39.61, 0.99)\\

 & Union & -- & 0.585 & 8.86 & --\\

\multirow{-4}{16em}{\raggedright\arraybackslash Analysis 1b: W is gestational age at prior PTB, Z is number of prior PTBs} & Mean & 0.43 (-3.92, 4.78) & 0.549 & 4.78 & 0.11 (-9.97, 0.93)\\
\cmidrule{1-6}
 & Meis & -0.46 (-3.70, 2.78) & 0.573 & 3.70 & 0.03 (-4.34, 0.82)\\

 & PROLONG & 0.20 (-7.68, 8.08) & 0.508 & 8.08 & 0.58 (-20.81, 0.99)\\

 & Union & -- & 0.573 & 8.08 & --\\

\multirow{-4}{16em}{\raggedright\arraybackslash Analysis 2: W is gestational age at prior PTB, Z is pre-pregnancy BMI} & Mean & 0.33 (-3.67, 4.33) & 0.538 & 4.33 & 0.31 (-7.47, 0.94)\\
\cmidrule{1-6}
 & Meis & -0.40 (-3.51, 2.72) & 0.562 & 3.51 & 0.17 (-3.06, 0.83)\\

 & PROLONG & 0.25 (-7.52, 8.02) & 0.513 & 8.02 & 0.48 (-27.35, 0.99)\\

 & Union & -- & 0.562 & 8.02 & --\\

\multirow{-4}{16em}{\raggedright\arraybackslash Analysis 2b: W is pre-pregnancy BMI, Z is gestational age at prior PTB} & Mean & 0.32 (-3.60, 4.25) & 0.537 & 4.25 & 0.33 (-7.46, 0.95)\\
\cmidrule{1-6}
 & Meis & -0.09 (-3.57, 3.38) & 0.499 & 3.57 & 0.81 (-1.37, 0.98)\\

 & PROLONG & -0.21 (-7.42, 6.99) & 0.510 & 7.42 & 0.55 (-20.91, 0.99)\\

 & Union & -- & 0.510 & 7.42 & --\\

\multirow{-4}{16em}{\raggedright\arraybackslash Analysis 3a: W is number of prior PTBs, Z is pre-pregnancy BMI} & Mean & 0.15 (-3.65, 3.96) & 0.509 & 3.96 & 0.68 (-3.34, 0.98)\\
\cmidrule{1-6}
 & Meis & -0.43 (-3.79, 2.93) & 0.564 & 3.79 & 0.10 (-3.50, 0.82)\\

 & PROLONG & 0.33 (-6.82, 7.48) & 0.521 & 7.48 & 0.31 (-25.42, 0.98)\\

 & Union & -- & 0.564 & 7.48 & --\\

\multirow{-4}{16em}{\raggedright\arraybackslash Analysis 3b: W is pre-pregnancy BMI, Z is number of prior PTBs} & Mean & 0.38 (-3.27, 4.03) & 0.550 & 4.03 & 0.21 (-7.28, 0.92)\\
\cmidrule{1-6}
 & Meis & -0.38 (-0.77, -0.00) & 0.890 & 0.77 & 0.20 (-2.14, 0.79)\\

 & PROLONG & 0.14 (-0.33, 0.60) & 0.552 & 0.60 & 0.72 (-1.05, 0.96)\\

 & Union & -- & 0.890 & 0.77 & --\\

\multirow{-4}{16em}{\raggedright\arraybackslash Analysis 4: W is number of prior PTBs, X is maternal age, Z is all other covariates} & Mean & 0.26 (-0.01, 0.54) & 0.832 & 0.54 & 0.46 (-0.41, 0.79)\\
\cmidrule{1-6}
 & Meis & -0.44 (-0.76, -0.12) & 0.959 & 0.76 & 0.08 (-0.83, 0.54)\\

 & PROLONG & 0.41 (0.10, 0.72) & 0.949 & 0.72 & 0.15 (-1.47, 0.70)\\

 & Union & -- & 0.959 & 0.76 & --\\

\multirow{-4}{16em}{\raggedright\arraybackslash Analysis 5: W is gestational age at prior PTB, X is maternal age, Z is all other covariates} & Mean & 0.43 (0.16, 0.69) & 0.976 & 0.69 & 0.11 (-0.45, 0.46)\\
\cmidrule{1-6}
 & Meis & -0.51 (-0.80, -0.21) & 0.987 & 0.80 & -0.05 (-0.99, 0.44)\\

 & PROLONG & 0.41 (0.10, 0.72) & 0.952 & 0.72 & 0.14 (-1.62, 0.72)\\

 & Union & -- & 0.987 & 0.80 & --\\

\multirow{-4}{16em}{\raggedright\arraybackslash Analysis 6: W is pre-pregnancy BMI, X is maternal age, Z is all other covariates} & Mean & 0.46 (0.20, 0.72) & 0.988 & 0.72 & 0.04 (-0.51, 0.39)\\
\bottomrule
\end{tabular}}
\end{table}

\begin{table}[!h]
\centering
\caption{\label{tab:relevance_mult}Supplementary measures of proxy relevance for different analyses using naive standard errors (SEs) reported directly from the second-stage regression on the multiplicative scale. Joint-test p-values are from F-tests of the null that all proxy-relevance coefficients are zero.}
\centering
\resizebox{\ifdim\width>\linewidth\linewidth\else\width\fi}{!}{
\begin{tabular}[t]{>{\raggedright\arraybackslash}p{12em}>{\raggedright\arraybackslash}p{11em}>{\raggedright\arraybackslash}p{8em}>{\raggedright\arraybackslash}p{8em}>{\raggedright\arraybackslash}p{8em}}
\toprule
 & Coefficients & Estimate & SE & p-value\\
\midrule
 & $\widehat{\beta}_{u,0}^*$ & 0.620 & 1.095 & 0.571\\

 & $\widehat{\beta}_{u,1}^* - \widehat{\beta}_{u,0}^*$ & -1.014 & 1.374 & 0.461\\

 & $\widehat{\beta}_{au,0}^*$ & 0.197 & 1.415 & 0.889\\

 & $\widehat{\beta}_{au,1}^* - \widehat{\beta}_{au,0}^*$ & 1.778 & 1.789 & 0.320\\

\multirow[t]{-5}{12em}{\raggedright\arraybackslash Analysis 1: W is number of prior PTBs, Z is gestational age at prior PTB} & Joint test & -- & -- & 0.167\\
\cmidrule{1-5}
 & $\widehat{\beta}_{u,0}^*$ & 0.268 & 0.122 & 0.028\\

 & $\widehat{\beta}_{u,1}^* - \widehat{\beta}_{u,0}^*$ & 0.004 & 0.143 & 0.976\\

 & $\widehat{\beta}_{au,0}^*$ & -0.055 & 0.178 & 0.756\\

 & $\widehat{\beta}_{au,1}^* - \widehat{\beta}_{au,0}^*$ & -0.015 & 0.232 & 0.948\\

\multirow[t]{-5}{12em}{\raggedright\arraybackslash Analysis 1b: W is gestational age at prior PTB, Z is number of prior PTBs} & Joint test & -- & -- & 0.000\\
\cmidrule{1-5}
 & $\widehat{\beta}_{u,0}^*$ & -0.662 & 0.286 & 0.021\\

 & $\widehat{\beta}_{u,1}^* - \widehat{\beta}_{u,0}^*$ & 0.164 & 0.369 & 0.657\\

 & $\widehat{\beta}_{au,0}^*$ & 1.268 & 0.400 & 0.002\\

 & $\widehat{\beta}_{au,1}^* - \widehat{\beta}_{au,0}^*$ & -0.859 & 0.486 & 0.077\\

\multirow[t]{-5}{12em}{\raggedright\arraybackslash Analysis 2: W is gestational age at prior PTB, Z is pre-pregnancy BMI} & Joint test & -- & -- & 0.004\\
\cmidrule{1-5}
 & $\widehat{\beta}_{u,0}^*$ & 0.013 & 0.124 & 0.918\\

 & $\widehat{\beta}_{u,1}^* - \widehat{\beta}_{u,0}^*$ & 0.143 & 0.154 & 0.352\\

 & $\widehat{\beta}_{au,0}^*$ & 0.083 & 0.157 & 0.597\\

 & $\widehat{\beta}_{au,1}^* - \widehat{\beta}_{au,0}^*$ & -0.150 & 0.195 & 0.441\\

\multirow[t]{-5}{12em}{\raggedright\arraybackslash Analysis 2b: W is pre-pregnancy BMI, Z is gestational age at prior PTB} & Joint test & -- & -- & 0.253\\
\cmidrule{1-5}
 & $\widehat{\beta}_{u,0}^*$ & 2.060 & 2.707 & 0.447\\

 & $\widehat{\beta}_{u,1}^* - \widehat{\beta}_{u,0}^*$ & -2.528 & 3.310 & 0.445\\

 & $\widehat{\beta}_{au,0}^*$ & 0.339 & 2.880 & 0.906\\

 & $\widehat{\beta}_{au,1}^* - \widehat{\beta}_{au,0}^*$ & 0.291 & 3.545 & 0.934\\

\multirow[t]{-5}{12em}{\raggedright\arraybackslash Analysis 3a: W is number of prior PTBs, Z is pre-pregnancy BMI} & Joint test & -- & -- & 0.171\\
\cmidrule{1-5}
 & $\widehat{\beta}_{u,0}^*$ & -0.065 & 0.177 & 0.715\\

 & $\widehat{\beta}_{u,1}^* - \widehat{\beta}_{u,0}^*$ & 0.100 & 0.199 & 0.617\\

 & $\widehat{\beta}_{au,0}^*$ & -0.057 & 0.188 & 0.760\\

 & $\widehat{\beta}_{au,1}^* - \widehat{\beta}_{au,0}^*$ & -0.092 & 0.222 & 0.677\\

\multirow[t]{-5}{12em}{\raggedright\arraybackslash Analysis 3b: W is pre-pregnancy BMI, Z is number of prior PTBs} & Joint test & -- & -- & 0.200\\
\cmidrule{1-5}
 & $\widehat{\beta}_{u,0}^*$ & 1.840 & 0.532 & 0.001\\

 & $\widehat{\beta}_{u,1}^* - \widehat{\beta}_{u,0}^*$ & -0.605 & 0.646 & 0.349\\

 & $\widehat{\beta}_{au,0}^*$ & -1.356 & 0.743 & 0.068\\

 & $\widehat{\beta}_{au,1}^* - \widehat{\beta}_{au,0}^*$ & -0.036 & 0.927 & 0.969\\

\multirow[t]{-5}{12em}{\raggedright\arraybackslash Analysis 4: W is number of prior PTBs, X is maternal age, Z is all other covariates} & Joint test & -- & -- & 0.000\\
\cmidrule{1-5}
 & $\widehat{\beta}_{u,0}^*$ & 0.047 & 0.144 & 0.742\\

 & $\widehat{\beta}_{u,1}^* - \widehat{\beta}_{u,0}^*$ & 0.177 & 0.162 & 0.275\\

 & $\widehat{\beta}_{au,0}^*$ & 0.130 & 0.184 & 0.479\\

 & $\widehat{\beta}_{au,1}^* - \widehat{\beta}_{au,0}^*$ & -0.350 & 0.219 & 0.110\\

\multirow[t]{-5}{12em}{\raggedright\arraybackslash Analysis 5: W is gestational age at prior PTB, X is maternal age, Z is all other covariates} & Joint test & -- & -- & 0.031\\
\cmidrule{1-5}
 & $\widehat{\beta}_{u,0}^*$ & 0.099 & 0.036 & 0.006\\

 & $\widehat{\beta}_{u,1}^* - \widehat{\beta}_{u,0}^*$ & -0.019 & 0.044 & 0.666\\

 & $\widehat{\beta}_{au,0}^*$ & -0.183 & 0.059 & 0.002\\

 & $\widehat{\beta}_{au,1}^* - \widehat{\beta}_{au,0}^*$ & 0.069 & 0.076 & 0.369\\

\multirow[t]{-5}{12em}{\raggedright\arraybackslash Analysis 6: W is pre-pregnancy BMI, X is maternal age, Z is all other covariates} & Joint test & -- & -- & 0.001\\
\bottomrule
\end{tabular}}
\end{table}

\end{document}